\documentclass[apj]{emulateapj-rtx4}

\usepackage{times,epsfig,natbib,amssymb,amsmath,graphics,longtable,lscape,threeparttable}

\usepackage{rotating,hyperref}

\def \msun{\ifmmode{{\rm\ M}_\odot}\else{${\rm\ M}_\odot$}\fi}
\newcommand{\hei}{He\,{\sc i}{}}

\newcommand{\ha}{H$\alpha${}}

\slugcomment{Accepted for publication in ApJ journal}

\shorttitle{SNe~II $V$-band light-curves}
\shortauthors{Anderson et al.}

\begin{document}


\title{Characterizing the $V$-band light-curves of hydrogen-rich\\ 
type II supernovae\footnote{B\lowercase{ased on observations obtained with the
du-}P\lowercase{ont and} S\lowercase{wope telescopes at} LCO, \lowercase{and the} 
S\lowercase{teward}
O\lowercase{bservatory's} CTIO60, SO90 \lowercase{and} CTIO36 \lowercase{telescopes.}}}


\author{Joseph P. Anderson\altaffilmark{1,2}, 
Santiago Gonz\'alez-Gait\'an\altaffilmark{1}, 
Mario Hamuy\altaffilmark{1,3},
Claudia P. Guti\'errez\altaffilmark{3,1},
Maximilian D. Stritzinger\altaffilmark{4},
Felipe Olivares E.\altaffilmark{5}, 
Mark M. Phillips\altaffilmark{6},
Steve Schulze\altaffilmark{7},
Roberto Antezana\altaffilmark{1},
Luis Bolt\altaffilmark{8},
Abdo Campillay\altaffilmark{6},
Sergio Castell\'on\altaffilmark{6},
Carlos Contreras\altaffilmark{4},
Thomas de Jaeger\altaffilmark{3,1},
Gast\'on Folatelli\altaffilmark{9},
Francisco F\"orster\altaffilmark{1},
Wendy L. Freedman\altaffilmark{10},
Luis Gonz\'alez\altaffilmark{1},
Eric Hsiao\altaffilmark{6},
Wojtek Krzemi\'nski\altaffilmark{11},
Kevin Krisciunas\altaffilmark{12},
Jos\'e Maza\altaffilmark{1},
Patrick McCarthy\altaffilmark{10},
Nidia I. Morrell\altaffilmark{6},
Sven E. Persson\altaffilmark{10},
Miguel Roth\altaffilmark{6},
Francisco Salgado\altaffilmark{13},
Nicholas B. Suntzeff\altaffilmark{12},
Joanna Thomas-Osip\altaffilmark{6}
}
\email{janderso@eso.org}
\altaffiltext{1}{Departamento de Astronom\'ia, Universidad de Chile, Casilla 36-D, 
Santiago, Chile}
\altaffiltext{2}{European Southern Observatory, Alonso de Cordova 3107, Vitacura, Santiago, Chile}
\altaffiltext{3}{Millennium Institute of Astrophysics, Casilla 36-D, 
Santiago, Chile}
\altaffiltext{4}{Department of Physics and Astronomy, Aarhus University, Ny
Munkegade 120, DK-8000 Aarhus C, Denmark}
\altaffiltext{5}{Departamento de Ciencias Fisicas, Universidad Andres Bello, Avda. Republica 252, Santiago, Chile}
\altaffiltext{6}{Carnegie Observatories, Las Campanas Observatory, Casilla 601, La Serena, Chile}
\altaffiltext{7}{Instituto de Astrof\'isica, Facultad de F\'isica, Pontif\'icia
Universidad Cat\'olica de Chile, 306, Santiago 22, Chile}
\altaffiltext{8}{Argelander Institut f\"{u}r Astronomie, Universit\"{a}t Bonn, Auf dem Hügel 71, D-53111 Bonn, Germany}
\altaffiltext{9}{Institute for the Physics and Mathematics of the Universe (IPMU), University of Tokyo, 5-1-5 Kashiwanoha, Kashiwa, Chiba 277-8583, Japan}
\altaffiltext{10}{Observatories of the Carnegie Institution for Science, Pasadena, CA 91101, USA}
\altaffiltext{11}{N. Copernicus Astronomical Center, ul. Bartycka 18, 00-716 Warszawa,
Poland}
\altaffiltext{12}{George P. and Cynthia Woods Mitchell Institute for
  Fundamental Physics and Astronomy, Department of Physics and Astronomy,
  Texas A\&M University, College Station, TX 77843, USA}
\altaffiltext{13}{1 Leiden Observatory, Leiden University, P.O. Box 9513, NL-2300 RA Leiden, The Netherlands}

\begin{abstract}
We present an analysis of the diversity of
$V$-band light-curves of hydrogen-rich type II supernovae. 
Analyzing a sample of 116 supernovae, 
several magnitude measurements are defined, together with decline rates at
different epochs, and time durations of different phases.
It is found that magnitudes measured at maximum light correlate more strongly with decline rates 
than those measured 
at other epochs: brighter supernovae at maximum generally have faster declining light-curves
at all epochs.
We find a relation between the decline rate during the `plateau' phase and
peak magnitudes, which has a dispersion
of 0.56 magnitudes, offering the prospect of using type II supernovae as
purely photometric distance indicators.   
Our analysis suggests that the type II 
population spans a continuum from low-luminosity events which have flat light-curves during the
`plateau' stage, through to the brightest events which decline much faster. 
A large range in optically thick phase durations is observed, implying a range in progenitor envelope
masses at the epoch of explosion. During the radioactive tails, we find many 
supernovae with faster declining light-curves than
expected from full trapping of radioactive emission, implying low mass ejecta. 
It is suggested that the main driver of light-curve
diversity is the extent of hydrogen envelopes retained before explosion.
Finally, a new
classification scheme is introduced where hydrogen-rich events are typed as simply `SN~II' with an `$s_2$'
value giving the decline rate during the `plateau' phase, indicating its
morphological type.
\end{abstract}


\keywords{(stars:) supernovae: general}



\section{Introduction}
Supernovae (SNe) were initially classified into types I and
II by \cite{min41}, dependent on the absence or presence of hydrogen in
their spectra. It is now commonly assumed that hydrogen-rich type II SNe (SNe~II
henceforth) arise from the core-collapse of massive ($>$8-10\msun) stars that
explode with a significant fraction of their hydrogen envelopes retained. 
A large diversity in the photometric and
spectroscopic properties of SNe~II is observed, which leads to many questions 
regarding the physical
characteristics of their progenitor scenarios and explosion properties.\\
\indent The most abundant of the SNe~II class (see e.g.\ \citealt{li11} for
rate estimates) are the SNe~IIP which show a long duration plateau in
their photometric evolution, understood to be the consequence of the hydrogen
recombination wave propagating back through the massive SN ejecta. SNe~IIL
are so called due to their `linear' declining light-curves (see
\citealt{bar79} for the initial separation of hydrogen-rich events into these
two sub-classes). A further two sub-classes exist in the form of 
SNe~IIn and SNe~IIb. SNe~IIn show narrow emission lines within their spectra
\citep{sch90}, but present a large diversity of photometric and spectral
properties (see e.g.\ \citealt{kie12,tad13}), which clouds interpretations of
their progenitor systems and how they link to the `normal' SNe~II
population. (We note that a progenitor detection of the SN~IIn:
2005gl does exist, and points
towards a very massive progenitor: \citealt{gal09}, at least in that particular case).
SNe~IIb appear to be observationally transitional events 
as at early times they show hydrogen features, while
later such lines disappear and their spectra
appear similar to SNe~Ib \citep{fil93}. These events appear to show more
similarities with the hydrogen deficient SN~Ibc objects (see \citealt{arc12},
Stritzinger et al. in prep.). As these last two sub-types are distinct from the classical
hydrogen-rich SNe~II, they are no longer discussed in the current paper. An even rarer sub-class
of type II events, are those classed as similar to SN~1987A. While SN~1987A is generally 
referred to as a type IIP, its light-curve has a peculiar shape (see e.g.
\citealt{ham88}), making it distinct from classical type IIP or IIL. A number of `87A-like' events were identified in the
current sample and removed, with those from the CSP being published in \cite{tad12} (see also \citealt{kle11}, and 
\citealt{pas12}, for detailed investigations of other 87A-like events).\\
\indent The progenitors of SNe~II are generally assumed to be stars of ZAMS mass in 
excess of 8-10\msun, which have retained a significant fraction of their hydrogen envelopes before explosion.
Indeed, initial light-curve modeling of SNe~IIP implied that red-supergiant progenitors with massive
hydrogen envelopes were required to reproduce typical light-curve morphologies (\citealt{gra71,che76,fal77}). 
These assumptions and predictions have been shown to be consistent with detections of 
progenitor stars on pre-explosion images, where progenitor detections of SNe~IIP have been constrained to be
red supergiants in the 8-20\msun\ ZAMS range (see \citealt{sma09} for a review, and \citealt{van12b} for a 
recent example). It has also been suggested that SN~IIL progenitors may be more massive than their type IIP counterparts 
(see \citealt{eli10,eli11}).\\
\indent Observationally, hydrogen-rich SNe~II are characterized by showing
P-Cygni hydrogen features in their spectra\footnote{While as shown in \cite{sch96} there are a number
of SNe~II which show very weak \ha\ absorption (which tend to be of the type IIL class), the vast majority 
of events do evolve to have significant absorption features. Indeed this will be shown to be
the case for the current sample in Guti\'errez et al. (submitted).}, while displaying a range of
light-curve morphologies and spectral profiles. Differences that exist between the photometric evolution
within these SNe are most likely related to the mass extent and density profile
of the hydrogen envelope of the progenitor star at the time of explosion. 
In theory, SNe with less prominent and shorter
`plateaus' (historically classified as SNe~IIL) are believed to have smaller hydrogen
envelope masses at the epoch of explosion (\citealt{pop93}, also see \citealt{lit83} for
generalized model predictions of relations between different SNe~II properties). 
Further questions such as how the nickel mass and extent
of its mixing affects e.g.\ the plateau luminosity and length have also been
posed (e.g. \citealt{kas09,ber11}).\\
\indent While some further classes of SN~II events with similar properties have been
identified (e.g. sub-luminous SNe~IIP, \citealt{pas04,spi14}; luminous SNe~II,
\citealt{ins13}; `intermediate' events, \citealt{gan13,tak14}), analyses of statistical samples of SN~II light-curves are,
to date uncommon in the literature, with researchers often publishing
in-depth studies of individual SNe. While this affords detailed knowledge of
the transient evolution of certain events, and thus their explosion and
progenitor properties, often it is difficult to put each event into
the overall context of the SNe~II class, and how events showing peculiarities
relate.\\
\indent Some exceptions to the above statement do however exist: 
\cite{psk67} compiled photographic plate SN photometry for all supernova types, finding in the 
case of 
SNe~II (using a sample of 18 events), that the rate of decline appeared to correlate with peak brightness,
together with the time required to observe a `hump' in the light-curve (see also \citealt{psk78}).
All available SN~II photometry at the time of publication (amounting to 23 SNe) was presented by
\cite{bar79}, who were the first to separate events into SNe~IIP and SNe~IIL, on the
basis of $B$-band light-curve morphology. 
\cite{you89} discussed possible differences in the $B$-band absolute
magnitudes of different SNe~II, analyzing a sample of 15 events. A large `Atlas' of historical photometric data
of 51 SNe~II was first presented and then analyzed by \cite{pat93} and \cite{pat94}
respectively. These data (with significant photometry available in the $B$ and $V$ bands), revealed a number of photometric and
spectroscopic correlations: more steeply
declining SNe~II appeared to be more luminous events, and also of bluer
colors than their plateau companions. Most recently, \cite{arc12} published
an analysis of $R$-band light-curves (21 events, including 3 SNe~IIb), concluding that SNe~IIP and SNe~IIL
are distinct events which do not show a continuum of
properties, hence possibly pointing towards distinct progenitor
populations. We also note that bolometric light-curves of 
a significant fraction of the current sample were presented and analyzed by \cite{ber13}, 
where similar light-curve characterization to that outlined below was presented.\\ 
\indent The aim of the current paper is to present a statistical analysis
of SN~II $V$-band light-curve properties that will significantly add weight to the
analysis thus far presented in the literature, while at the same time 
introduce new
nomenclature to help the community define SN~II photometric properties in a
standardized way. Through this we hope to increase 
the underlying physical understanding of SNe~II.
To proceed with this aim, we present analysis of the $V$-band
light-curves of 116 SNe~II, obtained over the last three decades. We
define a number of absolute magnitudes, light-curve decline rates, and time
epochs, and use
these to search for correlations in order to characterize the diversity of
events.\\
\indent The paper is organized as follows: in the following Section we outline the data sample,
and briefly summarize the reduction and photometric procedures employed. 
In \S\ 3 we define the photometric
properties for measurement, outline our explosion epoch, extinction, and error estimation
methods, and present light-curve fits to SN~II photometry. 
In \S\ 4 results on various correlations between photometric properties, together with their
distributions are presented. In \S\ 5 we discuss the
most interesting of these correlations in detail, and try to link these to
physical understanding of the SN~II phenomenon. Finally, several concluding
remarks are listed.\\ 
\indent In addition, an appendix is included where 
detailed light-curves (together with their derived parameters) 
and further analysis and figures not included in the main body of
the manuscript are presented. The keen reader is encouraged to delve into
those pages for a full understanding of our analysis and results.

\section{Data sample}
The sample of $V$-band light-curves is compiled from data obtained between 1986
and 2009 from 5 different systematic SN follow-up programs. These are: 1) the
Cerro Tololo SN program (CT, PIs: Phillips \&\ Suntzeff, 1986-2003); 2) the
Cal\'an/Tololo SN program (PI: Hamuy 1989-1993); 3) the Optical and Infrared
Supernova Survey 
(SOIRS, PI: Hamuy, 1999-2000);
4) the Carnegie Type II Supernova Program (CATS, PI: Hamuy, 2002-2003); and 5) the
Carnegie Supernova Project (CSP, \citealt{ham06}, PIs: Phillips \&\ Hamuy, 2004-2009).
SN~II photometry for these
samples have in general not yet been published. These follow-up campaigns
concentrated on obtaining high cadence and quality light-curves and spectral
sequences of nearby SNe (z$<$0.05). The 116 SNe from those campaigns which form the
current sample are listed in Table 1 along with host
galaxy information. Observations were obtained with a range of 
telescopes and instruments, but data were processed in very similar ways, as
outlined below. SNe types IIn and IIb were excluded from the sample (the CSP SN~IIn sample was recently 
published in \citealt{tad13}). This exclusion was based on information from sample spectra and light-curves. 
Initial classification references are listed in Table 3, together with details of
sample spectroscopy used to confirm these initial classifications. Optical 
spectroscopy for the currently discussed SNe will be presented in a future publication. In
addition, SNe showing similar photometric behavior to SN~1987A are also removed from the 
sample, based on their light-curve morphologies. We expect contamination from unidentified (because of 
insufficient constraints on their transient behavior) SNe types IIb, IIn and 87A-like
events into the current sample
to be negligible or non-existent. This is expected because of a) the data quality cuts
which have been used to exclude non-normal SN~II events, and b) the intinsic rarity of those
sub-types.
Finally, a small number of events which are likely to be of the hydrogen-rich 
group analyzed here are also excluded because of a combination of insufficient photometry, a lack of spectral information
and unconstrained explosion epochs.\\
\indent To proceed with initial characterization of the diversity of SN~II
presented in this paper, we 
chose to investigate $V$-band light-curve morphologies. This is due to a
number of factors. Firstly, from a historical point of view, the $V$ band has
been the most widely used filter for SN studies, and hence an investigation
of the behavior at these wavelengths facilitates easy comparisons with
other works. Secondly, the SNe within our sample have better coverage in the
$V$ band than other filters, therefore we are able to more easily measure the
parameters which we wish to investigate. Finally, it has been suggested \citep{ber09} 
that the $V$-band light-curve is a reasonable
proxy for the bolometric light-curve (with exception at
very early times)\footnote{However, it is noted that those authors did not analyze photometry obtained 
with $R$-band filters.}.\\
\indent We note that the SNe currently discussed were discovered by many different searches, which
were generally of targeted galaxies. Hence, this sample is heterogeneous
in nature. Follow-up target selection for the various programs was essentially determined by a SN being
discovered that was bright enough to be observed using the follow-up telescopes: 
i.e. in essence magnitude limited follow-up campaigns. 
The SN and host galaxy samples are further characterized in the appendix.

\begin{table*}
\centering
\begin{threeparttable}
\caption{SN~II sample}
\begin{tabular}[t]{ccccccc}
\hline
SN & Host galaxy & Recession velocity (km s$^{-1}$) & Hubble type & $M_{\rm B}$ (mag)& $E(B-V)_{\rm MW}$ (mag) & Campaign \\
\hline	
\hline
1986L	& NGC 1559	& 1305 & SBcd  & --21.3& 0.026  & CT\\
1991al	& anon          & 4575$^1$ &?  &--18.8& 0.054  & Cal\'an/Tololo\\
1992ad	& NGC 4411B	& 1272 & SABcd & --18.3&0.026  & Cal\'an/Tololo\\
1992af	&ESO 340-G038	& 5541 & S  & --19.7   & 0.046  & Cal\'an/Tololo\\
1992am	&MCG -01-04-039	&14397$^1$ & S &--21.4& 0.046  & Cal\'an/Tololo\\
1992ba	&NGC 2082	&1185  & SABc  & --18.0&0.051  & Cal\'an/Tololo\\
1993A	&anon	        &8790$^1$  & ? &$\cdots$& 0.153  & Cal\'an/Tololo\\
1993K	&NGC 2223	&2724  & SBbc  & --20.9&0.056  & Cal\'an/Tololo\\
1993S	&2MASX J22522390-4018432&9903 & S&--20.6&0.014 & Cal\'an/Tololo\\
1999br	&NGC 4900	&960   & SBc   & --19.4&0.021  & SOIRS\\
1999ca	&NGC 3120	&2793  & Sc    & --20.4&0.096  & SOIRS\\
1999cr	&ESO 576-G034	&6069$^1$  & S/Irr& --20.4& 0.086  & SOIRS\\
1999eg	&IC 1861	&6708  & SA0   &--20.9& 0.104  & SOIRS\\
1999em	&NGC 1637	&717   & SABc  & --19.1&0.036  & SOIRS\\
0210$^*$	&MCG +00-03-054	&15420 & ? & --21.2 & 0.033  & CATS\\
2002ew	&NEAT J205430.50-000822.0&8975  & ?&$\cdots$ &0.091  & CATS\\
2002fa	&NEAT J205221.51+020841.9&17988 & ?  &$\cdots$ & 0.088  & CATS\\
2002gd	&NGC 7537	&2676  & SAbc  &--19.8& 0.059  & CATS\\
2002gw	&NGC 922	&3084  & SBcd  & --20.8&0.017  & CATS\\
2002hj	&NPM1G +04.0097	&7080  & ? & $\cdots$  & 0.102  & CATS\\
2002hx	&PGC 023727	&9293  & SBb   & $\cdots$ &0.048  & CATS\\
2002ig	&anon	        &23100$^2$ & ? & $\cdots$  & 0.034  & CATS\\
2003B	&NGC 1097	&1272  & SBb   & --21.4&0.024  & CATS\\
2003E	&MCG -4-12-004	&4470$^3$  &Sbc& --19.7 & 0.043  & CATS\\
2003T	&UGC 4864	&8373  & SAab  & --20.8&0.028  & CATS\\
2003bl	&NGC 5374	&4377$^3$  & SBbc& --20.6 & 0.024  & CATS\\
2003bn	&2MASX J10023529-2110531&3828  & ?& --17.7  & 0.057  & CATS\\
2003ci	&UGC 6212	&9111  & Sb    & --21.8&0.053  & CATS\\
2003cn	&IC 849	        &5433$^3$  &SABcd& --20.4& 0.019  & CATS\\
2003cx	&NEAT J135706.53-170220.0&11100 & ? & $\cdots$ & 0.083  & CATS\\
2003dq	&MAPS-NGP O4320786358      &13800 & ? &$\cdots$  & 0.016  & CATS\\
2003ef	&NGC 4708	&4440$^3$  & SAab  &--20.6& 0.041  & CATS\\
2003eg	&NGC 4727	&4388$^1$  & SABbc & --22.3&0.046  & CATS\\
2003ej	&UGC 7820	&5094  & SABcd & --20.1&0.017  & CATS\\
2003fb	&UGC 11522	&5262$^3$  & Sbc   &--20.9& 0.162  & CATS\\
2003gd	&M74	        &657   & SAc   & --20.6&0.062  & CATS\\
2003hd	&MCG -04-05-010	&11850 & Sb    & --21.7&0.011  & CATS\\
2003hg	&NGC 7771	&4281  & SBa   & --21.4&0.065  & CATS\\
2003hk	&NGC 1085	&6795  & SAbc &--21.3 & 0.033  & CATS\\
2003hl	&NGC 772	&2475  & SAb   & --22.4&0.064  & CATS\\
2003hn	&NGC 1448	&1170  & SAcd  & --21.1&0.013  & CATS\\
2003ho	&ESO 235-G58	&4314  & SBcd  &--19.8& 0.034  & CATS\\
2003ib	&MCG -04-48-15	&7446  & Sb    & --20.8&0.043  & CATS\\
2003ip	&UGC 327	&5403  & Sbc   & --19.4&0.058  & CATS\\
2003iq	&NGC 772	&2475  & SAb   & --22.4&0.064  & CATS\\
2004dy	&IC 5090	&9352  & Sa    & --20.9&0.045  & CSP\\
2004ej	&NGC 3095	&2723  & SBc   &--21.6& 0.061  & CSP\\
2004er	&MCG -01-7-24	&4411  & SAc   &--20.2& 0.023  & CSP\\
2004fb	&ESO 340-G7	&6100  & S     & --20.9&0.056  & CSP\\
2004fc	&NGC 701	&1831  & SBc   & --19.5&0.023  &  CSP\\  
2004fx	&MCG -02-14-3 	&2673  & SBc   &$\cdots$& 0.090  &  CSP\\
2005J	&NGC 4012	&4183  & Sb    &--20.4& 0.025  &  CSP\\
2005K	&NGC 2923	&8204  & ?   & --19.6&0.035  &  CSP\\
2005Z	&NGC 3363	&5766  & S     & --20.5&0.025  &  CSP\\
2005af	&NGC 4945	&563   & SBcd  & --20.5&0.156  &  CSP\\
2005an	&ESO 506-G11	&3206  & S0    &--18.6& 0.083  &  CSP\\
2005dk	&IC 4882	&4708  & SBb   & --19.8&0.043  &  CSP\\
2005dn	&NGC 6861	&2829  & SA0   & --21.0&0.048  &  CSP\\
2005dt	&MCG -03-59-6	&7695  & SBb  &--20.9 & 0.025  &  CSP\\
2005dw	&MCG -05-52-49	&5269  & Sab  &--21.1 & 0.020  &  CSP\\
2005dx	&MCG -03-11-9	&8012  & S   &--20.8  & 0.021  &  CSP\\
2005dz	&UGC 12717	&5696  & Scd   & --19.9&0.072  &  CSP\\
2005es	&MCG +01-59-79 	&11287 & S    &--21.1 & 0.076  &  CSP\\
2005gk	&2MASX J03081572-0412049&8773  & ?& $\cdots$  & 0.050  &  CSP\\
\hline
\hline
\end{tabular}
\begin{tablenotes}
\item$^*$This event was never given an official SN name, hence it is referred to as listed.
\end{tablenotes}
\end{threeparttable}
\end{table*}

\setcounter{table}{0}
\begin{table*}
\centering
\begin{threeparttable}
\caption{SN~II sample --\textit{Continued}}
\begin{tabular}[t]{ccccccc}
\hline
SN & Host galaxy & Recession velocity (km s$^{-1}$) & Hubble type &$M_{\rm B}$ (mag)& $E(B-V)_{\rm MW}$ (mag)& Campaign \\
\hline	
\hline
2005hd	&anon	        &8778$^2$  & ? &$\cdots$  & 0.054  &  CSP\\
2005lw	&IC 672	        &7710  & ?  &$\cdots$ & 0.043  &  CSP\\
2005me	&ESO 244-31 	&6726  & SAc &--21.4  & 0.022  &  CSP\\
2006Y	&anon	        &10074$^2$ & ? &$\cdots$ & 0.115  &  CSP\\
2006ai	&ESO 005-G009	&4571$^1$  & SBcd &--19.2 & 0.113  &  CSP\\
2006bc	&NGC 2397	&1363  & SABb  & --20.9&0.181  &  CSP\\
2006be	&IC 4582	&2145  & S     & --18.7&0.026  &  CSP\\
2006bl	&MCG +02-40-9	&9708  & ?   & --20.9&0.045  &  CSP\\
2006ee	&NGC 774	&4620  & S0    & --20.0&0.054  &  CSP\\
2006it	&NGC 6956	&4650  & SBb   & --21.2&0.087  &  CSP\\
2006iw	&2MASX J23211915+0015329&9226&?&--18.3&0.044 &  CSP\\
2006qr	&MCG -02-22-023	&4350  & SABbc & --20.2&0.040  &  CSP\\
2006ms	&NGC 6935	&4543  & SAa   & --21.3&0.031  &  CSP\\
2007P	&ESO 566-G36 	&12224 & Sa   &--21.1 & 0.036  &  CSP\\
2007U	&ESO 552-65	&7791  & S   &--20.5  & 0.046  &  CSP\\
2007W	&NGC 5105	&2902  & SBc &--20.9  & 0.045  &  CSP\\
2007X	&ESO 385-G32	&2837  & SABc &--20.5 & 0.060  &  CSP\\
2007aa	&NGC 4030	&1465  & SAbc  & --21.1&0.023  &  CSP\\
2007ab	&MCG -01-43-2 	&7056  & SBbc &--21.5 & 0.235  &  CSP\\
2007av	&NGC 3279	&1394  & Scd   & --20.1&0.032  &  CSP\\
2007hm	&SDSS J205755.65-072324.9&7540&?&$\cdots$&0.059&  CSP\\
2007il	&IC 1704	&6454  & S  &--20.7   & 0.042  &  CSP\\
2007it	&NGC 5530	&1193  & SAc& --19.6  & 0.103  &  CSP\\
2007ld	&anon&7499$^1$&?&$\cdots$&0.081&  CSP\\
2007oc	&NGC 7418	&1450  & SABcd &--19.9& 0.014  &  CSP\\
2007od	&UGC 12846	&1734  & Sm   &--16.6 & 0.032  &  CSP\\
2007sq	&MCG -03-23-5	&4579  & SAbc &--22.2 & 0.183  &  CSP\\
2008F	&MCG -01-8-15	&5506 & SBa & --20.5&0.044 & CSP\\
2008K	&ESO 504-G5	&7997 & Sb &--20.7 & 0.035 & CSP\\
2008M	&ESO 121-26	&2267 & SBc & --20.4&0.040 & CSP\\
2008W	&MCG -03-22-7 	&5757 & Sc &--20.7 & 0.086 & CSP\\
2008ag	&IC 4729	&4439  & SABbc & --21.5&0.074  &  CSP\\
2008aw	&NGC 4939	&3110  & SAbc  & --22.2&0.036  &  CSP\\
2008bh	&NGC 2642	&4345  & SBbc  & --20.9&0.020  &  CSP\\
2008bk	&NGC 7793	&227   & SAd   & --18.5&0.017  &  CSP\\
2008bm	&CGCG 071-101	&9563  & Sc  &--19.5  & 0.023  &  CSP\\
2008bp	&NGC 3095	&2723  & SBc &--21.6  & 0.061  &  CSP\\
2008br	&IC 2522	&3019  & SAcd &--20.9 & 0.083  &  CSP\\
2008bu	&ESO 586-G2	&6630  & S   &--21.6  & 0.376  &  CSP\\ 
2008ga	&LCSB L0250N	&4639 & ? &$\cdots$& 0.582 & CSP\\
2008gi	&CGCG 415-004	&7328 & Sc &--20.0 & 0.060 & CSP\\
2008gr	&IC 1579	&6831 & SBbc& --20.6&0.012 & CSP\\
2008hg	&IC 1720	&5684 & Sbc & --20.9&0.016 & CSP\\
2008ho	&NGC 922	&3082 & SBcd& --20.8&0.017 & CSP\\
2008if	&MCG -01-24-10	&3440 & Sb &--20.4 & 0.029 & CSP\\
2008il	&ESO 355-G4	&6276 & SBb & --20.7&0.015 & CSP\\
2008in	&NGC 4303	&1566 & SABbc&--20.4&0.020 & CSP\\
2009N	&NGC 4487	&1034 & SABcd& --20.2&0.019& CSP\\
2009ao	&NGC 2939	&3339 & Sbc & --20.5&0.034 & CSP\\
2009au	&ESO 443-21	&2819 & Scd & --19.9&0.081 & CSP\\
2009bu	&NGC 7408	&3494 & SBc & --20.9&0.022 & CSP\\
2009bz	&UGC 9814	&3231 & Sdm & --19.1&0.035 & CSP\\
\hline
\hline
\end{tabular}
\begin{tablenotes}
\item$^1$Measured using our own spectra.
\item$^2$Taken from the Asiago supernova catalog: \url{http://graspa.oapd.inaf.it/} \citep{bar99}.
\item$^3$From our own data \citep{jon09}.
\item\textit{Observing campaigns:} CT=Cerro Tololo Supernova Survey; Cal\'an/Tololo=Cal\'an/Tololo Supernova
Program; SOIRS=Supernova Optical and Infrared Survey; CATS=Carnegie Type II
Supernova Survey; CSP=Carnegie Supernova Project.
\end{tablenotes}
\setcounter{table}{0}
\tablecomments{SNe and host galaxy information. In the first column the SN
  name, followed by its host galaxy are listed. In column 3 we list the host
  galaxy heliocentric recession velocity. These are taken from the Nasa Extragalactic
  Database (NED: \url{http://ned.ipac.caltech.edu/}) unless indicated by a
  superscript (sources in table notes). In columns 4 and 5 we list the host
  galaxy morphological Hubble types (from NED) and their absolute $B$-band magnitudes (taken from the LEDA database: \url{http://leda.univ-lyon1.fr/})
  respectively.
  In column 6 we list the reddening due to dust in our Galaxy \citep{sch11}
taken from NED. Finally, the observing campaign from which each SN was
  taken are given in column 7, and acronyms are listed in the table notes.}
\end{threeparttable}
\end{table*}

\subsection{Data reduction and photometric processing}
A detailed description of the data reduction, host galaxy subtraction and
photometric processing for all SNe discussed, awaits the full data
release. 
Here we briefly summarize the general techniques used to
obtain host galaxy-subtracted, photometrically-calibrated $V$-band
light-curves for 116
SNe~II.\\
\indent Data processing techniques for CSP photometry were first outlined in \cite{ham06}
then fully described in \cite{con10} and \cite{str11}. The reader is referred to those
articles for additional information. Note, that those details are also relevant to the data obtained
in follow-up campaigns prior to CSP (listed above), which were
processed in a very similar fashion. One important 
difference between CSP and prior data is that the CSP magnitudes are in the 
natural system of the Swope telescope (located at Las Campanas Observatory, LCO), whereas previous data are calibrated to the 
Landolt standard system.
Briefly, $V$-band data were reduced
through a sequence of: bias subtractions, flat-field corrections, application
of a linearity correction and an exposure time correction for a shutter time
delay. Since SN measurements can be potentially affected by the underlying
light of their host galaxies, we exercised great care in subtracting 
late-time galaxy images from SN frames (see e.g. 
\citealt{ham93}).
This was achieved through obtaining host galaxy template images more than a year after the last
follow-up image, where templates were checked for SN residual flux (in the case of detected 
SN emission, additional templates were obtained at a later date). In the case of the CSP 
sample the majority of these images was obtained with the du-Pont telescope (the Swope telescope
was used to obtain the majority of follow-up photometry), and templates which were used for final
subtractions were always taken under seeing conditions either matching or exceeding those of science frames.
To proceed with host galaxy subtractions, the template images were geometrically transformed to each 
individual science frame, then convolved to match the point-spread functions, and finally scaled in flux. 
The template images were then subtracted from a circular region around the SN position on each science frame. 
This process 
was outlined in detail in \S\ 4.1 of \cite{con10} as applied to the CSP SN~Ia sample, where further 
discussion can be found on
the extent of possible systematic errors incurred from the procedure (which were found to be less
than 0.01 mag, and are not included in the photometric errors, also see \citealt{fol10}). A very 
similar procedure to the above was employed  
for the data obtained prior to CSP. \\
\indent SN magnitudes were then obtained differentially with respect to a set
of local sequence stars, where absolute photometry of local sequences was
obtained using our own photometric standard observations. 
$V$-band photometry for three example SNe is shown in Table 2, and the 
complete sample of $V$-band photometry can be downloaded from 
\scriptsize
\url{http://www.sc.eso.org/~janderso/SNII_A14.tar.gz} \normalsize also available at \scriptsize
\url{http://csp.obs.carnegiescience.edu/data/)},\normalsize 
or 
requested from the author. The .tar file also contains a list of all epochs and magnitudes
of upper limits for non-detections prior to SN discovery, together with a results file with all 
parameters from Table 6, plus other additional values/measurements made in the process of our
analysis.
Full multi-color optical and
near-IR photometry, together with that of local sequences,
will be published for all SNe included in this sample in the near future.\\ 

\begin{table}
\centering
\caption{SN~II photometry}
\begin{tabular}[t]{cccccc}
\hline
SN & JD date &$V$-band magnitude & Error \\
\hline	
\hline
1999ca&2451305.50 &15.959 &0.015\\
&2451308.56 &16.067 &0.015\\
&2451309.51 &16.108 &0.008\\
&2451313.47 &16.244 &0.015\\
&2451317.52 &16.371 &0.015\\
&2451317.54 &16.392 &0.015\\
&2451319.46 &16.425 &0.015\\
&2451321.46 &16.469 &0.015\\
&2451322.50 &16.510 &0.009\\
&2451327.46 &16.592 &0.015\\
&2451329.46 &16.612 &0.015\\
&2451331.46 &16.636 &0.015\\
&2451335.45 &16.701 &0.015\\
&2451340.46 &16.819 &0.015\\
&2451345.46 &16.868 &0.015\\
&2451351.47 &16.984 &0.015\\
&2451355.46 &17.100 &0.015\\
&2451464.86 &20.685 &0.141\\
&2451478.86 &20.857 &0.092\\
&2451481.83 &21.217 &0.110\\
&2451484.85 &21.097 &0.057\\
&2451488.83 &21.293 &0.043\\
&2451493.85 &21.291 &0.071\\
&2451499.86 &21.327 &0.065\\
&2451506.85 &21.393 &0.114\\
&&&\\
\hline
2003dq&2452754.6 &19.800 &0.019\\
&2452764.6 &20.241 &0.036\\
&2452777.6 &20.417 &0.083\\
&2452789.6 &20.645 &0.087\\
&2452794.5 &21.097 &0.046\\
&&&\\
\hline
2008aw &2454530.79 &15.776 &0.010\\
&2454538.70 &15.851 &0.006\\
&2454539.75 &15.904 &0.007\\
&2454540.76 &15.941 &0.008\\
&2454541.83 &15.958 &0.007\\
&2454543.80 &16.032 &0.007\\
&2454545.82 &16.100 &0.009\\
&2454552.83 &16.329 &0.006\\
&2454558.76 &16.498 &0.010\\
&2454560.78 &16.554 &0.007\\
&2454562.79 &16.582 &0.010\\
&2454568.75 &16.741 &0.009\\
&2454570.76 &16.784 &0.008\\
&2454571.74 &16.790 &0.008\\
&2454572.77 &16.816 &0.009\\
&2454573.75 &16.835 &0.007\\
&2454574.72 &16.848 &0.008\\
&2454576.71 &16.868 &0.013\\
&2454580.74 &16.984 &0.008\\
&2454587.72 &17.163 &0.006\\
&2454591.69 &17.282 &0.007\\
&2454595.68 &17.442 &0.008\\
&2454624.62 &19.229 &0.018\\
&2454628.67 &19.278 &0.029\\
&2454646.63 &19.666 &0.029\\
&2454653.60 &19.760 &0.032\\
&2454654.62 &19.825 &0.038\\
\hline
\hline
\end{tabular}
\setcounter{table}{1}
\tablecomments{$V$-band apparent magnitudes of three example SNe from the sample. The full 
sample of $V$-band photometry can be downloaded from \url{http://www.sc.eso.org/~janderso/SNII_A14.tar.gz}
(also available at \url{http://csp.obs.carnegiescience.edu/data/)}, or obtained from the author
on request. (Note, the CSP magnitudes are in the 
natural system of the Swope telescope, whereas previous data are calibrated to the 
Landolt standard system).}
\end{table}

\begin{figure*}
\includegraphics[width=16cm]{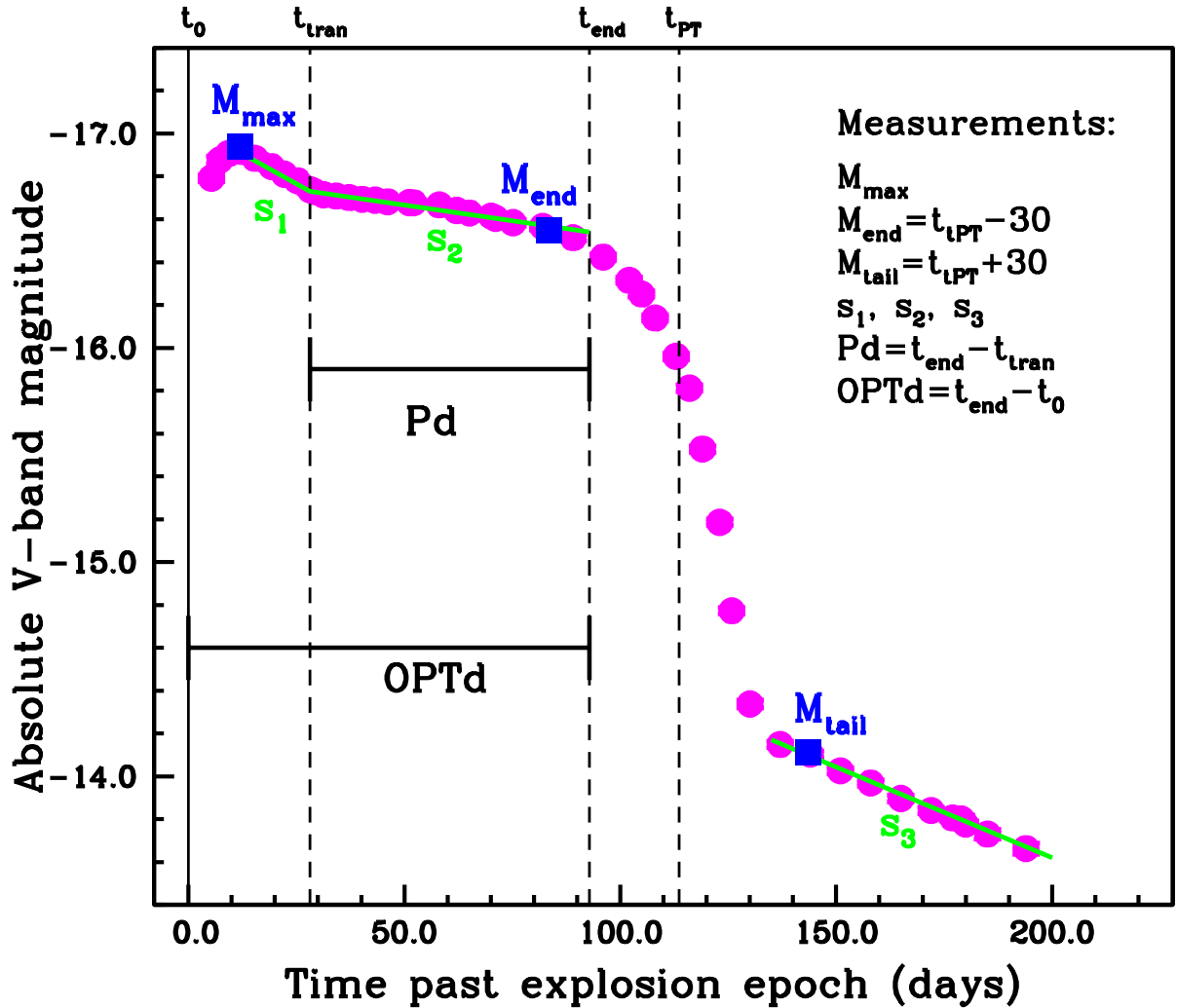}
\caption{An example of the light-curve parameters measured for each SN. 
Observed magnitudes at peak, $M_\text{max}$, end of `plateau', $M_\text{end}$,
and beginning of linear decline, $M_\text{tail}$ are shown in blue, as applied to the
example dummy data points (magenta). 
The positions of the three measured
slopes: $s_1$, $s_2$, and $s_3$ are shown in
green. The time durations: `plateau' length, \textit{Pd}, and optically
thick phase duration, \textit{OPTd} are indicated in black. Four time epochs
are labeled: $t_{0}$, the explosion epoch; $t_{tran}$, the transition from
$s_1$ to $s_2$; $t_{end}$, the end of the optically thick phase; and $t_{PT}$,
the mid
point of the transition from `plateau' to radioactive tail.}
\end{figure*}

\section{Light-curve measurements}
In Fig.\ 1 we show a schematic of the $V$-band light-curve parameters chosen for measurement. 
On inspection of the light-curves it was immediately evident that many SNe~II
within the sample show evidence for an initial decline from maximum 
(which to our knowledge is not generally discussed in detail in the literature , 
although see \citealt{clo96}, with respect to SN~1992H),
before settling onto a second slower decline rate, normally defined as the
plateau. Hence, we proceeded to define and measure two decline rates in the
early light-curve evolution, as will be outlined below. 
For the time origin we employ
both the explosion epoch (as estimated by the process outlined in \S\ 3.1) and 
$t_\text{PT}$: the mid-point of the transition between plateau
and linear decline epochs obtained through fitting SN~II light-curves with the
sum of three functions: a Gaussian which fits to the early time peak/decline;
a Fermi Dirac function which
provides a description of the transition between
the plateau and radioactive phases; and a straight line which accounts for the
slope due to the radioactive decay (see \citealt{oli10} for further
description). It is important to note here, 
while the fitting process of $t_\text{PT}$ appears to give 
good objective estimations of the time epoch of transition between plateau and
later radioactive phases, its fitting of precise parameters such as decline rates,
together with magnitudes and epochs of maximum light is less satisfactory. Therefore, we employ
this fitting procedure solely for the measurement of the epoch $t_\text{PT}$ (from which
other time epochs are defined). In the future,
it will be important to build on current template fitting techniques of SNe~II light-curves,
in order to measure all parameters in a fully automated way. For the current study, we continue as outlined
below. \\
With the above epochs in hand, the measured parameters are:\\
\begin{itemize}
\item $M_\text{max}$: defined as the initial peak in the $V$-band light-curve. Often this is
not observed, either due to insufficient early time data or poorly sampled
photometry. 
In these cases we take
the first photometric point to be $M_\text{max}$. When a true peak is observed, it is
measured by fitting a low order (four to five) polynomial to the photometry in close
proximity to the brightest photometric point (generally $\pm$5 days).\\  
\item $M_\text{end}$: defined as the absolute $V$-band magnitude measured 30 days before
$t_\text{PT}$. If $t_\text{PT}$ cannot be defined, and the photometry shows a single declining
slope, then $M_\text{end}$ is measured to be the last point of the light-curve. If the end of the
plateau can be defined (without a measured $t_\text{PT}$) 
then we measure the epoch and corresponding magnitude manually.\\ 
\item $M_\text{tail}$: defined as the absolute $V$-band magnitude measured 30 days after
$t_\text{PT}$. If $t_\text{PT}$ cannot be estimated, but it is clearly observed that the SN has fallen onto
the radioactive decline, then $M_\text{tail}$ is measured taking the magnitude at the
nearest point after transition. \\
\item $s_1$: defined as the decline rate in magnitudes per 100 days of the
initial, steeper slope of the light-curve. This slope is not always observed
either because of a lack of early time data, or because of insufficiently sampled light-curves
However, 
in some instances a lack of detection may simply imply a lack of any true peak in the light-curve,
together with an intrinsic lack of an early decline phase.\\
\item $s_2$: defined as the decline rate ($V$-band magnitudes per 100 days) of the
second, shallower slope in the light curve. This slope is
that referred to in the literature as the `plateau'. We note here, there are
many SNe within our sample which have light-curves which decline at a rate
which is ill-described by the term `plateau'. 
However, in the majority SNe~II in our sample (with sufficiently
sampled photometry) there is suggestive evidence for a `break' in the light-curve 
before a transition to the radioactive tail (i.e., an end to a `plateau' or optically thick phase). 
Therefore,
hereafter we use the term `plateau' in quotation marks to refer to this phase
of nearly constant decline rate (yet not necessarily a phase of constant magnitude)
for all SNe.\\
\item $s_3$: defined as the linear decline rate ($V$-band magnitudes per 100 days) of
the slope reached by each transient after its transition from the previous
`plateau' phase. This is commonly referred to in the literature as the
radioactive tail.\\
\end{itemize}
\indent To measure light-curve parameters photometry is analyzed with the
\textit{curfit} package within IRAF.\footnote{IRAF is distributed 
by the National Optical Astronomy Observatory, which is operated by the 
Association of Universities for Research in Astronomy (AURA) under 
cooperative agreement with the National Science Foundation.} 
In the case of measurements of $M_\text{end}$ and $M_\text{tail}$ we interpolate to the desired
epoch when $t_\text{PT}$ is defined, or define an epoch by eye when this
information is not available. $M_\text{max}$, as defined above is either the maximum
magnitude as defined by fitting a low order polynomial to the maximum of the
light-curve (only possible in 15 cases), or is simply taken as the magnitude
of the first epoch of $V$-band photometry.\\
\indent Photometric decline rates are all measured by fitting a straight line to each
of the three defined phases, taking into account photometric errors. 
To measure $s_1$ and $s_2$ we fit a piecewise linear model with four parameters: 
the rates of decline (or rise), i.e. the slopes $s_1$ and $s_2$, 
the epoch of transition between the two slopes, $t_\text{tran}$, and the magnitude
offset.
This process requires that the start of $s_1$ and end of $s_2$ are
pre-defined, in order to exclude data prior to maximum or once the SN starts
to transition to the radioactive tail. The
values of $s_1$, $s_2$ and their transition point are then determined
through weighted least squares 
minimization\footnote{This process was also checked using a more `manual'
approach, with very consistent slopes measured, and our overall results and
conclusions remain the same independent of the method employed.}. We then
determine whether the light-curve is better fit with one slope (just $s_2$), or
both $s_1$ and $s_2$ using the Bayesian Information Criterion (BIC)
\citep{sch78}. This statistical analysis uses the best fit chi-square,
together with the number of free parameters to determine whether the 
data are better fit by increasing the slopes from one to two, assuming 
that all our measurements are independent and follow a Gaussian distribution.
It should be noted that although this procedure works extremely well, there
are a few cases where one would visually expect two slopes and only one is
found. The possible biases of including data where one only measures an $s_2$,
but where intrinsically there are two slopes, are discussed later in
the paper.\\
\indent Measurements of $s_3$ are relatively straight forward, as it is easy to identify
when SNe have transitioned to the radioactive tail. Here, we require three data
points for a slope measurement, and simply fit a straight line to the
available photometry.\\

\subsection{Explosion epoch estimations}
While the use of $t_\text{PT}$ allows one to measure parameters at consistent epochs with
respect to the transition from `plateau' to tail phases, much physical understanding
of SNe~II rests on having constraints on the epoch of explosion.
The most accurate method for determining this epoch for any given SN
is when sufficiently deep pre-explosion images are available close to the time
of discovery. However, in many cases in the current
sample, such strong constraints are not available. Therefore, to further
constrain this epoch, matching of spectra to those of a library of spectral
templates was used through employing the Supernova Identification (SNID) code
\citep{blo07}. The earliest spectrum of each SN within our sample was run
through SNID and top spectral matches inspected. The best fit was then
determined, which gives an epoch of the spectrum with respect to maximum
light of the comparison SN. Hence, using the published explosion epochs for
those comparison spectra with respect to maximum light, 
one can determine an explosion epoch for each SN in the sample. 
Errors were estimated using the deviation in time
between the epoch of best fit to those of other good fits
listed, and combining this error with that of the epoch of explosion of the
comparison SN taken from the literature for each object.\\
\indent In the case of SNe with non-detections between 1-20 days before
discovery, we use explosion epochs as the mid point between those two epochs,
with the error being the explosion date minus the non-detection date. For
cases with poorer constraints from non-detections, explosion epochs
from the spectral matching outlined above are employed. The validity of this spectral matching technique is
confirmed by comparison of estimated epochs with SNe non-detections (where
strong constraints exist). 
Where the error on the non-detection explosion epoch estimation is less than
20 days, the mean \textit{absolute} difference between the explosion epoch calculated using the
non-detection and that estimated using the spectral matching is 4.2
days (for the 61 SNe where this comparison is possible). 
The mean offset between the two methods is 1.5 days, in the sense that
the explosion epochs estimated through spectral matching are on average 1.5
days later than those estimated from non-detections. In the appendix these issues are
discussed further, and we show that the inclusion of parameters which
are dependent on spectral matching
explosion epochs make no difference to our results and conclusions.
Finally, we note that a similar analysis was also achieved by \cite{har08_2}.

\subsection{`Plateau' and optically thick phase durations}
A key SN~II light-curve parameter often discussed in the literature is the
length of the `plateau'. This has been claimed to be linked to the
mass of the hydrogen envelope, and to a lesser extent the mass of $^{56}$Ni synthesized in the
explosion \citep{lit85,pop93,you04,kas09,ber11}. Hence, we also attempt to measure this
parameter. Before doing so it is important to note how one defines the
`plateau' length in terms of current nomenclature in the literature. It is
common that one reads that a SN~IIP is defined as having `plateau of almost constant
brightness for a period of several months'. Firstly, it is unclear how many of
these types of objects actually exist in nature, as we will show later. 
Secondly, this phase of constant brightness (or at least constant
change in brightness for SNe where significant $s_2$ values are measured), should
be measured starting \textit{after} initial decline from maximum, a phase
which is observed in a significant fraction of hydrogen rich SNe~II (at least
in the $V$- and bluer bands).\\
\indent To proceed with adding clarity to this issue, the start of the
`plateau' phase is defined to be $t_\text{tran}$ (the $V$-band transition between $s_1$ and $s_2$
outlined in \S\ 3). 
The end of the `plateau' is defined as when the 
extrapolation of the straight line $s_2$ becomes 0.1
magnitudes more luminous than the light-curve, $t_\text{end}$\footnote{Note, this
time epoch is very similar to the epoch where $M_\text{end}$ is measured
(i.e. 30 days before $t_\text{PT}$). However, given that in some cases we
measure an $t_\text{PT}$  but the 0.1 mag criterion is not met, we choose to define
these epochs separately for consistency purposes.}. This 0.1 mag criterion is somewhat
arbitrary, however it ensures that both the light-curve has definitively started
to transition from `plateau' to later phases, and that we do
not follow the light-curve too far into the transitional phase.
Using these time epochs
we define two time durations: 
\begin{itemize}
\item the `plateau' duration: $Pd$ = $t_\text{end}$--$t_\text{tran}$ 
\item the optically thick phase duration: $OPTd$ = $t_\text{end}$--$t_\text{0}$ 
\end{itemize}
\indent These parameters are labeled in the light-curve parameter
schematic presented in Fig.\ 1. In addition, all derived light-curve
parameters: decline rates, magnitudes, time durations, are depicted on their respective 
photometry in the appendix.

\subsection{Extinction estimates}
All measurements of photometric magnitudes are first corrected for extinction
due to our own Galaxy, using the re-calibration of dust maps provided by
\cite{sch11}, and assuming an $R_{\rm V}$ of 3.1 \citep{car89}. We then
correct for host galaxy extinction using measurements of the equivalent width (EW) of
sodium absorption (NaD) in low resolution spectra of each object. For each spectrum within the sequence
(of each SN), the presence of NaD, shifted to the
velocity of the host galaxy is investigated. If detectable line absorption is
observed a mean EW from all spectra within the sequence is calculated, and we take the EW
standard deviation to be the 1$\sigma$ uncertainty of these values. 
Where no evidence
of NaD is found we assume zero extinction. In these cases, the error is
taken to be that calculated for a 2$\sigma$ EW upper limit on the
non-detection of NaD. 
Host galaxy $A_{\rm V}$ values are then estimated using the relation taken
from \cite{poz12},
assuming an $R_{\rm V}$ of 3.1 \citep{car89}.\\
\indent The validity of using NaD EW line measurements in low resolution spectra,
as an indicator of dust
extinction within the host
galaxies of SNe~Ia has been 
recently questioned (\citealt{poz11,phi13}). 
Even in the Milky Way where a clear correlation between NaD EW measurements
and $A_{\rm V}$ is observed, the RMS scatter is large. 
At the same time, absence of NaD is generally a first order
approximation of low level or zero extinction, while high EW of NaD 
possibly implies some degree of host galaxy reddening \citep{phi13}. 
The use of the \cite{poz12} relation for the current sample is complicated as 
it has been
shown (e.g. \citealt{mun97}) to saturate for NaD EWs approaching and
surpassing 1 \AA. Indeed, 10 SNe within the current sample have NaD EWs of more
than 2 \AA \footnote{The example of 2 \AA\ is shown here to outline the issues
of saturation as one measures large EWs. As measured values approach 1 \AA\
the significance of saturation becomes much smaller.}. The \cite{poz12} relation gives an $A_{\rm V}$
(assuming $R_{\rm V}$ = 3.1) of 9.6 for an EW of 2 \AA, i.e. implausibly high. Therefore,
we choose to eliminate magnitude measurements of SNe~II in our sample 
with EW measurements higher than 1 \AA, due to the uncertainty in any
$A_{\rm V}$ corrections. In addition, when measuring 2$\sigma$ upper
limits for NaD non-detections, we also eliminate SNe from magnitude analysis
if the limits are higher than 1 \AA\ (i.e. spectra are too noisy to detect
significant NaD absorption). 
Finally, those SNe which do not have
spectral information are also cut from magnitude analysis.\\
\indent The above situation is far from satisfying, however for SNe~II there is currently
no accepted method which accurately corrects for host galaxy extinction.
\cite{oli10} used the $(V-I)$ color excess at the end of the plateau to
correct for reddening, with the assumption that all SNe~IIP evolve to similar
temperatures at that epoch (see also \citealt{nug06,poz09,kri09,dan10}). 
Those authors found that using such a
reddening correction helped to significantly reduce the scatter in Hubble
diagrams populated with SNe~IIP.
However, there are definite outliers from this
trend; e.g.\ sub-luminous SNe~II tend to have red intrinsic colors at the end 
of the plateau (see e.g. \citealt{pas04,spi14}), which,
if one assumed were due to extinction would lead to corrections for
unreasonable amounts of reddening (e.g.\ in the case of SN~2008bk; Pignata private
communication). 
With these issues in mind we proceed, listing our adopted 
$A_{\rm V}$ values measured from NaD EWs in Table 6\footnote{In three
cases where accurate extinction estimates are available in the literature we
use those values, as noted in Table 6.}.\\ 
We note that while the current extinction estimates
are uncertain, all of
the light-curve relations that will be presented below hold even if we assume
zero extinction corrections. 

\subsection{Distances}
Distances are calculated using CMB-corrected 
recession velocities if this value is
higher than 2000 km s$^{-1}$, together with an $H_0$ of 73 km s$^{-1}$ Mpc$^{-1}$ (and
$\Omega_{m}$ = 0.27, $\Omega_\Lambda$ = 0.73, \citealt{spe07}), assuming a 300 km s$^{-1}$ velocity error due to galaxy peculiar velocities. 
For host galaxies with recession velocities less than
2000 km s$^{-1}$ peculiar velocities 
make these estimates unreliable. For these cases `redshift independent' distances
taken from NED are employed, where the majority are Tully-Fisher estimates, but Cepheid values
are used where available. Errors are taken to be the standard deviation of the mean value of distances
in cases with multiple, e.g.\ Tully-Fisher values, or in cases with single
values the literature error on that value. Distance moduli, together
with their associated errors are listed in Table 6.\\

\subsection{Light-curve error estimation}
Errors on absolute magnitudes are a combination of uncertainties in: 1) photometric
data; 2) extinction estimations and; 3) distances. 
In 1)
when the magnitude is taken from a single photometric point we take the error
to be that of the individual magnitude. When this magnitude is obtained
through interpolation we combine the errors in quadrature of the two
magnitudes used. 
Published photometric errors are the sum in quadrature of two error
components: 1) the uncertainty in instrumental magnitudes estimated from 
the Poisson noise model of the flux of the SN and background regions, and 2) the 
errors on the zero point of each image (see \citealt{con10}, as applied 
to the CSP SN~Ia sample).
In the case of 2) errors in $A_{\rm V}$ are
taken as the standard deviation of the mean measurement of the NaD EWs,
together with the error on the relation used \citep{poz12}. However, it is
believed that the error on the relation provided by those authors is significantly
underestimated. Using the dispersion of individual measurements on the NaD
EW-$A_{\rm V}$ relation as seen in \cite{phi13}, we
obtain an additional error of 47\%\ of $A_{\rm V}$ estimations, which
is added to the error budget. 
In the case of 3) errors
for SNe within host galaxies with recession velocities above 2000 km s$^{-1}$
are derived from assuming peculiar velocity errors of 300 km s$^{-1}$, while
for recession velocities below that limit errors are those published along
with the distances used (as outlined above). These three errors are combined in quadrature and are
listed for each of the three estimated absolute magnitudes in Table 6.\\ 
\indent Decline rate uncertainties come from the linear fits to $s_1$,
$s_2$ and $s_3$. The uncertainty in $Pd$ is 
the error on the epoch of the transition between $s_1$ and $s_2$, as
estimated in \S\ 3, while the error on $OPTd$ is that estimated for the
explosion epoch, as outlined in \S\ 3.1. We also combine with the uncertainty in 
$Pd$/$OPTd$,  an additional 
4.25 days to the error budgets to account for the
uncertainty in the definition of $t_{end}$. This error is estimated in the following way.
For each SN we calculate the average cadence of photometry
at epochs in close proximity to $t_{end}$. The mean of these cadences for all SNe is 8.5 days.
Given that $t_{end}$ is measured as an interpolation between photometric points (following $s_2$
together with the morphology of the changing light-curve), we assume the error in defining
any epoch at this phase to be half the cadence, 
i.e. 4.25 days.\\
\indent An additional magnitude error inherent in all SNe measurements is that
due to $K$-terms (the wavelength shift of the spectral energy distribution 
with respect to the observer's band-pass, \citealt{oke68}). 
For the current analysis we do not make such corrections owing to the 
low-redshift of our sample. 
To test this assumption we follow the technique employed in \cite{oli10}, and described in \cite{oli08}.
This involved synthesizing $K$-terms from a library of synthetic SN spectra, resulting
in a range of corrections as a function of $(B-V)$ SN color. At the mean
redshift of our sample of 0.018, the average $K$-correction during the plateau is estimated
to be 0.02 mag. At the redshift limit of our sample: 0.077, this mean correction is 0.07 mag.
Hence, our neglection of this term is justified.
While we also choose to ignore $S$-corrections (magnitude corrections between different photometric systems), 
we note again that the CSP data are tied to the natural system of the Swope telescope (see \citealt{con10} for details),
while all previous data are calibrated to the Landolt standard system. This may bring 
differences to the photometry of each sub-sample. For a typical SN, the difference in $V$-band magnitudes between
photometry in the CSP and standard system is less than 0.1 mag at all epochs (with mean corrections of around
0.03 mag). The differences between $S$-corrections at $M_\text{max}$ and $M_\text{end}$ are around $\sim$0.03 mag.
Therefore the influence of this difference on decline rate estimations will be negligible (they will be less than
this difference).
In conclusion, uncertainties in our measurements are dominated by those from distances and extinction estimates, and our neglection of these
corrections is very unlikely to affect our overall results and conclusions.

\subsection{SN~II $V$-band light-curves}
After correcting $V$-band photometry for both MW and host extinction we
produce absolute $V$-band light-curves by subtracting the distance
modulus from SNe extinction corrected apparent magnitudes. 
\indent In Fig.\ 2 results of Legendre polynomial fits to the absolute
light-curves of all those with explosion epochs defined, and
$A_{\rm V}$ corrections possible are presented. This shows the large
range in both absolute magnitudes and light-curve morphologies, the analysis of
which will be the main focus of the paper as presented below. In addition, in
the appendix light-curves are presented in more detail for all SNe to show the
quality of our data, its cadence, and derived SN parameters.
Note that these Legendre polynomial fits are merely used to show the data in a more
presentable fashion. They are not used for any analysis undertaken.

\begin{figure*}
\includegraphics[width=18cm]{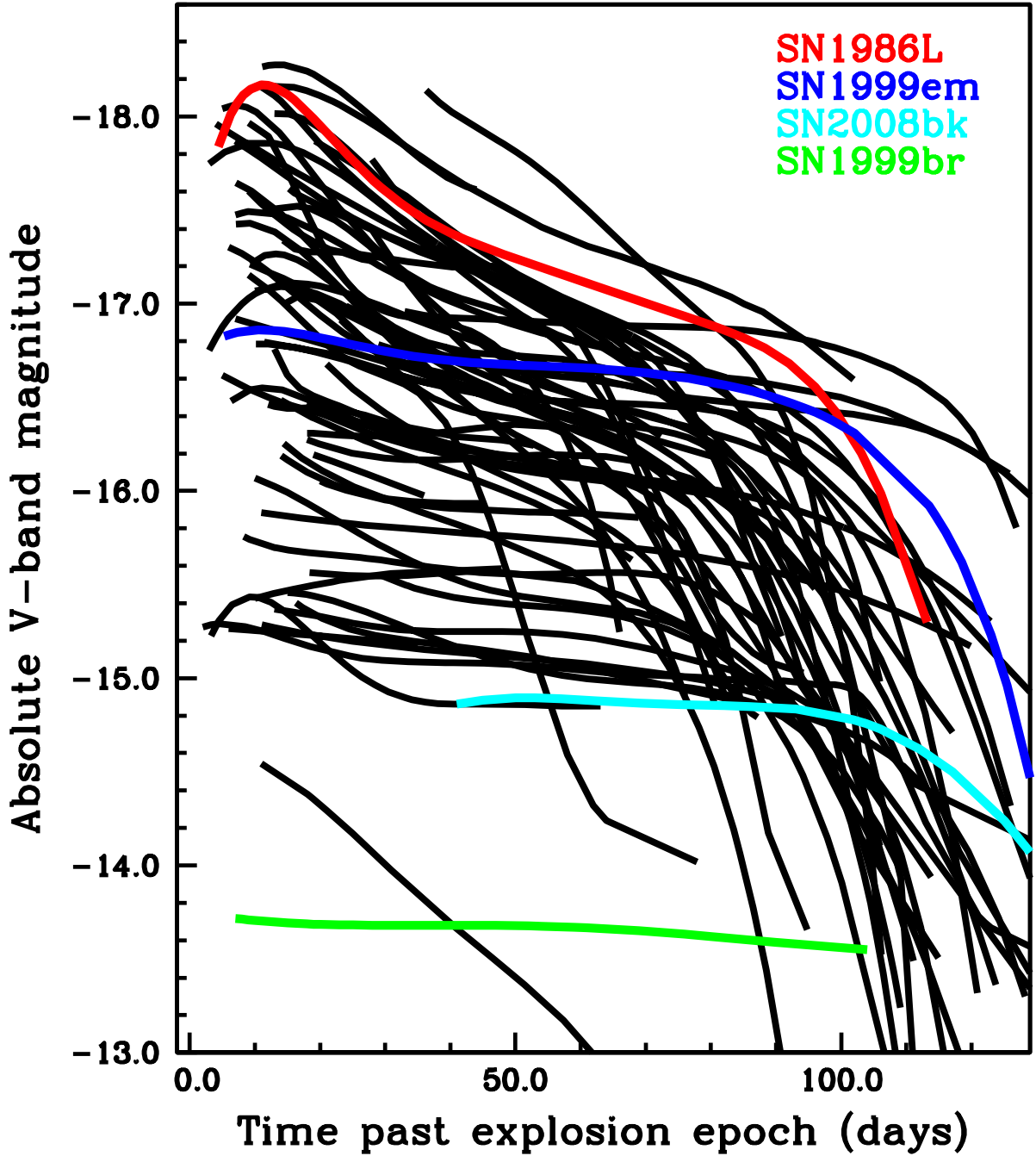}
\caption{SNe~II absolute $V$-band light-curves of the 60 events with explosion
epochs and $A_{\rm V}$ corrections. Light-curves are displayed as
Legendre polynomial fits to the data, and are presented by black lines. For
reference we also show in colors the fits to our data for 4 SNe~II: 1986L, 1999em,
2008bk, and 1999br.}
\end{figure*}

\section{Results}
In Table 6 we list the measured $V$-band parameters as defined above for each SN,
together with the SN distance modulus, host extinction estimate, and explosion
epochs.
Given eight measured parameters there are a large number of different correlations
one can search for and investigate. In this Section figures
and statistics of correlations are presented, choosing those we deem of
most interest. In the appendix additional figures not included in the main
text are presented, which may be of interest.\\
\indent Throughout the rest of the paper, correlations are tested for significance
using the Pearson test for correlation. We employ Monte Carlo
bootstrapping to further probe the reliability of such tests. For each of the
10,000 simulations (with random parameter pairs drawn from our measured
values) a Pearson's $r$-value is estimated. The mean $r$ of these 10,000 simulations is then calculated
and is presented on each figure, together
with its standard deviation. The mean is presented together with
an accompanying lower limit to the probability of finding such a correlation
strength by chance\footnote{Calculated using the on-line statistics tool found
at: http://www.danielsoper.com/statcalc3/default.aspx \citep{coh03}.}.
Where correlations are presented with parameter pairs ($N$)
higher than 20, binned data points are also displayed, with error bars taken
as the standard deviation of values within each bin. 

\subsection{SN~II parameter distributions}
In Fig.\ 3 histograms of the three absolute $V$-band magnitude
distributions: $M_\text{max}$, $M_\text{end}$ and $M_\text{tail}$ are presented. These distributions
evolve from being brighter at maximum, to lower luminosities at the
end of the plateau, and further lower values on the
tail. Our SN~II sample is characterized, after correction for extinction, 
by the following mean values: $M_\text{max}$ = --16.74 mag ($\sigma$ = 1.01, 68 SNe); 
$M_\text{end}$ = --16.03 mag ($\sigma$ = 0.81, 69 SNe);
$M_\text{tail}$ = --13.68 mag ($\sigma$ = 0.83, 30 SNe).\\
\indent The SN~II family spans a large range of $\sim$4.5
magnitudes at peak,
ranging from --18.29 mag (SN~1993K) through --13.77 mag (SN~1999br). 
At the end of their `plateau' phases the sample ranges from --17.61 to
--13.56 mag. SN~II maximum light absolute magnitude distributions have
previously been presented by \cite{tam90} and \cite{ric02}. Both of these were
$B$-band distributions. \citeauthor{tam90} presented a distribution for 23 SNe~II of
all types of $M_{\rm B}$ = --17.2 mag ($\sigma$ = 1.2), while \citeauthor{ric02}
found $M_{\rm B}$ = --17.0
mag ($\sigma$ = 1.1) for 29 type IIP SNe and $M_{\rm B}$ = --18.0 mag ($\sigma$
= 0.9) for 19  type IIL events. Given
that our distributions are derived from the $V$ band, a direct comparison to
these works is not possible without knowing the intrinsic colors of each SN
within both samples. However, our derived $M_\text{max}$ distribution is 
reasonably consistent with those
previously published (although slightly lower), with very similar standard deviations.\\
\indent At all epochs our sample shows a continuum of
absolute magnitudes, and the $M_\text{max}$ distribution shows a low-luminosity tail as
seen by previous authors (e.g.\ \citealt{pas04,li11}). All three epoch magnitudes
correlate strongly with each other: when a SN~II is bright at maximum light it
is also bright at the end of the plateau and on the radioactive
tail.

\begin{figure}
\includegraphics[width=9cm]{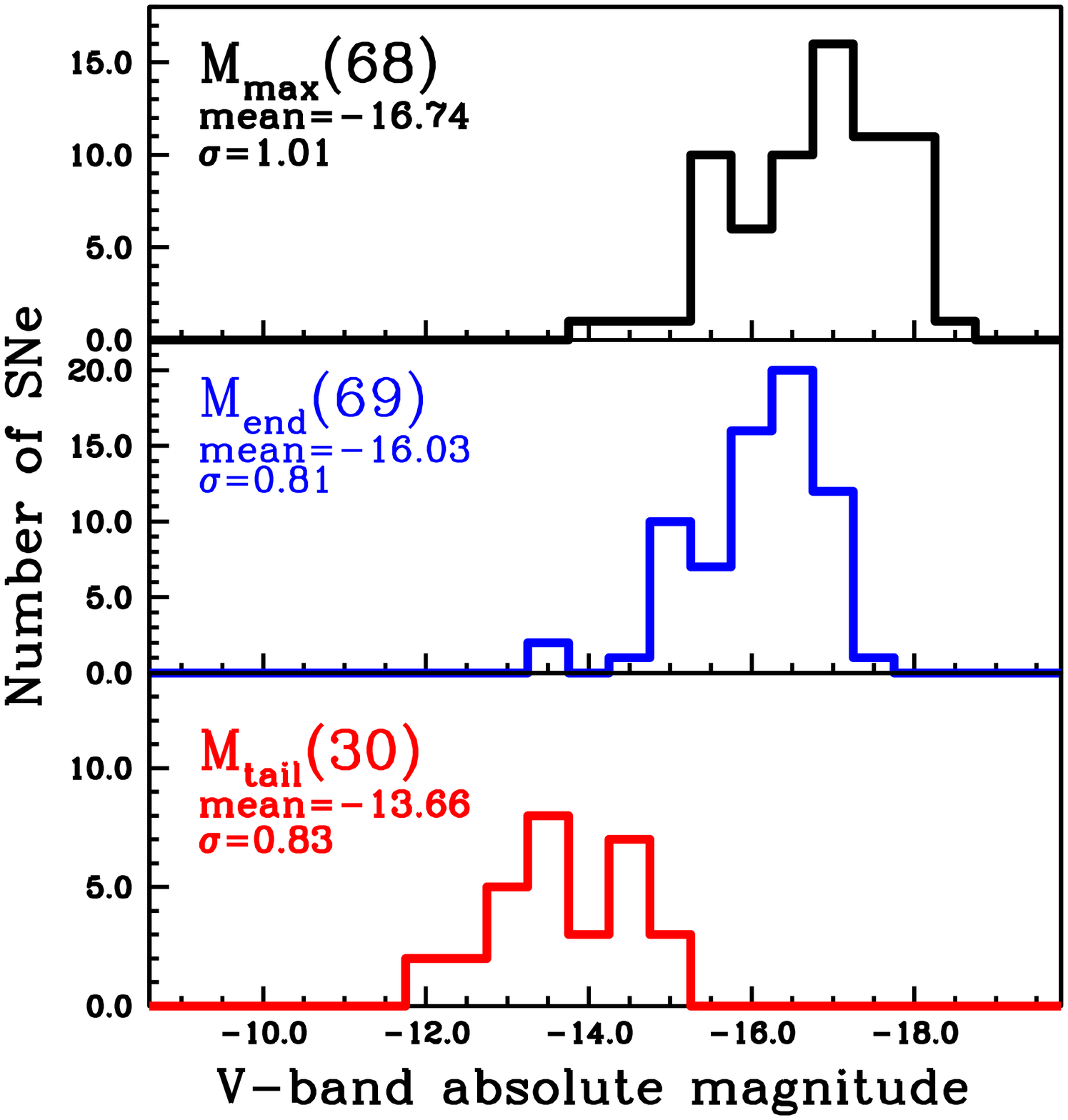}
\caption{Histograms of the three measured absolute magnitudes of
SNe~II. \textit{Top}: peak absolute magnitudes;
$M_\text{max}$. \textit{Middle}: absolute magnitudes at the end of the
plateau; $M_\text{end}$. \textit{Bottom}: absolute magnitudes at the
start of the radioactive tail; $M_\text{tail}$. In each panel the number of SNe is
listed, together with the mean absolute $V$-band magnitude and the standard
deviation on that mean. }
\end{figure}

\indent Fig.\ 4 presents histograms of the distributions of the three $V$-band decline rates,  
$s_1$, $s_2$ and $s_3$, together with their means and standard
deviations. SNe decline from maximum ($s_1$) at an average rate
of 2.65 mag per 100 days, before declining more slowly on the `plateau' ($s_2$) at a
rate of 1.27 mag per 100 days. Finally, once a SN completes its transition to
the radioactive tail ($s_3$) it declines with a mean value of 1.47
mag per 100 days. This last decline rate is higher than that
expected if one assumes full trapping of gamma-ray photons from the
decay of $^{56}$Co (0.98 mag per 100 days, \citealt{woo89}). This gives interesting
constraints on the mass extent and density of SNe ejecta, as will be
discussed below. 
We observe more variation in decline rates at earlier times ($s_1$) 
than during the `plateau' phase ($s_2$).\\
\indent As with the absolute magnitude distributions discussed above, 
the $V$-band decline rates appear to show a continuum in their distributions. The
possible exceptions are those SNe declining extremely
quickly through $s_1$: the fastest decliner SN~2006Y with an
unprecedented rate of 8.15 mag per 100 days. 
In the case of $s_2$ the fastest decliner is SN~2002ew with a decline rate of 3.58 mag per
100 days, while SN~2006bc shows a \textit{rise} during this phase, at a rate of --0.58
mag per 100 days. The $s_2$ decline rate distribution has a tail out to higher
values, while a sharp edge on the left hand side is seen, with
only 6 SNe having negative decline rates during this epoch.\\
\indent In Fig.\ 5 we show how the decline rates correlate.
A correlation between $s_1$ and $s_2$ is observed despite a
handful of outliers. $s_2$ and
$s_3$ also appear to show the same trend: a fast decliner at one epoch is
usually a fast decliner at other epochs. This suggestion that more steeply
declining SNe~II (during $s_2$) also have faster declining radioactive tails 
was previously suggested by \cite{dog85}.
In both of these plots there is at
least one obvious outlier: SN~2006Y. We mark the position of this event here 
and also on subsequent figures where it also often appears as an
outlier to any trend observed. Further analysis of this highly unusual event
will be the focus of future work.\\ 
\indent In summary of the overall distributions of decline rates:
SNe~II which decline more quickly at early epochs also generally decline more
quickly both during the plateau and on the radioactive tail.\\

\begin{figure}
\includegraphics[width=9cm]{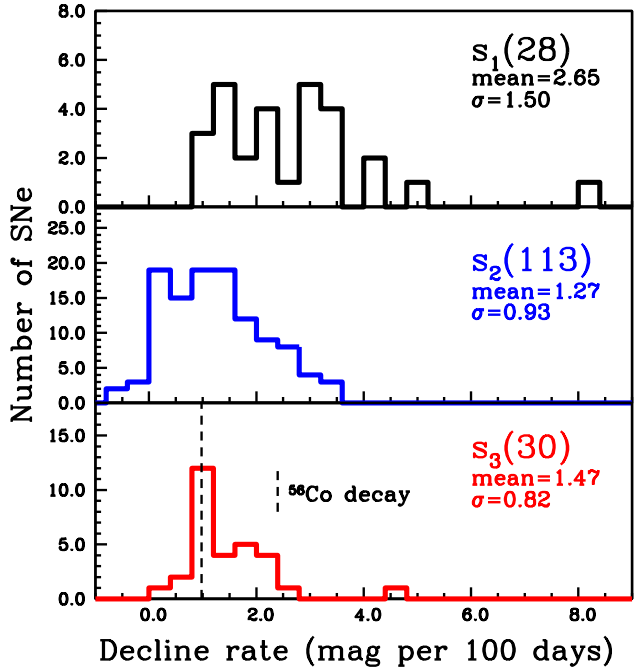}
\caption{Histograms of the three measured decline rates of
SNe~II. \textit{Top}: initial decline rates from maximum;
$s_1$. \textit{Middle}: decline rates on the `plateau' $s_2$. 
\textit{Bottom}: decline rates on the radioactive tail $s_3$. In this
last plot a vertical dashed line indicates the expected decline rate for full
trapping of emission from $^{56}$Co decay. In each panel the number of SNe is
listed, together with the mean decline rate and the standard
deviation on that mean. }
\end{figure}

\begin{figure*}
\includegraphics[width=9cm]{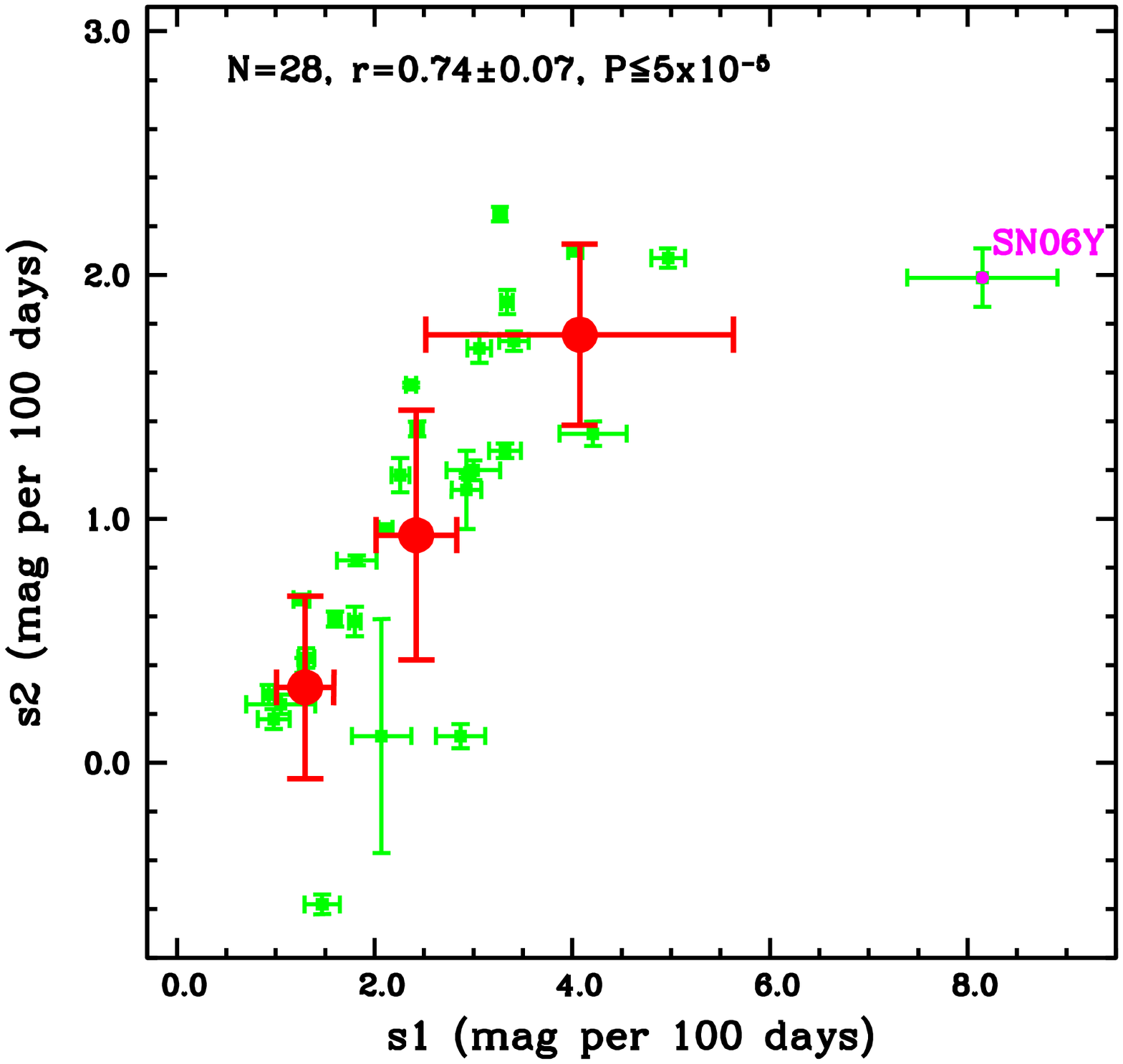}
\includegraphics[width=9cm]{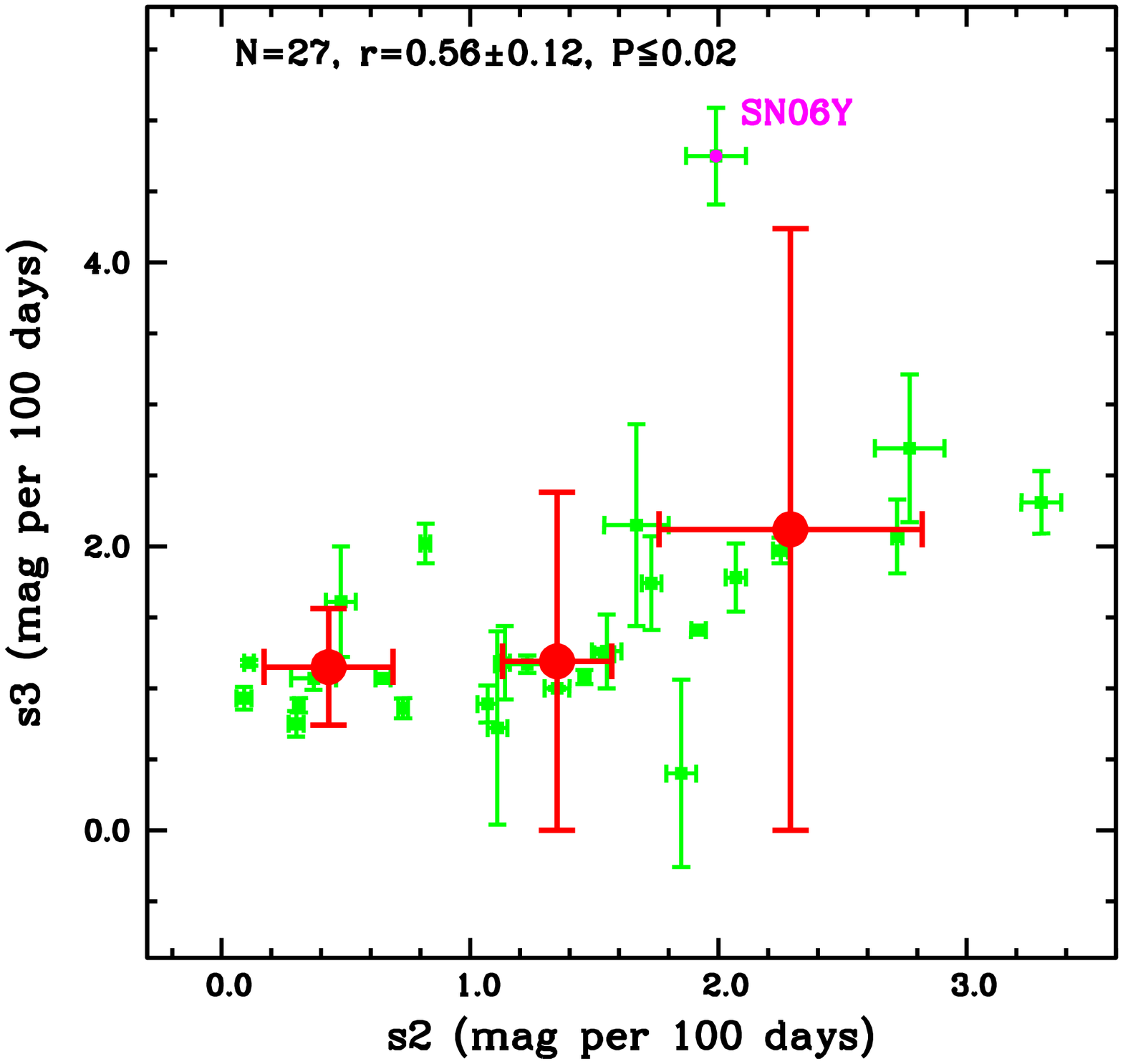}
\caption{Correlations between $s_1$ and $s_2$ \textit{(Left)}, and
$s_2$ against $s_3$ \textit{(Right)}. The results of Monte Carlo simulations on the
statistics of these two variables are noted at the top of the figure: $N$ =
number of events, $r$ = Pearson's correlation coefficient,
$P$ = probability of detecting a correlation by chance. Binned
data are shown in red circles, both here and throughout the paper. 
The outlier SN~2006Y noted throughout different figures throughout the rest of the paper 
is shown
in magenta. }
\end{figure*}

\subsection{Brightness and decline rate correlations}
\begin{figure}
\includegraphics[width=8.5cm]{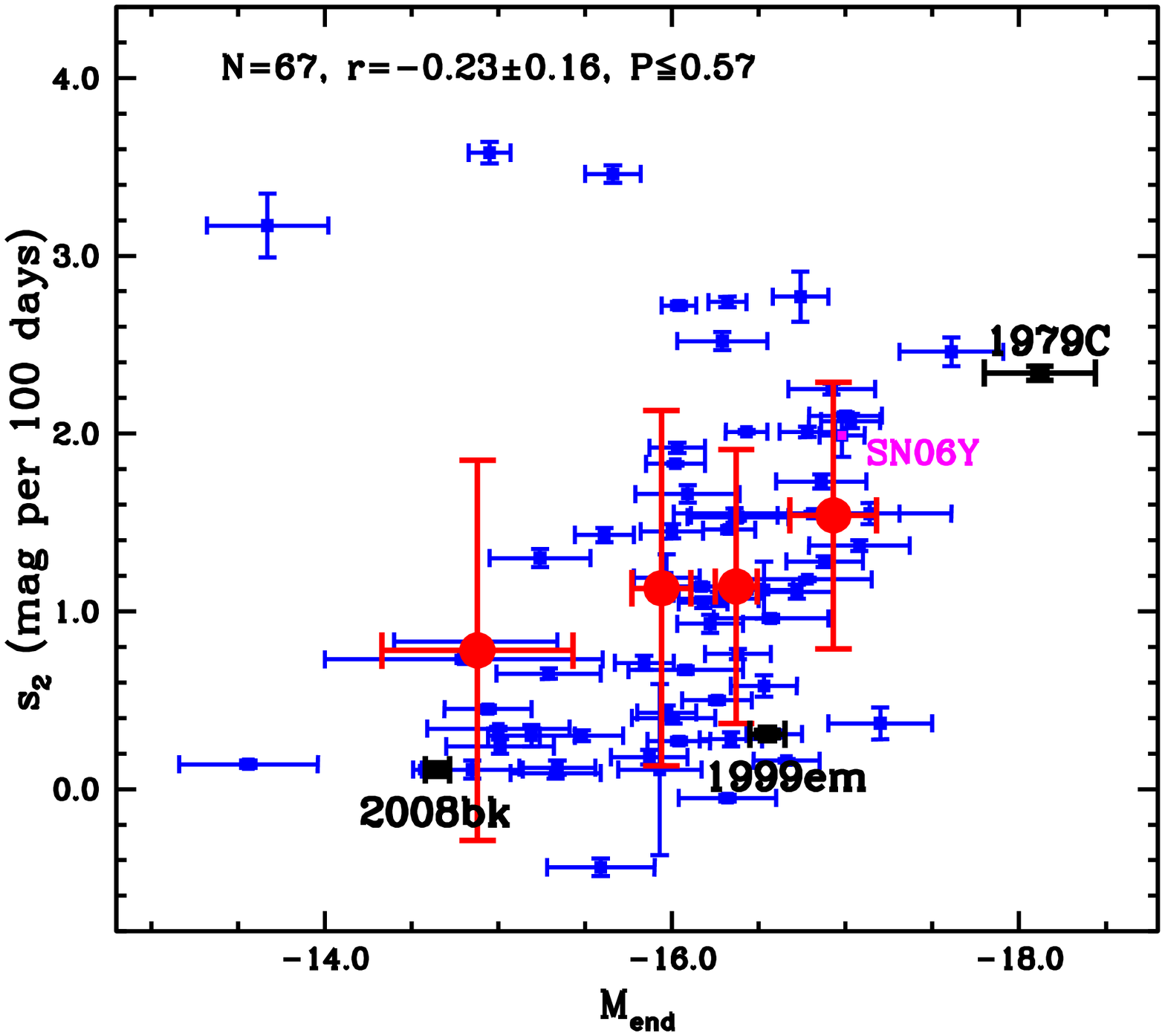}
\caption{The magnitude at the end of the `plateau', $M_\text{end}$, plotted against the decline
rate during the `plateau' $s_2$. Individual data points are shown as blue
squares.
The positions of three individual SNe~II are noted: the sub-luminous type IIP,
SN~2008bk, the prototype type IIP, SN~1999em (both from our own sample), and the
prototype type IIL, SN~1979C \citep{dev81}. The results of Monte Carlo simulations on the
statistics of these two variables are noted as in Fig.\ 5.}
\end{figure}

\begin{figure*}
\includegraphics[width=17cm]{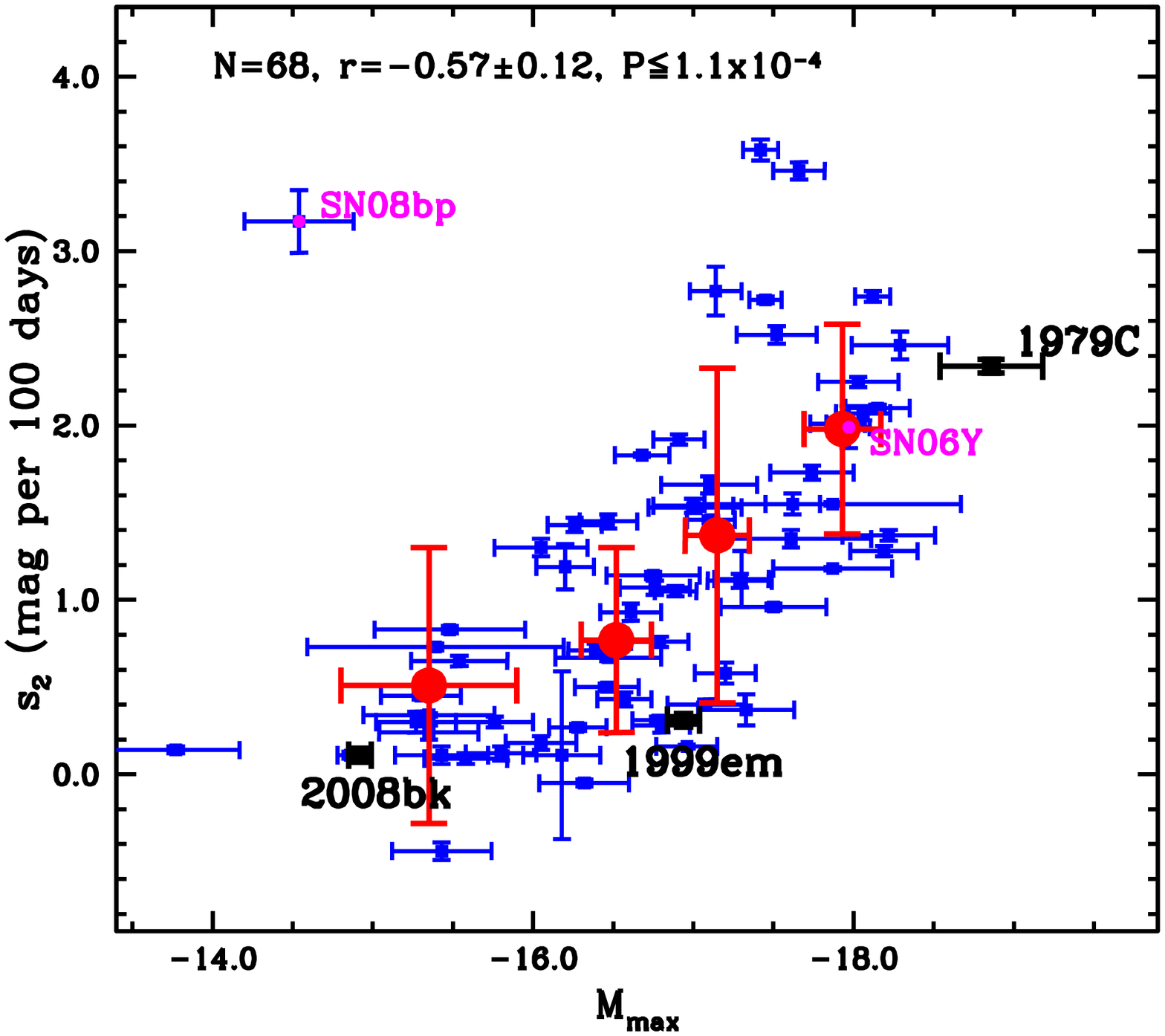}
\caption{Peak magnitude, $M_\text{max}$, plotted against the decline
rate during the `plateau' $s_2$. Individual data points are shown as blue
squares. The positions of three individual SNe~II are noted: the sub-luminous
type IIP,
SN~2008bk, the prototype type IIP, SN~1999em (both from our own sample), and the
prototype type IIL, SN~1979C \citep{dev81}. The results of Monte Carlo simulations on the
statistics of these two variables are noted as in Fig.\ 5. SN~2008bp, a major outlier within this correlation
is shown in magenta. }
\end{figure*}

Having presented parameter distributions and correlations in the previous
Section, 
here we turn our attention to
investigating whether SN brightness and rate of decline of their light-curves
are connected. In some of the plots presented in this Section photometric measurements of
the prototypical type IIL SN~1979C are also included for comparison (\citealt{dev81}). 
The positions within the
correlations of the prototypical type IIP SN~1999em, plus the sub-luminous
type IIP SN~2008bk (both from our own sample) are also indicated.\\
\indent Given that peak epochs of SNe~II are often hard to define,
literature measurements of SN~II brightness have generally concentrated on some
epoch during the `plateau' (e.g.\ \citealt{ham03,nug06}). Therefore, we start by
presenting the correlation of $M_\text{end}$ (brightness at the end of the
`plateau') with $s_2$ in Fig.\ 6. 
No correlation is apparent which is confirmed by a Pearson's
test. Given that $M_\text{end}$ does not seem to correlate with decline
rates (Fig.\ 30 shows correlations against $s_1$ and $s_3$), we move to
investigate $M_\text{max}$-decline rate correlations.\\
\indent In Fig.\ 7 $M_\text{max}$ against $s_2$ is plotted. It is found that these
parameters show a trend with lower luminosity SNe declining more slowly (or even
rising in a few cases), and more luminous events declining more rapidly
during the `plateau'. One possible bias in this correlation is that SNe are included if
only one slope is measured (just $s_2$) together with those events with two. 
However, making further cuts to the sample (only
including those with $s_1$ \textit{and} $s_2$ measurements) does not
significantly affect the results and conclusions presented here. Later in
\S\ 5.6 we discuss this
issue further and show how using a sub-sample of events with both slopes
measured enables us to further refine the predictive power of $s_2$.\\
\indent The above finding is consistent with previous results showing that
SNe~IIL are generally more luminous than SNe~IIP \citep{you89,pat94,ric02}. It is
also interesting to note the positions of the example SNe displayed in
Fig.\ 7: the sub-luminous type IIP SN~2008bk has a low luminosity and a
very slowly declining light-curve; the prototype type IIL SN~1979C is brighter than
all events in our sample, and also has one of the highest $s_2$ values. 
The prototype type IIP SN~1999em has, as expected, a small $s_2$ value, 
and most (78\%) of the remaining SNe II 
decline more quickly. In terms of $M_\text{max}$, on the other hand, 
SN~1999em does not stand out as a particularly bright or faint object.
We note one major outlier to the trend presented in Fig.\ 7:
SN~2008bp with a high $s_2$ value (3.17 mag per 100 days) but a very low
$M_\text{max}$ value (--14.54 mag). It is possible that the extinction has been significantly
underestimated for this event based on its weak 
interstellar absorption NaD, and indeed the SN has a red color during
the `plateau' (possibly implying significant reddening). Further analysis will be left for future work.\\
\indent In Fig.\ 8 we plot $M_\text{max}$ against $s_1$. While the
significance of correlation is not as strong as
seen with respect to $s_2$, it still appears that more luminous SNe at maximum light
decline more quickly also at early times.\\
\indent Finally, in Fig.\ 9 $M_\text{max}$ against $s_3$ is presented, and the
expected decline rate (dashed horizontal line) if
one assumes full trapping of gamma-ray photons from the decay of $^{56}$Co to $^{56}$Fe
is displayed. There appears
to be some correlation in that more luminous SNe~II at maximum have
higher $s_3$ values. However, more relevant than a strict one to one correlation, 
it is remarkable that none of the fainter SNe deviate \textit{significantly} 
from the slope expected from full trapping, while at brighter magnitudes
significant deviation is observed. The physical implications 
of this will be discussed below. Finally, we note that while the striking
result from Figs 4 and 9 is the number of SNe with $s_3$ values higher than
0.98 mag per 100 days, and the fact that nearly all of these are overluminous
(compared to the mean) SNe~II, it is also observed that there are a number of SNe which
have $s_3$ values \textit{lower} than the expected value. These SNe also
generally have lower peak luminosities. Indeed this has been noted before, especially
for sub-luminous SNe~IIP by \cite{pas09} and \cite{fra11}. In the case
of SN~1999em, \cite{utr07} argued that the discrepancy (from the rate expected
due to radioactivity) was due to radiation from
the inner ejecta propagating through the external layers and providing
additional energy (to that of radioactivity) to the light-curve, naming this
period the `plateau tail phase'.\\

\begin{figure}
\includegraphics[width=8.5cm]{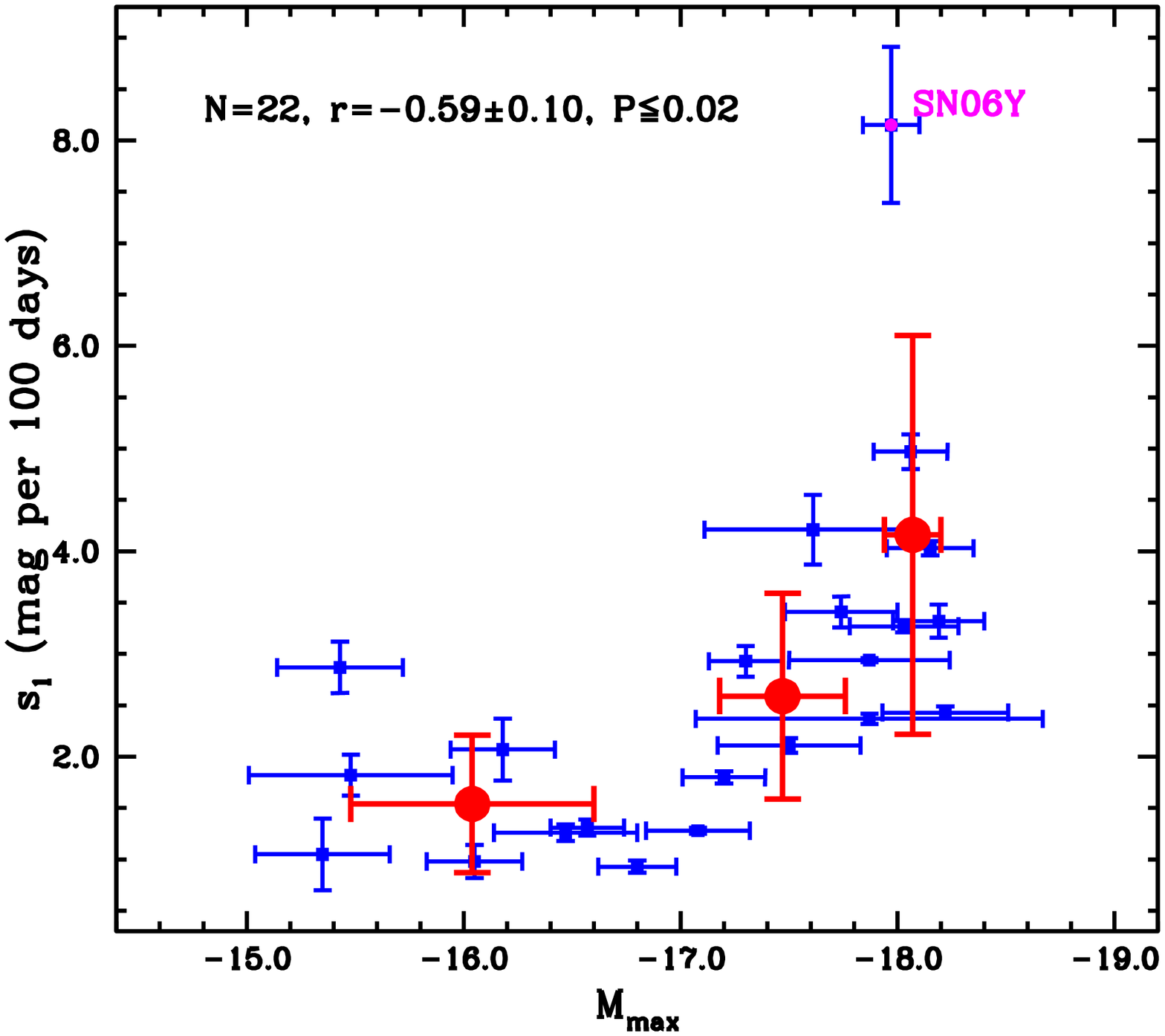}
\caption{Peak magnitude $M_\text{max}$, plotted against the initial decline
rate $s_1$. Individual data points are shown as blue
squares. The results of Monte Carlo simulations on the
statistics of these two variables are noted as in Fig.\ 5.}
\end{figure}

\begin{figure}
\includegraphics[width=8.5cm]{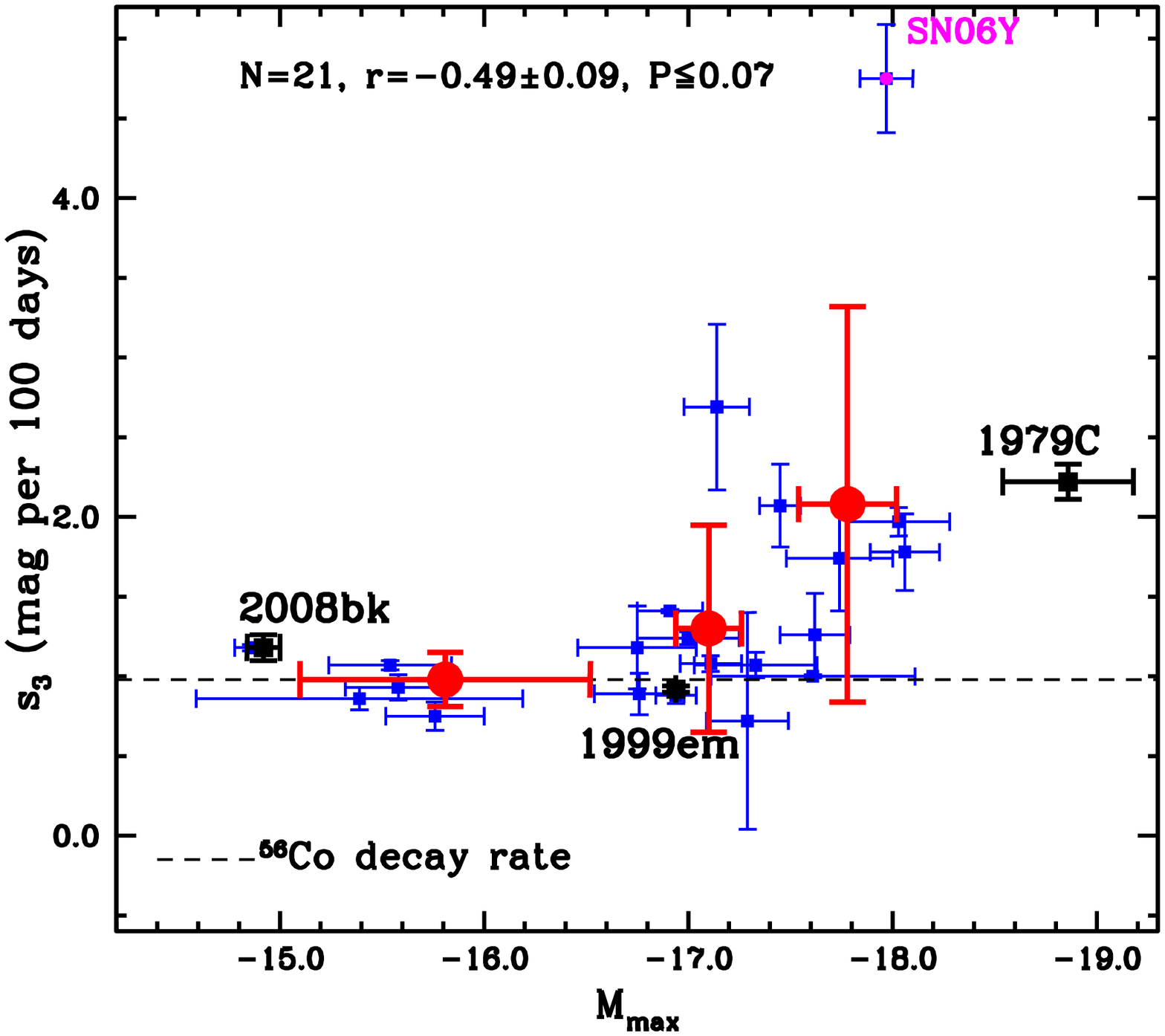}
\caption{SN peak magnitudes, $M_\text{max}$, plotted against the decline
rate during the radioactive tail $s_3$. Individual data points are shown as blue
squares. The dashed horizontal line shows the
expected decline rate on the radioactive tail, assuming full trapping of
gamma-rays from $^{56}$Co to $^{56}$Fe decay. The positions of three different
SNe~II are noted: the sub-luminous type IIP,
SN~2008bk, the prototype type IIP, SN~1999em (both from our own sample), and the
prototype type IIL, SN~1979C \citep{dev81}. The results of Monte Carlo simulations on the
statistics of these two variables are noted as in Fig.\ 5. }
\end{figure}

\subsection{Plateau duration}
In Fig.\ 10 both the $V$-band `plateau' duration ($Pd$) distribution, together with its
correlation with $M_\text{max}$ are displayed. A large range of $Pd$ values is
observed, from the shortest of $\sim$27 days (again the outlier SN~2006Y
discussed above) to the longest of $\sim$72 days (SN~2003bl), with no signs of
distribution bimodality. The mean $Pd$ for the 19 events where a
measurement is possible is 48.4 days, with a
standard deviation of 12.6. While statistically there is only small evidence for a
trend, it is interesting to
note that the two SNe with the longest $Pd$ durations are also of low-luminosity,
while that with the shortest $Pd$ is one of the most luminous
events.\\
\indent In Fig.\ 11 both the $V$-band optically thick phase duration ($OPTd$) distribution, 
together with its
correlation with $M_\text{max}$ are displayed. Again, a large range in values is
observed with a mean $OPTd$ of 83.7 days, and $\sigma$= 16.7. 
The mean error on estimated $OPTd$ values is 7.8$\pm$3.0 days. Hence, the
standard deviation of $OPTd$ values is almost twice as large as the typical error on any given 
individual SN. This argues that the large range in observed $OPTd$ values 
is a true intrinsic property of the analyzed sample.
The shortest duration of this
phase is SN~2004dy with $OPTd$= 25 days, while the largest $OPTd$ is for
SN~2004er, of 120 days. 
SN~2004dy is a large outlier within the OPTd distribution, marking it out as a very peculiar SN. Further comment
on this will be left for future analysis, however we do note that \cite{fol04} observed strong \hei\ 
emission in the spectrum of this event, also marking out the
SN as spectroscopically peculiar.
Again, within the $OPTd$ distribution there appears to be a continuum of
events. Although there is no statistical evidence for a correlation between the
$OPTd$ and $M_\text{max}$, the most sub-luminous
events have some of the longest $OPTd$, while those SNe with the shortest $OPTd$
durations are more luminous SNe~II. 
These large continuous ranges in 
$Pd$ and $OPTd$ are in contrast to the claims of  \cite{arc12} 
who suggested that \textit{all} SNe~IIP have `plateau' durations of $\sim$100 days (however we note
that the \citeauthor{arc12} study investigated $R$-band photometry, rather than the
$V$-band data presented in the current work).
Further comparison of the current results to those of \citeauthor{arc12} will
be presented below.\\
\indent In Fig.\ 12 correlations between both $Pd$ and $OPTd$ with $s_2$ are presented. 
While there is much scatter in the relations, these results are 
consistent with the picture of faster declining SNe~II
having shorter duration `plateau' phases (see e.g. \citealt{bli93}). Indeed,
such a trend was first observed by \cite{psk67}.
A similar, but stronger trend is observed when $Pd$ and $OPTd$
are correlated against $s_1$, as presented in Fig.\ 13
\\
\indent Finally, in Fig.\ 14 we present correlations between $Pd$ and $OPTd$ with
$s_3$. As will be discussed in detail below, both $OPTd$ and $s_3$
provide independent evidence for significant variations in the mass of the
hydrogen envelope/ejecta mass at the epoch of explosion. 
The trend shown in Fig.\ 14 in that SNe with longer $OPTd$ values have smaller
$s_{3}$ 
slopes, is consistent with the claim that such variations can be attributed to envelope mass.\\
\indent In summary, generally SNe which have shorter duration $Pd$ and $OPTd$
tend to decline more quickly at all epochs.\\

\begin{figure}
\includegraphics[width=8.5cm]{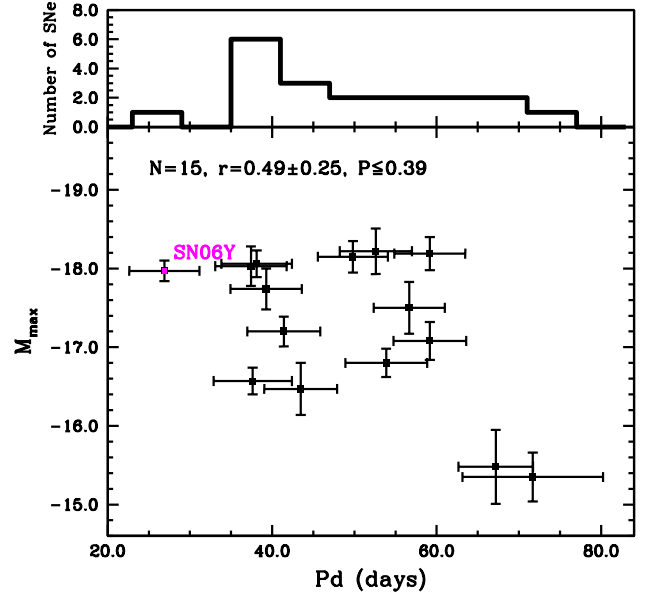}
\caption{\textit{Top panel:} histogram of distribution of
  $Pd$. \textit{Bottom panel:} 
$Pd$ against
$M_\text{max}$. (Note, in the histogram there are more events than in the
correlation due to the removal of those SNe without constrained extinction
values from the correlation). The results of Monte Carlo simulations on the
statistics of these two variables are noted as in Fig.\ 5.
in magenta. }
\end{figure}

\begin{figure}
\includegraphics[width=8.5cm]{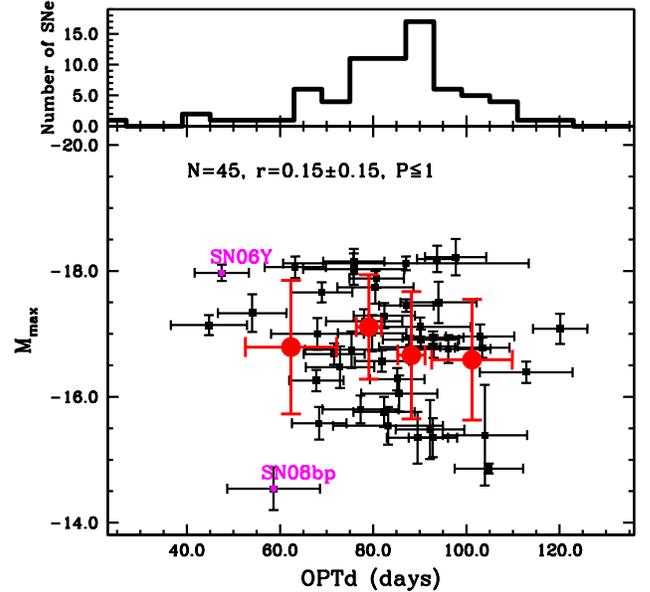}
\caption{\textit{Top panel:} histogram of
  $OPTd$. \textit{Bottom panel:} 
$OPTd$ 
against $M_\text{max}$. (Note, in the histogram there are more events than in the
correlation due to the removal of those SNe without constrained extinction
values from the correlation). The results of Monte Carlo simulations on the
statistics of these two variables are noted as in Fig.\ 5.
In addition, the outlier in Fig.\ 7: SN~2008bp is also seen as an outlier in this plot.}
\end{figure}

\begin{figure}
\includegraphics[width=8.5cm]{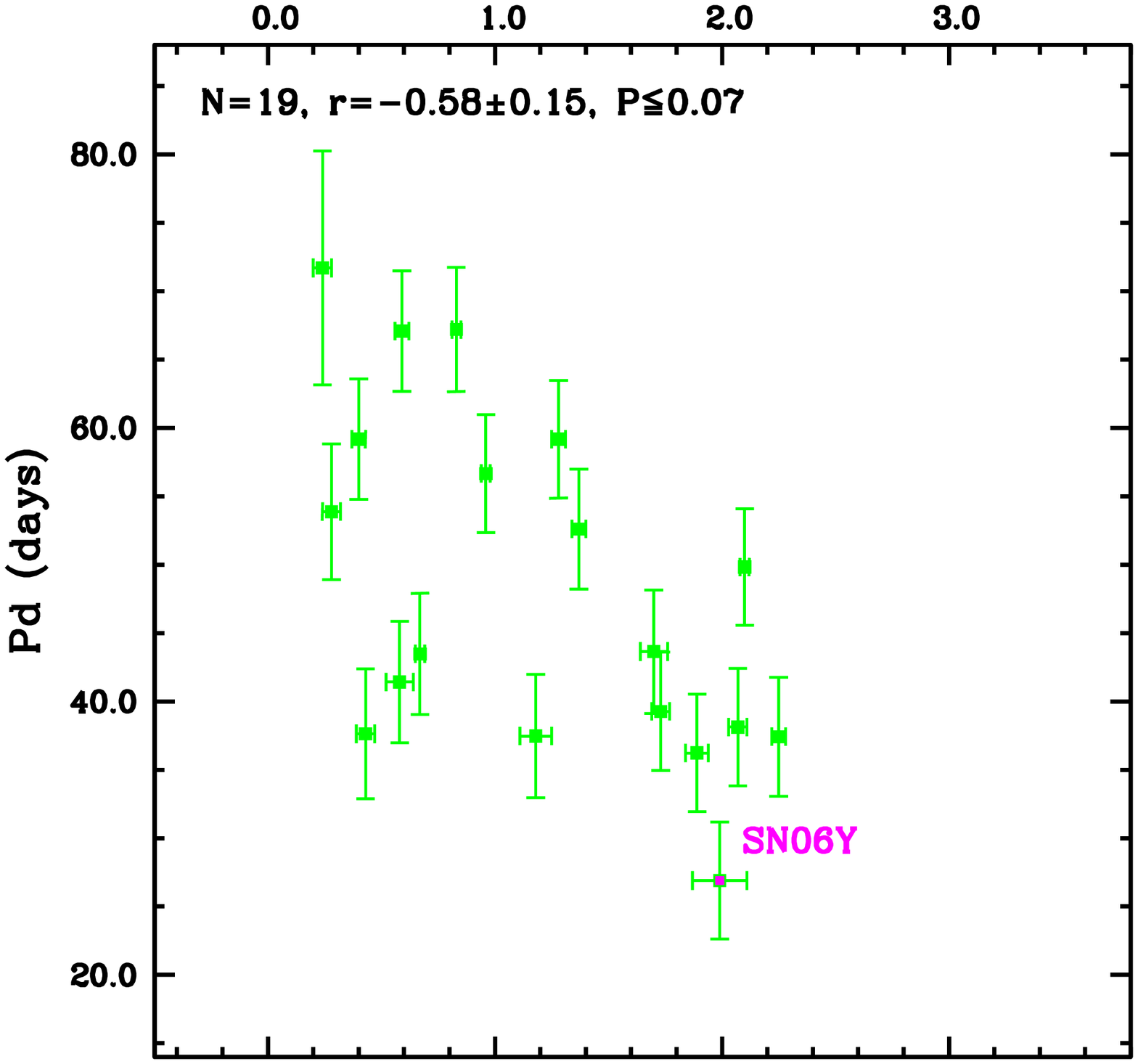}
\includegraphics[width=8.5cm]{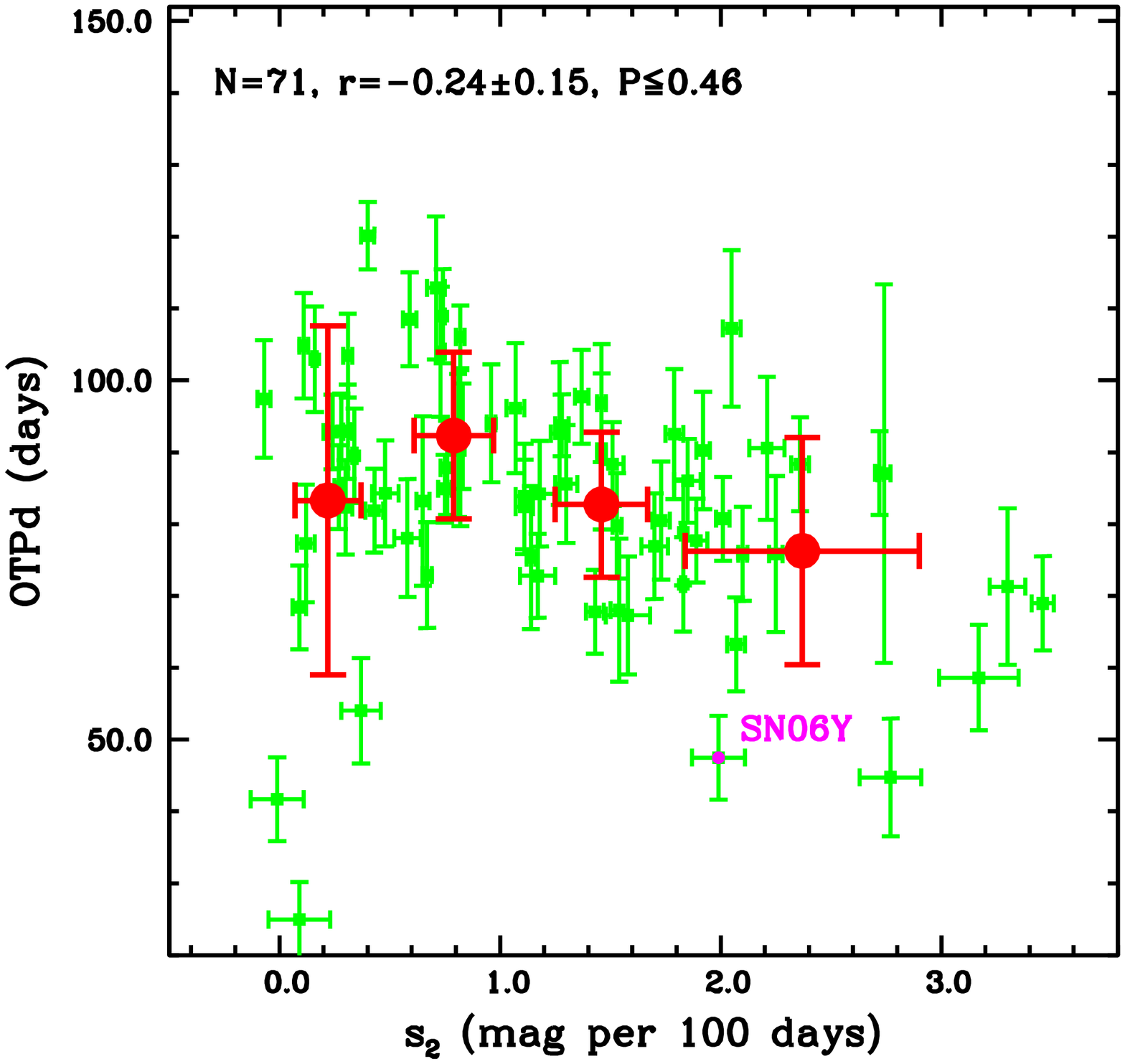}
\caption{\textit{Top panel:} SN `plateau' durations ($Pd$) plotted against
$s_2$. \textit{Bottom panel:} SN optically thick durations ($OPTd$) plotted against
$s_2$. The results of Monte Carlo simulations on the
statistics of these two variables are noted as in Fig.\ 5. }
\end{figure}

\begin{figure}
\includegraphics[width=8.5cm]{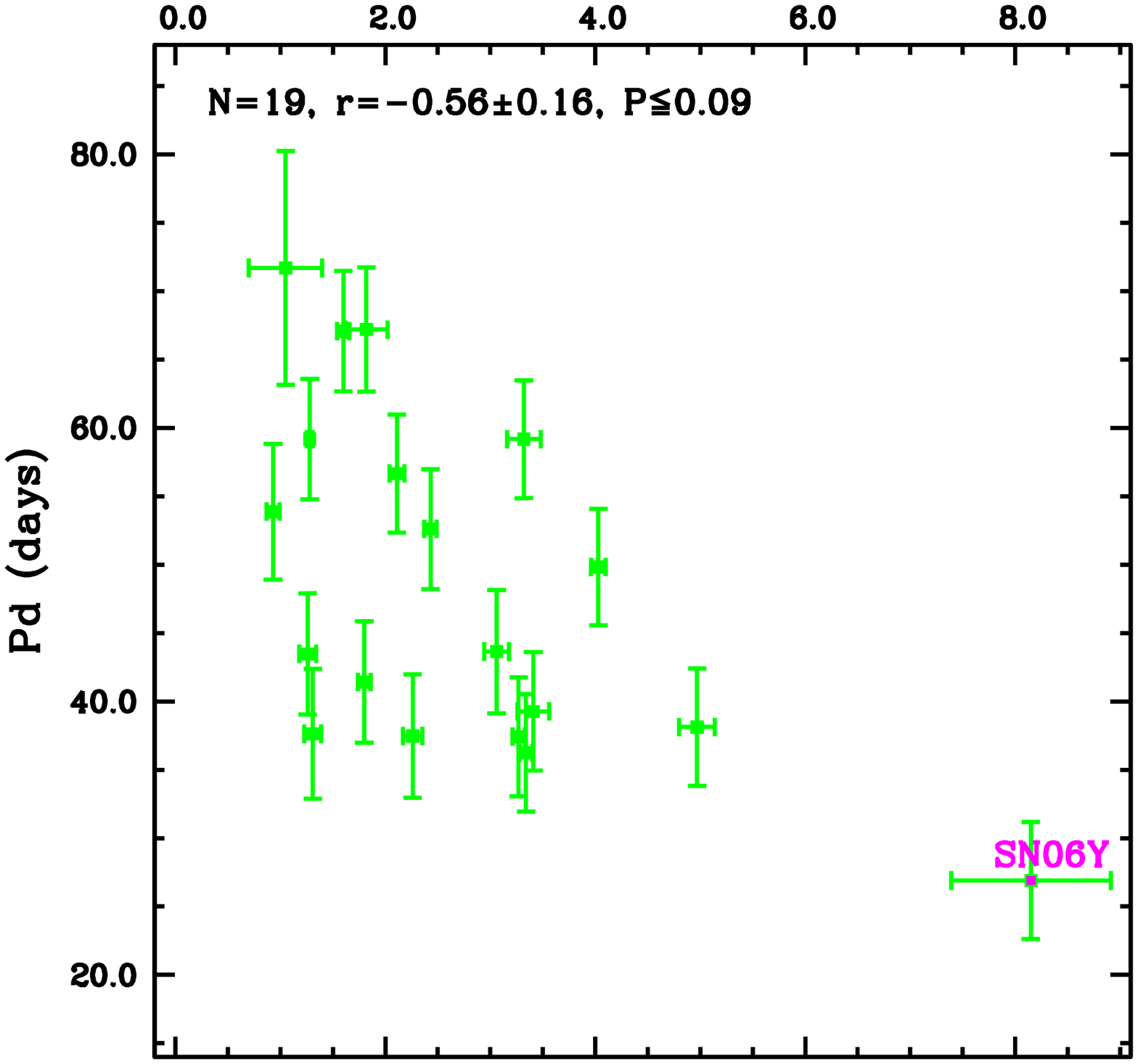}
\includegraphics[width=8.5cm]{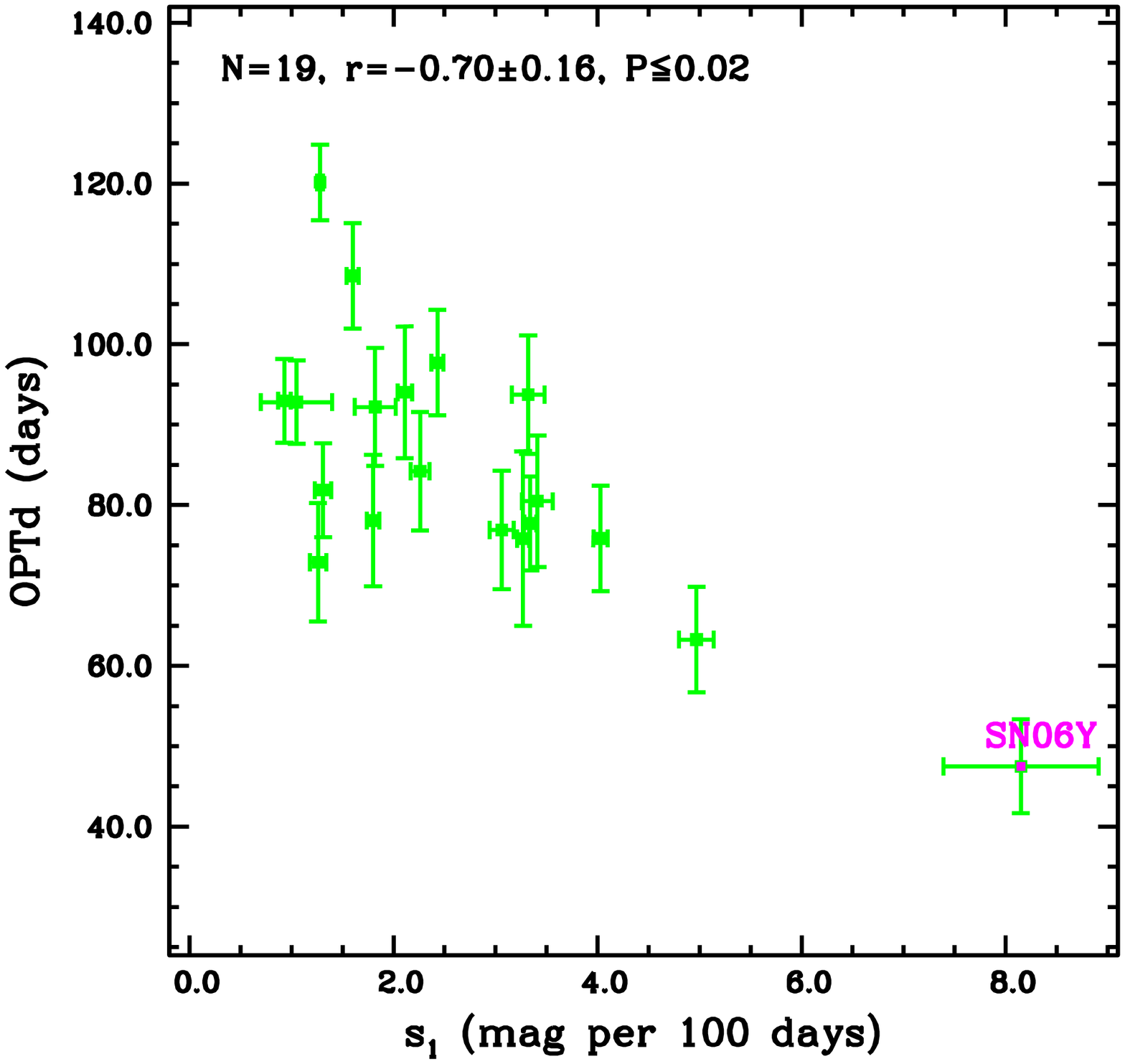}
\caption{\textit{Top panel:} SN `plateau' durations ($Pd$) plotted against
$s_1$. \textit{Bottom panel:} SN optically thick durations ($OPTd$) plotted against
$s_1$. The results of Monte Carlo simulations on the
statistics of these two variables are noted as in Fig.\ 5. }
\end{figure}

\begin{figure}
\includegraphics[width=8.5cm]{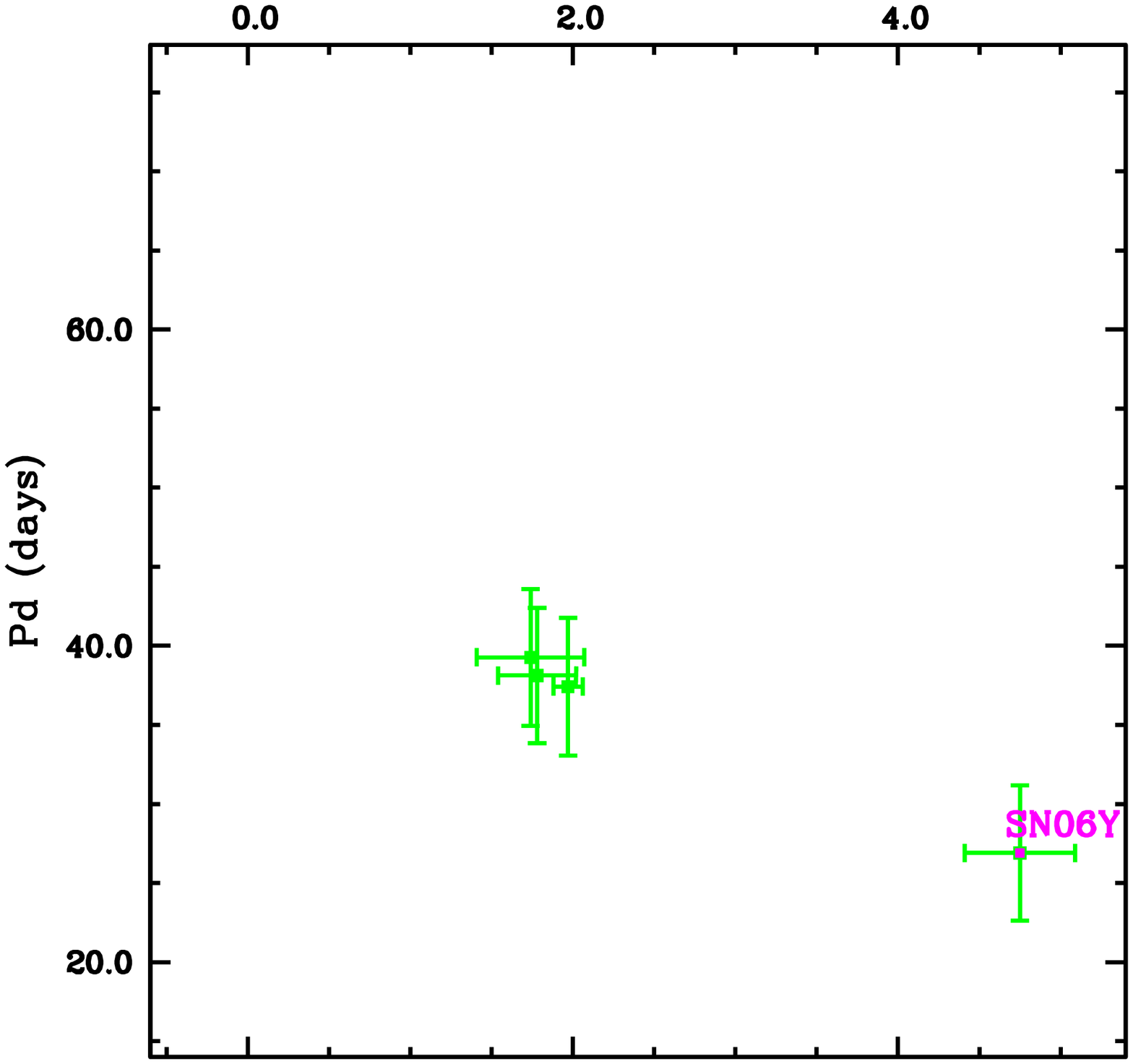}
\includegraphics[width=8.5cm]{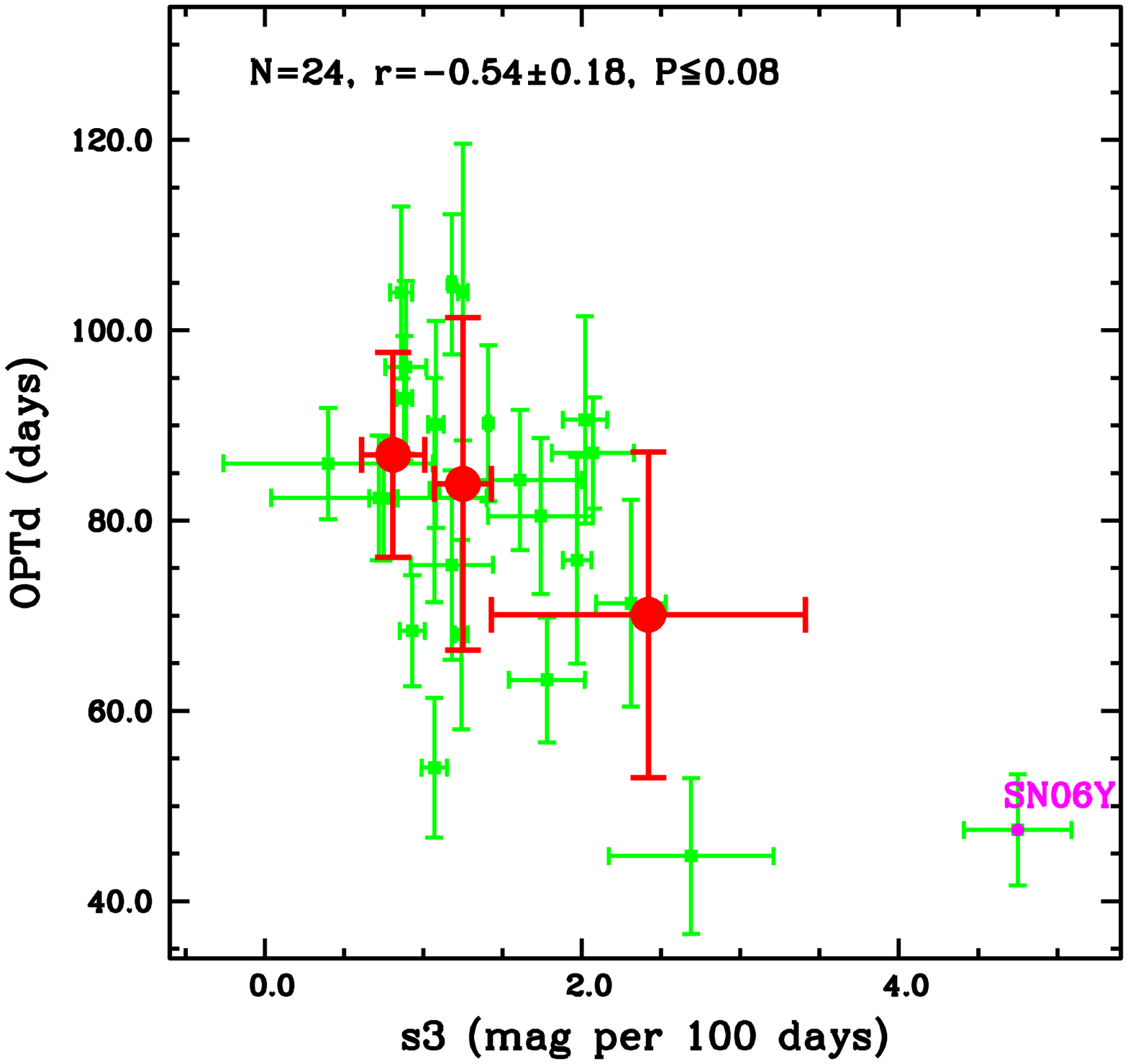}
\caption{\textit{Top panel:} SN `plateau' durations ($Pd$) plotted against
$s_3$. \textit{Bottom panel:} SN optically thick durations ($OPTd$) plotted against
$s_3$. The results of Monte Carlo simulations on the
statistics of these two variables are noted as in Fig.\ 5. }
\end{figure}

\subsection{$^{56}$Ni mass estimates}
Given that the exponential tail (phase $s_3$) of the SN II light-curves are
presumed to be powered
by the radioactive decay of $^{56}$Co (see e.g. \citealt{woo88}), 
one can use the brightness at these epochs to determine the mass
of $^{56}$Ni synthesized in the explosion. 
However, as we have shown above, 
there is a significant spread in the distribution of $s_3$ values, 
implying that not all gamma-rays released during to the decay of $^{56}$Co to 
$^{56}$Fe are fully trapped by the ejecta during the exponential tail, 
rendering the determination of the $^{56}$Ni mass quite uncertain.
Therefore, we
estimate $^{56}$Ni masses only in cases where we have evidence that $s_3$
is consistent with the value expected for full trapping (0.98 mag per 100 days). For
all other SNe with magnitude measurements during the tail, but either $s_3$
values significantly higher (0.3 mag per 100 days higher) than 0.98 (mag per 100 days), or
SNe with less than three photometric points (and therefore no $s_3$ value can be estimated), lower
limits to $^{56}$Ni masses are calculated. In addition, for those SNe without
robust host galaxy extinction estimates (see \S\ 3.3) we also calculate lower limits.\\
\indent To estimate $^{56}$Ni masses and limits, the procedure presented in
\cite{ham03} is followed. $M_\text{tail}$ $V$-band magnitudes are
converted into bolometric luminosities using the bolometric correction
derived by \cite{ham01}, together with the distance moduli and extinction values
reported in Table 6\footnote{The validity of using this bolometric correction for the
entire SN~II sample was checked through comparison with the color dependent
bolometric correction presented by \cite{ber09}. Consistent luminosities and hence $^{56}$Ni masses were found between the two methods.}. 
Given our estimated explosion epochs, mass estimates
are then calculated and are listed in Table 6.\\
\indent In Fig.\ 15 we compare $OPTd$ and $^{56}$Ni masses. \cite{you04} and \cite{kas09} claimed that heating from the radioactive decay
of $^{56}$Ni should further extend the plateau duration. We do not find
evidence for that trend in the current sample (consistent with the finding of \citealt{ber13}). In fact,
if any trend is indeed observed it is the opposite direction to that predicted.\\
\indent The range of
$^{56}$Ni masses is from 0.007\msun\ (SN~2008bk) to 0.079\msun\ (SN~1992af), and
the distribution (shown in Fig.\ 15) has a mean value of 0.033 ($\sigma$= 0.024)\msun.
Given the
earlier trends observed between $s_3$ and other parameters it is probable that
there 
is a systematic effect, and those SNe~II where only lower limits are
possible are probably not simply randomly distributed within the rest of the
population. Therefore, we caution that this $^{56}$Ni synthesized mass
distribution is probably biased compared to the true intrinsic range.\\

\begin{figure}
\includegraphics[width=8.5cm]{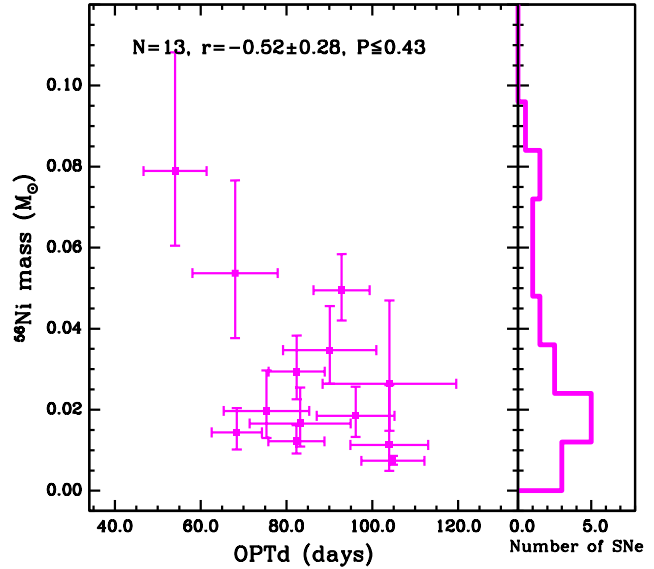}
\caption{\textit{Left:} $OPTd$ against
derived $^{56}$Ni masses. \textit{Right:} Distribution of $^{56}$Ni
masses (lower mass limits are not included in the figure). (Note, in the histogram there are more events than in the
correlation, due to the removal of those SNe without constrained $OPTd$
values). The results of Monte Carlo simulations on the
statistics of these two variables are noted as in Fig.\ 5.}
\end{figure}

\subsection{Transition steepness}
\cite{elm03} reported a correlation between the steepness of $V$-band
light-curves during the transition from plateau to tail phases (named the
inflection time in their work), and estimated $^{56}$Ni masses. With the
high quality data presented in the current work, we are in an excellent
position to test this correlation. The \citeauthor{elm03} method entailed
weighted least squares minimization fitting of a defined function (their equation 1),
to the $V$-band light-curve in the transition
from plateau to radioactive phases in the period of $\pm$50 days around the
inflection point of this transition. The steepness parameter is then defined
as the point at which the derivative of magnitude with time is maximal. We
attempt this same procedure for all SNe in the current sample. We find that it
is possible to define a reliable steepness ($S$) term in only 21 cases. 
This is because one needs extremely well sampled
light curves at epochs just before, during and after the
transition. Even for well observed SNe this type of cadence is uncommon, and
we note that in cases where one can define an $S$-value, additional data
points at key epochs could significantly change the results. 
We do not find any evidence for a correlation as seen by \citeauthor{elm03},
and there does
not appear to be any trend that could be used for cosmological purposes
to standardize SNe~II light-curves (see the appendix for further details).\\

\section{Discussion}
Using the $V$-band light-curve parameters as defined in \S\ 3 and
displayed in Fig.\ 1 we have presented a thorough characterization of 
116 SN~II light curves, both in terms of morphologies and absolute magnitudes, 
and explored possible correlations among parameters.
Our main result is that SNe~II which are brighter
at maximum light ($M_\text{max}$) decline more quickly at all three phases of their
$V$-band light-curve evolution. In addition, our data imply a continuum of $V$-band
SN~II properties such as absolute magnitude, decline rates, and length of the
`plateau' and optically thick phase durations. 
In this Section we further discuss the most interesting 
of our results, in comparison to previous observational and
theoretical SN~II work, and outline possible physical explanations to
explain the diversity of SN~II events found.

\subsection{$M_\text{max}$ as the dominant brightness parameter}
$M_\text{max}$ shows the highest degree of correlation with decline rates 
of all defined SN magnitudes. From an
observational point of view this is maybe somewhat surprising, given the
difficultly in defining this parameter: a true maximum is often not observed
in our data, and hence in the majority of cases we are forced to simply use
the first photometric point available for our estimation. This uncertainty in
$M_\text{max}$ indeed implies that the intrinsic correlation between $M_\text{max}$ and the two
initial decline rates, $s_1$ and $s_2$, is probably even stronger. Paying close
attention to Fig.\ 7 together with Fig.\ 27, 
it is easy to see why this could be the
case. In general a low-luminosity SN has a slow decline rate at both initial
and `plateau' epochs. Therefore, if one were to extrapolate back to the `true'
$M_\text{max}$ magnitude, its measured value would change very little. However, this is
not the case for the most luminous SNe. In general these have much faster
declining light-curves. Hence, if we extrapolate these back to their `true'
peak values, these SNe will have even brighter $M_\text{max}$ values, and hence the strength of the
correlation could be even higher than that presently measured.\\
\indent The statement that faster declining SNe (type IIL) are brighter
than slower declining ones (type IIP) is not a new result. This has been seen
previously in the samples of e.g.\ \cite{psk67,you89,pat94} and \cite{ric02}. However, (to our
knowledge) this is the first time that a wide-ranging correlation as shown in
Fig.\ 7 has been presented with the supporting statistical analysis.

\subsection{A continuum of SN~II properties in the $V$-band}
\cite{arc12} recently claimed that SNe~IIP and SNe~IIL appear to be separated into
two distinct populations in terms of their $R$-band light-curve behavior, possibly
suggesting distinct progenitor scenarios in place of a continuum of events. In
the current paper we have made no attempt at definitive classifications
of events into SNe~IIP and SNe~IIL (ignoring whether such an \textit{objective} classification
actually exists). However, in the above presented distributions and
figures we see no evidence for a separation of events into distinct
categories, or a suggestion of bimodality.
While it is important to note that these separate analyses were undertaken 
using different optical filters, these results are intriguing.\\
\indent In Fig.\ 16 Legendre polynomial fits to all light-curves of SNe with explosion epoch
constraints are presented, the same as in Fig.\ 2, but now with all SNe
normalized to $M_\text{max}$, and also including SNe where host extinction
corrections were not possible. While this figure shows the
wide range in decline rates and light-curve morphologies discussed above,
there does not appear to be any suggestion of a break in morphologies. Given
all of the distributions we present, together with the qualitative arguments
of Fig.\ 16, it is concluded that this work implies a continuum of
hydrogen-rich SNe~II events, with no clear separation between SNe~IIP and SNe~IIL 
(at least in the $V$ band).\\
\indent To further test this suggestion of an observational continuum, in Fig.\ 17 we 
present the same light-curves as in Fig.\ 16, but now presented as a density plot. This is
achieved in the following way. Photometry for each SN was interpolated linearly to the time interval of
each bin (which separates the overall time axis into 20 bins). This estimation starts with the first bin where the 
first photometric point lies for each SN, and ends at the bin containing the last photometric point used to make
Fig.\ 16. This interpolation is done in order to compensate for data gaps and to homogenize the observed
light-curve to the given time intervals. In Fig.\ 16 it is hard to see whether there are certain parts of the 
light-curve parameter space that are more densely populated than others, given that light-curves 
are simply plotted on top of one another. If the historically defined SNe types IIP and IIL showed
distinct morphologies, then one may expect such differences to be more easily observed
in a density plot such as that presented in Fig.\ 17. However, we do not observe any such 
well defined distinct morphologies in Fig.\ 17 (although we note that even the large sample of more
than 100 SN light-curves is probably insufficient to elucidate these differences
through such a plot). In conclusion, Fig.\ 17 gives further weight to the argument that in the 
currently analyzed sample there is no evidence for multiple distinct $V$-band light-curve morphologies,
which clearly separate SNe~IIP from SNe~IIL.\\
\indent The suggestion of a large scale
continuum derived from our analysis is insightful for the physical
understanding of hydrogen-rich explosions. A continuum could imply that the
SN~II population is formed by a continuum of progenitor properties, such as
ZAMS mass, with the possible conclusion that more massive progenitors lose
more of their envelopes prior to explosion, hence exploding with lower mass
envelopes, producing faster evolving light-curves than their
lower progenitor mass companions. 
We note that the claim of a continuum of properties for
explaining differences in hydrogen-rich SNe~II light-curves is also supported by previous
modeling from e.g. \cite{bli93}. These authors suggest that the
diversity of events could be explained through differing pre-SNe radii, the
power of the pre-SN wind, and the
amount of hydrogen left in the envelope at the epoch of explosion.\\

\begin{figure*}
\includegraphics[width=16cm]{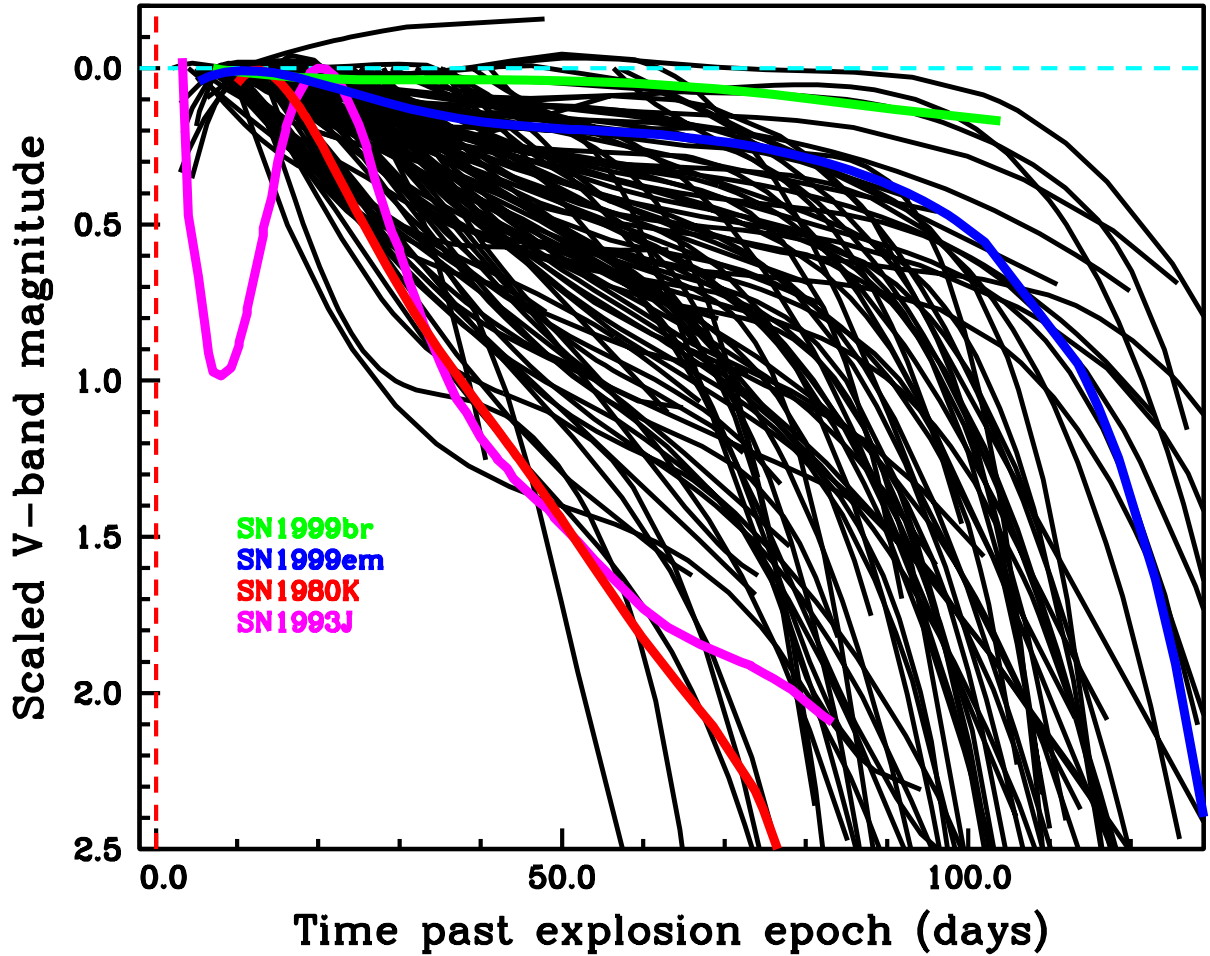}
\caption{A continuum of SNe~II $V$-band light-curve morphologies. 
Similar to Fig.\ 2, however here magnitudes are normalized to peak SN magnitudes, 
and we also include
SNe where host extinction corrections were not possible. For reference, fits to the
data of SN~1980K (\citealt{but82,bar82}, as no explosion date estimate is
available for this event, we assume an epoch of 10 days before the first
photometric point), SN~1999em and SN~1999br are displayed in
colored lines. In addition, we show a fit to the $V$-band light-curve of the type IIb,
SN~1993J \citep{ric94}.}
\end{figure*}

\begin{figure}
\includegraphics[width=8.5cm]{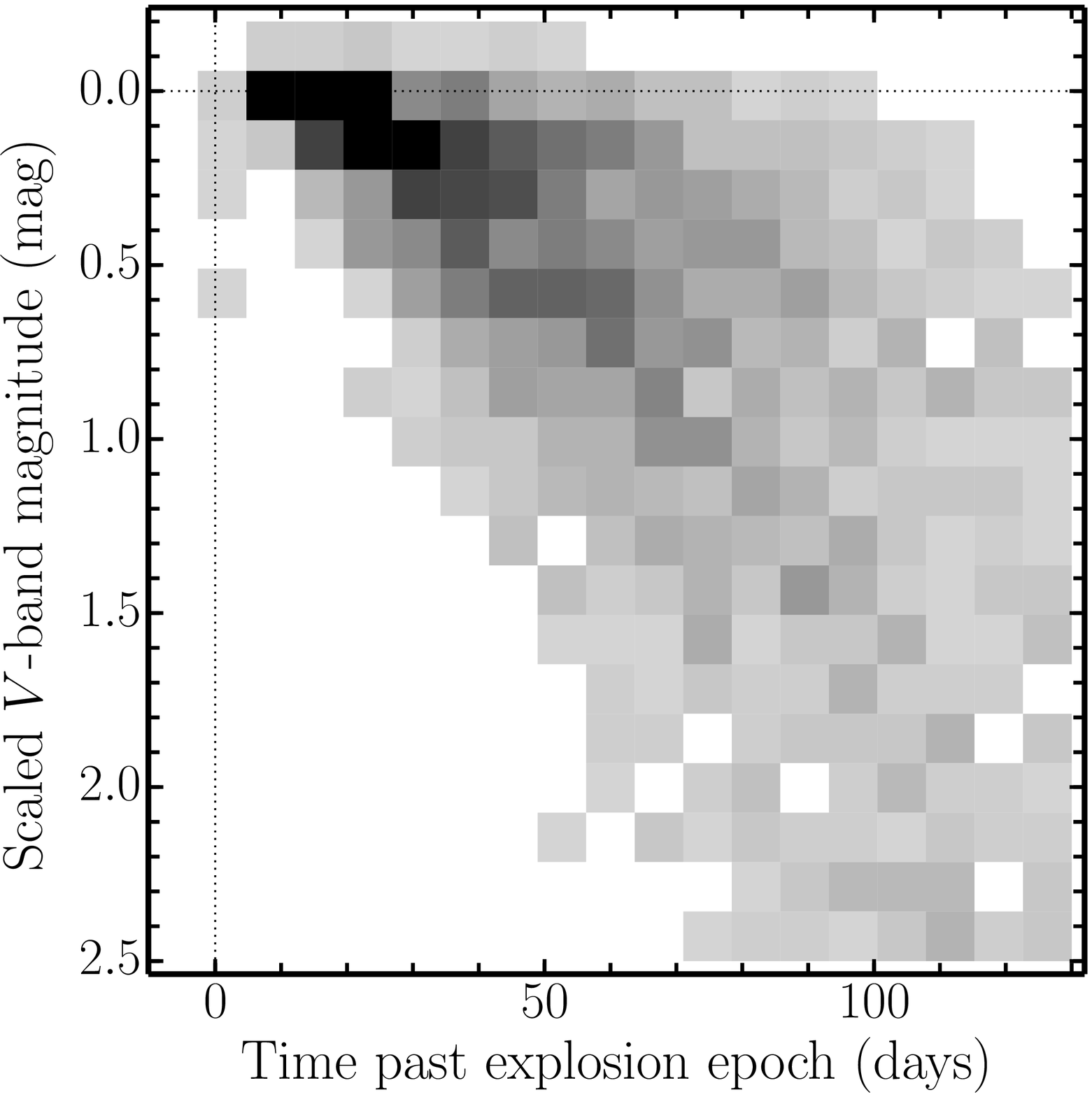}
\caption{2d density histogram of the normalized light-curves presented in Fig.\ 16. 
Photometry is interpolated to the time intervals of the grid, which separate the light-curve
parameter space into $20\times20$ bins.}
\end{figure}

\subsection{A lack of true `linear' SNe~II?}
In the process of this work it has been noted that there is no 
clear objective classification system for defining a SN~IIP from a SN~IIL. 
The
original classification was made by \cite{bar79} who from $B$-band 
light curves, stated that (1) SNe~IIP are characterized 
by a rapid decline, a plateau, a second rapid decline, 
and a final linear phase, which in our terminology correspond 
to $s_1$, $s_2$, transition between `plateau' and tail phases (the mid point
being $t_\text{PT}$), and $s_3$ respectively; 
(2) SNe~IIL are characterized by an almost linear decline up to around 100
days post maximum light (also see \citealt{dog85}). This qualitative classification then appears to have been
used for the last 3 decades without much discussion of its meaning or indeed
validity. A `plateau' in the truest sense of the word would imply something
that does not change, in this case the light-curve luminosity 
remaining constant for a period of at least a few months. However, if we apply this meaning to
the current sample, claiming that anything which changes at a rate of less
than 0.5 magnitudes per 100 days during $s_2$ is of SN~IIP class, then this includes only
$\sim$25\%\ of SNe~II in the current sample, much less than the percentages
of 75\%\ and 95\%\ estimated by \cite{li11} and \cite{sma09} respectively 
(or the 65\%\ in the original publication of \citealt{bar79}). Even if we relax this criterion to 1 magnitude per 100
days, then still only 41\%\ of the current sample would be classified as
SNe~IIP.\\
\indent Meanwhile, as outlined above, the generally accepted terminology for SNe~IIL appears to be that
these events have fast linear decline phases post maximum until they evolve to the
radioactive tail. 
However, it is unclear how many SNe actually exist either in
the literature or within our own sample which fulfill this criterion. Two
literature prototype type IIL events are SN~1979C and SN~1980K. However, while both of these SNe have
relatively fast declining light-curves, they both show evidence for an end to
their `plateau' phases, or `breaks' in their light-curves (i.e. an end to $s_2$ or the optically thick phase, before transition to $s_3$). This is shown in
Fig.\ 18, where the $V$-band photometry for both objects (data from
\citealt{dev81} and \citealt{but82,bar82} respectively) are plotted, together with the data for the three SNe within our sample which
have the highest $s_2$ values. In the case of four of the five SNe plotted in Fig.\ 18 there is
evidence for a `break' in the light-curves, i.e. an end to an $s_2$ phase, while in the remaining
case the data are insufficient to make an argument either way. (We note that 
in the case of SN~1979C one observes a smooth transition between phases, more than a sharp change
in light-curve shape, while in the case of SN~2003ej the evidence for a `break' 
is supported by only one data point after the $s_2$ decline).\\
\indent It could be that SNe~IIL are not solely those that evolve the quickest,
but those where there is no evidence for a `plateau' or `break' in
their light-curves. Therefore, we visually search through our sample
attempting to identify those SNe which are most likely to be classified as
SNe~IIL in the historically defined terminology. Only six cases are identified, together with 
SN~2002ew already presented in Fig.\ 18, and
their $V$-band light-curves are displayed in Fig.\ 19. This figure shows some
interesting features. At least half of the SNe displayed show signs of a maximum around 10-20 days post explosion, 
a characteristic
that is rare in the full SN~II sample. After early epochs it would appear
that these events could be defined as having `linear' morphologies. 
However, there are two important caveats before one
can label these as SNe~IIL in the historical sense: 1) none of the seven SNe presented in
Fig.\ 19 has photometry after day 100 post explosion (with only 2 SNe
having data past 80 days), hence it could be that
a `break' before the radioactive tail is merely unobserved due to the lack of
late time data, and 2) for about half of the events the cadence at
intermediate epochs is such that one may have simply missed any `break'
phase during these epochs. Putting these caveats aside, even if we assume that
these 7 events should be classified as SNe~IIL, this would amount to a
relative fraction of SNe~IIL events in the current sample of $\sim$6\% (of
hydrogen rich SNe~II),
i.e. at lowest limit of previously estimated percentages \citep{bar79,sma09,li11}. Hence, we conclude that if
one uses the literal meaning of the historically defined SNe~IIL class, then these
events are intrinsically extremely rare.\\

\begin{figure}
\includegraphics[width=8.5cm]{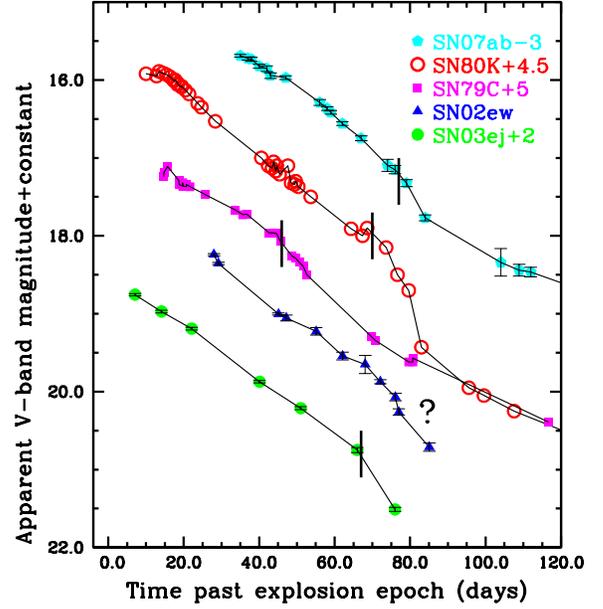}
\caption{$V$-band light-curves of the two prototypical SNe~IIL:
1979C \citep{dev81} and 1980K \citep{but82,bar82}, together with light-curves of the 3 SNe within our sample
with the highest $s_2$ values: SNe 2002ew, 2003ej and 2007ab. On each
light-curve we indicate where there is a possible `break' (at the end of $s_2$), before the SN
transitions to the radioactive tail (for SN~2002ew a '?' indicates that any such epoch is unclear). }
\end{figure}

\begin{figure}
\includegraphics[width=8.5cm]{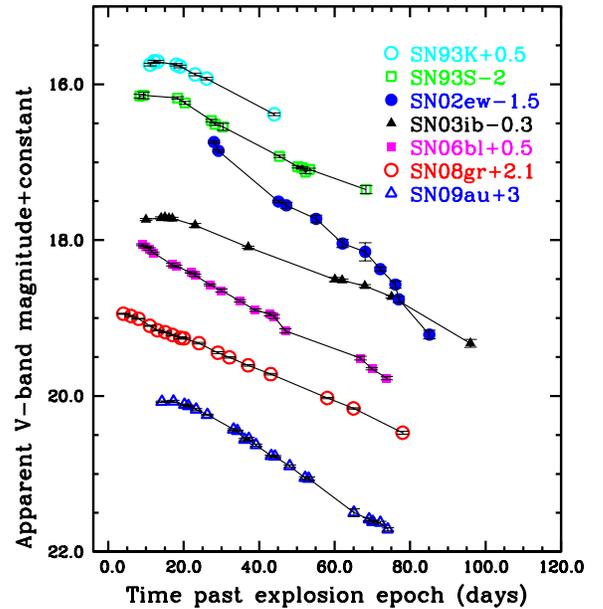}
\caption{Seven SNe~II $V$-band light-curves taken from our sample to be the most
probable SNe~IIL events.}
\end{figure}

\subsection{Light-curve classifications of SNe~II}
Given the arguments presented in the previous two sub-sections: a continuum of
events; no clear separation between the historically defined SNe~IIP and
SNe~IIL; a possible lack of any true steeply declining `linear' SNe~II; many SNe
having much faster declining light-curves than the prototypical type IIP SN~1999em, we believe
that it is time for a reappraisal of the SN~II light-curve classification
scheme. Indeed, the current terminology in the literature, and its use by
different researchers prohibits consistent statistical analyses and gives virtually no
quantitative information on light-curve morphology. It is
therefore proposed that given the easiest parameter to measure for any given
SN~II is the decline rate during the `plateau' phase, its $s_2$ value, that this
parameter should be used to define any given SN~II. This would follow the similar
procedure of SN~Ia classifications, where individual events are 
further classified by referring to their $\bigtriangleup$\textit{m}$_{15}$ value
(decline in brightness during the first 15 days post maximum light,
\citealt{phi93}) or `stretch' parameter \citep{per97}.
Hence, we propose that to standardize terminology, future SNe~II are
referred to as e.g. `SN1999em, a SN~II with $s_2$= 0.31'. This will enable
investigators to define SN samples in a consistent manner which is not
currently possible. This is particularly pertinent for future surveys (such as
that of the Large Synoptic Survey Telescope, LSST, \citealt{ive08}) 
which will produce thousands of SNe~II light-curves and where simply
referring to a SN as type IIP or type IIL gives little quantitative information. Using a
standardized terminology will allow one to define e.g. a range of SNe~II which
are good standard candles, i.e. possibly only those traditionally classified
as SNe~IIP. However, now one could quantify this by giving a sample range of
e.g. $-0.5<s_2<1$.

\subsection{Physical interpretations of SN~II $V$-band light-curve diversity}

\subsubsection{The prevalence of $s_1$: extended cooling in SNe~II?}
While many historical SNe~II have been published with sufficient
data to observe the initial maximum and decline from maximum (see
e.g. SN~1999em photometry published in \citealt{ham01} and \citealt{leo02}), the initial $s_1$
decline rate has attracted little attention in SN~II discussion (although its
presence was indeed noted in the original SN~II classification publication of \citealt{bar79}, and 
was discussed in more detail for the case of SN~1992H by \citealt{clo96}). In the
current work we have objectively measured $s_1$ values for 28 SNe or
$\sim$24\%\ of our sample, while there is visual evidence for such slopes in
many more SNe. Physically, this initial steeper decline phase is most probably
related to the remaining early-time cooling following shock break-out
(i.e. related to the early-time declining bolometric light-curve: see e.g. \citealt{gra71,fal77}). 
The extent in time, post explosion of this cooling is predicted to
be directly related to the radius of the pre-SN progenitor star (see
\citealt{ber12}, for recent application of this hypothesis to a SN~IIb).
It is therefore intriguing to note that these $s_1$ durations continue for several weeks
post explosion. It is also interesting that the $s_1$
distribution spans a wide range in decline rates, and shows
trends with $Pd$ and $OPTd$ as displayed in Fig.\ 13. A full analysis of early-time bolometric
light-curves of CSP and literature SNe, and their comparison to model
light-curves, together with the implications for progenitor properties is underway.\\

\subsubsection{Probing ejecta/envelope properties with $s_3$ and $OPTd$}
Prior to this work, it has generally been assumed that the decline rates of
the majority of SNe~II at late times follow that predicted by the decay of
$^{56}$Co, at least until 300-400 days post explosion (see
e.g. \citealt{you04} for predictions of deviations at very late times). 
Indeed this was claimed to be the case in \cite{pat94} (although
see \citealt{dog85} for a suggestion that faster declining SNe have late time
light-curves which deviate from those expected). 
This has been based on the assumption that
SNe~II have sufficiently massive and dense ejecta to fully trap the
radioactive emission. However, as seen in Fig.\ 4 there are
significant deviations from the expected decay rate of $s_3$= 0.98 mag per
100 days, at epochs only $\sim$80-150 days post explosion. 
Indeed, 10 of the 30 SNe~II with measured $s_3$ have values significant
higher than 0.98 mag per 100 days (we note the 
caveat that these observational measurements are in the $V$ band, while theoretical
expected rates are for bolometric magnitudes). 
In the most extreme cases there is a SN with an $s_3$ value
higher than 3 mag per 100 days (SN~2006Y). This suggests
that in these cases the ejecta mass and/or density are too low for full trapping
of the gamma-ray emission resulting from the decay of $^{56}$Co. Given that $s_3$
shows significant correlation with $s_2$ (which can be used as a proxy to
differentiate between `plateau' and `linear' SNe), this would appear to be
direct evidence that faster declining SNe~II have smaller mass and less
dense ejecta, as suggested/predicted (see e.g. \citealt{bli93}).\\
\indent While a detailed analysis and discussion of
ejecta masses and densities implied by the above arguments is beyond the scope
of this paper, the above results and discussion of $s_3$ and its correlations
with other light-curve parameters implies that in a significant
fraction of SNe~II explosions the mass and/or density of the ejecta are
significantly smaller than previously assumed, thus allowing early leakage of
gamma-ray emission.\\
\indent The hydrogen envelope mass at the epoch of explosion has been claimed to directly influence
the optically thick
phase duration  (see \citealt{che76,lit83,lit85,pop93}). 
The standard historical picture is that as normal SNe~IIP transition to more
linear events, the subsequent SNe have shorter,
less pronounced `plateau'/optically thick phases (indeed the SN~1992H;
\citealt{clo96}, 
is likely to be an example
of such a transitional event). This is shown in Fig.\ 12,
where there is marginal evidence for a trend in that more steeply declining SNe have
smaller $Pd$ and $OPTd$ values implying that 
more `linear' events have smaller envelope masses at the time of explosion.
The large range in $OPTd$ values implies that SNe~II explode with a large
range of masses of retained hydrogen envelopes.\\

\subsubsection{Mass extent and density profile of SNe~II as the dominant
physical parameter}
Following the previous discussion, it appears that the majority of diversity
observed in SN~II light-curves and indeed their spectra (see below) can be
described through varying the extent (in mass and density profile) of hydrogen envelopes
retained prior to SN explosion. This is obviously a simplified picture and
there are other properties which will play a role, such as explosion energy,
pre-SN radius, and synthesized $^{56}$Ni mass. 
However, we suggest that the results of the above presented analysis 
are most easily explained by differing hydrogen masses. Indeed, this
would be completely consistent with the SNe~II model parameter space study of
\cite{you04}, who investigated the influence on the light-curve properties of
varying the progenitor radius, envelope mass, explosion energy, $^{56}$Ni
mass, and the extent of the mixing of the synthesized
$^{56}$Ni. 
\citeauthor{you04} concluded that the primary parameter (together
with progenitor radius) that affects the overall behavior of SNe~II
light-curves, and can explain their observed diversity is the observed 
hydrogen envelope mass at explosion epoch. He
predicted that decreasing the hydrogen envelope at the epoch of explosion
would lead to a number of changes in the light-curve properties. Firstly, a
larger mass envelope leads to a longer diffusion time, and hence radiation is
trapped for a longer time, leading to less luminous early time light-curves and
flatter `plateau' phases (i.e. lower $s_2$ values). Secondly, a
smaller mass leads to shorter duration plateaus (or $OPTd$) as the
recombination wave has less mass to travel back through. Finally, a
reduced hydrogen envelope leads light-curve tails to be steeper (higher $s_3$ values), which
is the result of leakage of radioactive emission. This can be seen in our
analysis as presented in Figs 4 and 14 (although we note the caveat that
deviation of $s_3$ values occur at much earlier times in our data than that
predicted).
We also note that \cite{des13} analyzed the dependencies of SN~IIP radiation on
progenitor and explosion properties. They found that explosions with higher kinetic energies lead
to SNe with brighter and shorter plateau phases. While that study did not analyze 
the effects of changes to the hydrogen envelope mass, Hillier et al. (in preparation) 
have included such changes, finding that explosions with lower envelope masses indeed are brighter,
have faster declining light-curves with shorter `plateau' phases, consistent with
our conclusions here (Dessart 2013 private communication).\\
\indent If the above hypothesis is true, together with our claim of a continuum of
SN~II properties, then the next question becomes: how do different
progenitors evolve to produce this diversity in final hydrogen envelope
profiles? The dominant process would appear to be the mass loss history of
each progenitor. The degree of mass loss suffered by high mass stars is
influenced by progenitor properties such as ZAMS mass, metallicity, binarity and
rotation (with mass loss rates increasing for increasing mass and
metallicity, and the presence of a close binary companion accelerating the
process). 
However, to date there are few constraints on how these parameters
change with respect to the diversity of events observed. Progenitor mass
constraints have been most accurately determined through direct detection of
progenitor stars on pre-explosion images. \cite{sma09} derived a progenitor
mass range for SNe~IIP of 8-16\msun. However, the upper limits of this range
has been questioned by \cite{wal12} due to the effects of circumstellar dust (although
see \citealt{koc12b}),
and in addition, \cite{fra12} and \cite{mau13} have published detections of
more recent SNe~IIP, suggesting higher masses for two further events. In the
case of
SNe~IIL there have been a couple of progenitor detections, with suggestions
that these could indeed arise from higher zero age main sequence (ZAMS) 
mass progenitor stars than SNe~IIP
(e.g. \citealt{eli10,eli11}). This would then be consistent with faster
declining SNe~II losing a higher degree of their envelopes prior to explosion,
through stronger mass-loss due to their more massive ZAMS progenitors, and hence
their light-curve behavior being similar to that presented above (also
consistent with the fact that historical SNe~IIL, e.g. SN~1980K, SN~1979C, have also been strong radio
emitters, \citealt{wei89}). This picture is also
supported by work on the environments of CC SNe, where SNe~IIL show a
higher degree of association to host galaxy on-going star formation
(as traced by \ha\ emission) than their SNe~IIP counterparts
\citep{and12}. Changes in progenitor metallicity (see \citealt{des14} for discussion on how progenitor
metallicities may be directly probed from SN spectra), binarity and rotation remain
relatively unconstrained, but are also likely to affect the transient behavior and
diversity of SNe~II events.\\
\indent Finally, at this stage it is important to mention the spectral
analysis currently being achieved on the same sample analyzed here. Guti\'errez
et al. (submitted) are focusing on the profile of the dominant
\ha\ P-Cygni profiles of hydrogen-rich SNe~II events. In that paper it is shown that
the events we discuss here as being bright, having steeply declining
light-curves, and shorter duration phases, also in general have less prominent
P-Cygni absorption features with respect to emission (together with higher
\ha\ measured velocities), a characteristic which again is likely to be linked
to smaller mass hydrogen envelopes at the epoch of explosion (see
e.g. \citealt{sch96}). Anderson et al. (submitted) have also shown that
the strength of the blueshift of the peak of emission of \ha\ correlates with both
$s_2$ and $M_\text{max}$, in the sense that brighter and faster declining SNe~II have 
higher blue-shifted velocities than their dimmer, slower declining counterparts.

\begin{figure*}
\includegraphics[width=17cm]{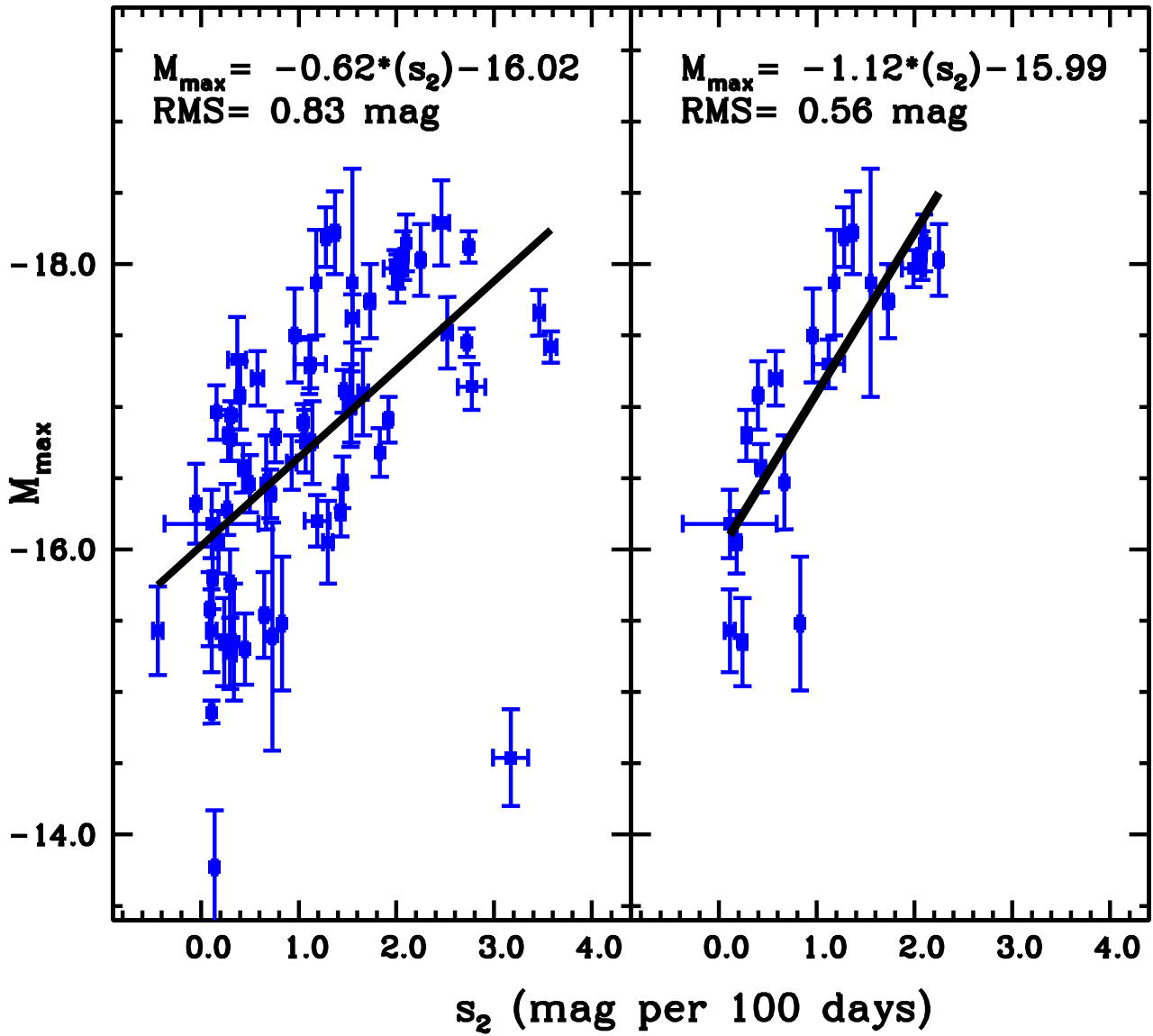}
\caption{Correlations between $s_2$ and $M_\text{max}$. \textit{Left} The full
sample as displayed earlier in Fig.\ 7. \textit{Right} SNe~II which have
measured $s_1$ and $s_2$ values. In each figure the line of best fit,
estimated through Monte Carlo bootstrapping is presented, together with the
linear relation between $s_2$ and $M_\text{max}$, and the RMS dispersion in
magnitudes of the derived relations.}
\end{figure*}

\subsection{SNe~II as photometric standardizable candles?}
SNe~II (or specifically SNe~IIP) have long been discussed as complementary (to SNe~Ia) distance
indicators, with a number of techniques being employed to standardize their
luminosities (e.g. the expanding photosphere method, EPM:
\citealt{kir74,eas96,des05,des08,jon09;bos14}, the spectral fitting expanding
photosphere method, SEAM: \citealt{mit02,bar04}, and the standard candle
method, SCM: \citealt{ham02,nug06,poz09,dan10,oli10}). 
However, a key issue that currently precludes their use for
higher redshift cosmology is the need for spectral measurements in these
techniques. Any photometric relation between SN~II luminosities and other
parameters, if found could enable SNe~II to be used as accurate high-redshift
distance indicators. This would be particularly pertinent when one keeps
pushing to higher redshift where the SN~Ia rate is expected to drop due to
their relative long progenitor lifetimes, while the SNe~II rate should remain
high as it follows the star formation rate of the Universe.\\
\indent In Fig.\ 7 a correlation between $M_\text{max}$ and $s_2$ was presented,
where more luminous events show steeper declining light-curves. In Fig.\ 20
we now invert the axis of the earlier figure in order to evaluate the
predictive power of $s_2$. The best fit line is
estimated by running 10,000 Monte Carlo bootstrap simulations and calculating the mean of
the slopes and y-intercept values obtained. This then leads to a relation between
$s_2$ and $M_\text{max}$ for the full sample, as presented in the left panel
of Fig.\ 20, which has a dispersion of 0.83 mag. However, it was earlier
noted in \S\ 4.2 that in this full sample we are including SNe where only a
measurement of $s_2$ was possible (i.e. no distinction between $s_1$ and
$s_2$). This may bias the sample, as if an $s_1$ is not detected 
but intrinsically is present, then it will merge with $s_2$, and
probably artificially increase the value of $s_2$ we present in Table 6.   
Following this we make a cut to our sample to only include SNe where both the
measurement of $s_1$ and $s_2$ are favored (see \S\ 3). 
We then obtain a sample of 22 events. The correlation between
$s_2$ and $M_\text{max}$ for this curtailed sample is stronger: running our
Monte Carlo simulated Pearson's test gives, $r= -0.82\pm0.05$ and $P\leq3\times10^{-5}$.
In the right panel of Fig.\ 20 this correlation is presented, and the
following relation is derived:
\begin{itemize}
\item[] $M_\text{max}= -1.12(s_2) -15.99$
\end{itemize}
This relation has a dispersion of only 0.56 mag. While this is still higher
than that published for SN~II spectroscopic distance methods (and considerably higher than
SN~Ia), the predictive power of $s_2$ is becoming very promising. If the accuracy
of parameters such
as host galaxy extinction can be improved, and SN color information included,
then it may be possible to bring this dispersion down to levels where SNe~II
are indeed viable \textit{photometric} distance indicators, independent of
spectroscopic measurements\footnote{A multi-color analysis of relations explored in this analysis is 
currently underway. This will go some way to determining the final usefulness of this
photometric distance indicator when applied to high redshift objects, where accurate redshifts needed for $K$-corrections
may not always be available.}.\\
\indent A model correlation between the SN luminosity at day 50 post
explosion (closest to our $M_\text{end}$ value) and the duration of the
plateau was predicted by \cite{kas09} (see also \citealt{ber13}),
which they claimed could
(if corroborated by observation) be used to standardize SNe~IIP luminosities. 
We only observe marginal evidence for any such trends.\\

\subsection{Future SNe~II studies}
While statistical studies of SNe~II as that presented here are to date rare 
or even non-existent, they promise to become much more prevalent in the near future,
with larger field of view and deeper transient searches planned or underway. 
During
the course of this investigation it has become obvious that a number of key
attributes are needed to efficiently further our understanding of
hydrogen-rich SNe~II through well planned observations. Firstly, strong
constraints on the explosion epoch is key, as this allows measurements to be
derived with respect to a well defined epoch (in the case of SNe~Ia one can
use the maximum of the light-curve, however for SNe~II such a method contains
higher levels of uncertainty). Secondly, high cadence photometry is warranted to be
able to differentiate between different slopes such as $s_1$ and $s_2$. Indeed, it
appears that it is often assumed that SNe~IIP light-curves are reasonably well
behaved and therefore one can obtained photometry in a more relaxed
fashion. However, it is clear from the current work that photometry obtained
every few days is needed to further our understanding of the diversity of
SNe~II: in general one does not know at what stage a SN~II will transition
from one phase to the other, and data at those points are vital for the phases to be
well constrained. Thirdly, detailed observations of the radioactive tails are needed to
confirm and further investigate deviations from $s_3$ values expected from full
trapping. Such observations contain direct information on the ejecta profiles
of each SN and are key to understanding the progenitor and pre-SN properties
of SNe~II. While the current study has opened many avenues for further
investigation and intriguing insights to the underlying physics of SNe~II
explosions, continued high quality data are needed to deepen our understanding
of hydrogen-rich SNe~II. As we have hinted throughout this paper, the CSP has
obtained multi-color optical and near-IR light-curves, together with high
quality spectral sequences of a large sample of SNe~II. The near-IR data could prove key to the
understanding of SNe~II, and their use as distance indicators, since they are
essentially unaffected by dust extinction.
The full analysis of those samples will present a
large increase in our understanding of the observational diversity of SNe~II events.

\section{Conclusions}
An analysis of $V$-band photometry of 116 SNe~II has been presented
with the aim of characterizing the diversity seen within their light-curves. 
This has been achieved through defining three magnitude
measurements at different epochs: $M_\text{max}$, $M_\text{end}$, 
$M_\text{tail}$, three photometric
decline rates: $s_1$, $s_2$ and $s_3$, together with the time durations $Pd$ and $OPTd$. 
We analyzed these distributions, and searched for possible correlations. 
Our main findings
are that the SN~II family forms a continuum of events in terms of their
light-curve morphologies (in the $V$ band), and that while large dispersion is observed, 
brighter SNe at maximum generally decline more quickly
at all epochs. We speculate that the majority of the diversity of SNe~II 
can be
explained through differences in their hydrogen envelope masses at the epoch
of explosion, a parameter which is most directly measured through observations of the
optically thick phase duration ($OPTd$).
Finally, we list our main
conclusions originating from this work.\\
\begin{itemize}
\item A continuum of SN~II $V$-band properties is observed in all measured parameters
(absolute magnitudes, decline rates, optically thick phase durations),
and we observe no clear bimodality or separation between the historically
defined SNe~IIP and SNe~IIL.
\item SNe which are brighter at maximum decline more quickly at all epochs.
\item After making a series of data quality cuts, it is found that the
dispersion in the relation between $s_2$ and $M_\text{max}$ can be reduced
to 0.56 mag, which opens the way to using SNe~II as photometric distance
indicators, independent of spectroscopic information.
\item While $M_\text{max}$ is more difficult to define and measure than other
magnitudes, it shows the highest degree of correlation with decline rates. 
Hence, it appears that $M_\text{max}$ is the dominant magnitude parameter
describing the diversity of SNe~II.
\item We find a large range in $V$-band optically thick, and `plateau' durations
($OPTd$, $Pd$) which implies a large range in hydrogen envelope masses at the
epoch of explosion. The fact that these parameters
show correlation with a number of other light-curve parameters suggests
that one of the most
dominant physical parameters that explains the diversity of SNe~II
light-curves is the envelope mass at the epoch of explosion.
\item There are a significant number of SNe~II which decline more quickly during
the radioactive tail, $s_3$, than the rate expected through full trapping of gamma-ray
emission. This implies a large range in ejecta masses, with many SNe~II having
low mass/density ejecta, through which emission can escape.
\item Given the qualitative nature of current discussion in the literature of
different SNe~II, we suggest the introduction of a new parameter, $s_2$: the
decline rate per 100 days of the $V$-band light-curve during the `plateau'
phase. This will enable future studies to make quantitative
comparisons between SNe and SNe samples in a standardized way.
\item The historically defined SN~IIL class does not appear to be
significantly represented within this sample, and therefore it is concluded
that truly `linearly' declining hydrogen-rich SNe~II are intrinsically
extremely rare events. 
\item SN~II $V$-band magnitudes show a dispersion at the end of
the `plateau', $M_\text{end}$ of 0.81 mag, 0.2 mag lower than
that at peak, $M_\text{max}$.
\end{itemize}

\acknowledgments
The referee is thanked for their thorough reading of the manuscript, which
helped clarify and improve the paper.
J.~A., S.~G., F.~F., and S.~S. acknowledge support from CONICYT through
FONDECYT grants 3110142, 3130680, 3110042, and 3140534 respectively.
J.~A., S.~G., M.~H., C.~G., F.~O., S.~S., F.~F., and T.~D. 
acknowledge support by projects IC120009 ``Millennium Institute
of Astrophysics (MAS)" and P10-064-F ``Millennium Center for Supernova
Science" of the Iniciativa Cient\'ifica Milenio del Ministerio Econom\'ia,
Fomento y Turismo de Chile. F.~F.
acknowledges partial support from Comite Mixto ESO-GOBIERNO DE CHILE.
The work of the CSP has been supported by the 
National Science Foundation 
under grants AST0306969, AST0607438, and AST1008343.
M.~D.~S. and C.~C. gratefully acknowledge generous support 
provided by the Danish Agency for Science and Technology and Innovation  
realized through a Sapere Aude Level 2 grant.
This research
has made use of the NASA/IPAC Extragalactic Database (NED) 
which is operated by the Jet 
Propulsion Laboratory, California
Institute of Technology, under contract with the National Aeronautics.

\bibliographystyle{aa}
\bibliography{Reference}

\appendix

In the above main body of this manuscript we have tried to present a broad
overview of our analysis and resulting conclusions, without discussing every
correlation or avenue of investigation that formed part of this SN~II
light-curve exploration. For completeness, in this appendix several more 
figures are presented to further elaborate on this theme, and provide further
examples of the methods used throughout the work earlier discussed. Finally,
Table 6 presents values for all measured parameters for the full SN sample.\\

\begin{table*}
\centering
\caption{Spectoscopic classification information of the SN~II sample}
\begin{tabular}[t]{ccccc}
\hline
SN & Classification& N. of spectra & Earliest and& Comments\\
& reference&                       & latest epochs&\\
\hline	
\hline
1986L & \cite{llo86}& 28 & +6, +118& Spectra analyzed in \cite{ham03}\\
1991al& \cite{bou91}& 8  & +30, +125& Spectra analyzed in \cite{ham03,oli10}\\
1992ad& \cite{mcn92}& 0  & $\cdots$& Classification from circular\\
1992af& \cite{del92}& 5  & +22, +136& Spectra analyzed in \cite{ham03,oli10}\\
1992am& \cite{phi92}& 2  & $\cdots$& Spectra published in \cite{sch94},\\ 
&&&&and further analyzed in \cite{ham03,oli10}\\
1992ba& \cite{eva92}& 8  & +9, +180 & Spectra analyzed in \cite{ham03,jon09,oli10}\\
1993A&  \cite{phi93_2}& 2& +20, +103& Spectra analyzed in \cite{ham03,oli10}\\
1993K&  \cite{ham93_2}& 10& +13, +363& \\
1993S&  \cite{ham93_3}& 4& +35, +94& Spectra analyzed in \cite{ham03}\\
1999br& \cite{gar99}& 8  & +16, +75& Spectra analyzed in \cite{ham03,jon09,oli10},\\
&&&&Literature data published in \cite{pas04}\\
1999ca& \cite{pat99}& 4  & +28, +41& Spectra analyzed in \cite{ham03,oli10}\\
1999cr& \cite{maz99}& 5  & +11, +57&Spectra analyzed in \cite{ham03,oli10} \\
1999eg& \cite{jha99_2}& 2& +27, +61& Spectra analyzed in \cite{ham03}\\
1999em& \cite{jha99_3}& 7& +10, +168& Spectra published in \cite{ham01}, and further\\
&&&& analyzed in \cite{ham03,jon09,oli10}.\\
&&&& Literature data published in \cite{bar00,leo02}\\
0210& $\cdots$& 6 & +56, +87& Spectra analyzed in \cite{ham03}\\
2002ew& \cite{cho02},& 7 & +30, +76& \\
      & \cite{fil02_2}&&&\\
2002fa& \cite{ham02_2}& 6 & +27, +74&Spectra analyzed in \cite{oli10} \\
2002gd& \cite{ham02_3}& 12& +4, +97& \\
2002gw& \cite{ham02_4}& 11& +14, +91& Spectra analyzed in \cite{jon09,oli10} \\
2002hj& \cite{cho02_2}& 7 & +24, +86& Spectra analyzed in \cite{oli10}\\
2002hx& \cite{mat02}& 9   & +25, +121& Spectra analyzed in \cite{oli10}\\
2002ig& \cite{mik02}& 5   & +17, +64& Spectra indicate hydrogen-rich type II event\\
2003B&  \cite{kir03}&  9  & +24, +282& Spectra analyzed in \cite{oli10}\\
2003E&  \cite{ham03_3}& 8 & +15, +131& Spectra analyzed in \cite{oli10}\\
2003T&  \cite{fol03}& 6   & +20, +111& Spectra analyzed in \cite{jon09,oli10}\\
2003bl& \cite{phi03}& 8   & +3, +96& Spectra analyzed in \cite{jon09,oli10}\\
2003bn& \cite{sal03}& 12  & +13, +127& Spectra analyzed in \cite{jon09,oli10}\\
2003ci& \cite{sal03}& 5   & +19, +87& Spectra analyzed in \cite{oli10}\\
2003cn& \cite{ham03_4}&5  & +11, +79& Spectra analyzed in \cite{oli10}\\
2003cx& \cite{ham03_5}&6  & +12, +93& Spectra analyzed in \cite{oli10}\\
2003dq& \cite{phi03_2}&3  & +34, +64& \\
2003ef& \cite{gan03}& 6   & +31, +107& Spectra analyzed in \cite{jon09,oli10}\\
2003eg& \cite{gan03}& 5   & +17, +99& \\
2003ej& \cite{mat03}& 3   & +15, +41& \\
2003fb& \cite{pap03}& 4   & +22, +96& Spectra analyzed in \cite{oli10}\\
2003gd& \cite{kot03}& 3   & +51, +141& Spectra analyzed in \cite{oli10}, literature\\
&&&& data published in \cite{hen05}\\
2003hd& \cite{ham03_6}& 9 & +10, +133& Spectra analyzed in \cite{oli10}\\
2003hg& \cite{eli03}&  5 & +25, +108&Spectra analyzed in \cite{oli10}\\
2003hk& \cite{fil03_2}& 4 & +34, +104& Spectra analyzed in \cite{oli10}\\
2003hl& \cite{fil03_2}& 6 & +34, +129& Spectra analyzed in \cite{jon09,oli10}\\
2003hn& \cite{sal03_2}&  9& +32, +175&Spectra analyzed in \cite{jon09,oli10}, \\
	  &               &   &          &and data published in \cite{kri09}\\
2003ho& \cite{ham03_7}& 5&  +43, +120& \\
2003ib& \cite{mor03}& 5   & +11, +77& \\
2003ip& \cite{fil03_2}& 4& +33, +94& Spectra analyzed in \cite{oli10}\\
2003iq& \cite{mat03_2}& 5& +10, +78& Spectra analyzed in \cite{jon09,oli10}\\
2004dy& \cite{fil04},& 3 & +14, +21, +27& Noisy spectra, however clear `plateau' in light-curve\\
&       \cite{fol04}&&&\\
2004ej& \cite{fol04}& 9& +38, +134& \\
2004er& \cite{mod04}& 10& +31, +175& \\
2004fb& \cite{mor04}& 4& +60, +101& \\
2004fc& \cite{sal04}& 10& +13, +124& \\
2004fx& \cite{sal04_2}& 10& +23, +110& Data published in \cite{ham06}\\
2005J&  \cite{mod05_2}& 11& +23, +97& \\
2005K&  \cite{mod05_2}& 2 & +40, +44& Clear `plateau' shaped light-curve\\
2005Z&  \cite{mor05}& 9& +12, +82& \\
2005af& \cite{fil05}& 9& +110, +176& \\
2005an& \cite{mod05_2}& 7& +15, +50& Spectra indicate hydrogen-rich type II event\\
\hline
\hline
\end{tabular}
\end{table*}
\setcounter{table}{2}
\begin{table*}
\centering
\caption{Spectoscopic classification information of the SN~II sample.}
\begin{tabular}[t]{ccccc}
\hline
SN & Classification& N. of spectra & Epochs of spectra & Comments\\
   & reference& & &\\
\hline	
\hline
2005dk& \cite{mor05_2}& 5& +40, +100& \\
2005dn& \cite{mor05_2}& 6& +38, +98& \\
2005dt& \cite{ser05}& 1 & +34& Light-curve indicates `plateau' classification\\
2005dw& \cite{ser05}& 3 & +36, +121& Spectra indicate hydrogen-rich type II event\\
2005dx& \cite{ser05}& 1& +14& Light-curve indicates `plateau' classification\\
2005dz& \cite{sta05}& 5& +20, +108& \\
2005es& \cite{mor05_3}& 1 & +14& Noisy spectrum, light-curve indicates probable `plateau'\\
2005gk& \cite{gan05}& 0 & $\cdots$& Clear `plateau' shaped light-curve\\
2005hd& \cite{ant05}& 0 & $\cdots$& Clear `plateau' shaped light-curve\\
2005lw& \cite{ham05}& 11& +5, +134& \\
2005me& \cite{leo05_2}& 1& +72& \\
2006Y&  \cite{mor06}& 10 & +13, +85& Spectra indicate hydrogen-rich type II event\\
2006ai& \cite{mor06}& 7& +17, +68& Spectra indicate hydrogen-rich type II event\\
2006bc& \cite{pat06}& 3& +9, +31& Literature data published in \cite{gal12}\\
2006be& \cite{pat06_2}& 4& +19, +45& \\
2006bl& \cite{blo06_2}& 3& +18, +28& \\
2006ee& \cite{fol06}& 6 & +42, +94& \\
2006it& \cite{sah06}& 4& +10, +36& \\
2006iw& \cite{mor06_2}& 4& +8, +77& \\
2006ms& \cite{mor06_3}& 3&+15, +28 & \\
2006qr& \cite{sil06}& 7& +20, +94& \\
2007P&  \cite{blo07}& 4& +21, +86& Spectra indicate hydrogen-rich type II event\\
2007U&  \cite{blo07_2}& 7& +8, +75& \\
2007W&  \cite{fol07_2}& 7& +14, +95& \\
2007X&  \cite{fol07_3}& 10& +6, +88& \\
2007aa& \cite{fol07_4}& 9& +15, +96& \\
2007ab& \cite{blo07_3}& 5& +40, +84& \\
2007av& \cite{har07}& 4& +10, +56& \\
2007hm& \cite{but07}& 6& +20, +86& Spectra indicate hydrogen-rich type II event\\
2007il& \cite{blo07_4}& 9& +12, +95& \\
2007it& \cite{con07}& 8& +6, +244& Literature data published in \cite{and11_2}\\
2007ld& \cite{bas07}& 6& +4, +38& Spectra indicate hydrogen-rich type II event\\
2007oc& \cite{oli07}& 9& +20, +64& \\
2007od& \cite{blo07_5}& 7& +6, +42& Literature data published in \cite{ins11}\\
2007sq& \cite{blo07_6}& 4& +31, +100& \\
2008F&  \cite{blo08}& 2& +12, +21& Noisy spectra and photometry\\
2008K&  \cite{str08}& 9& +9, +109& \\
2008M&  \cite{fol08}& 10& +19, +97& \\
2008W&  \cite{str08_2}& 8&+24, +123& \\
2008ag& \cite{str08_2}& 14& +30, +137& \\
2008aw& \cite{ben08}& 10& +20, +98& Spectra indicate hydrogen-rich type II event\\
2008bh& \cite{bar08}& 6& +10, +55& \\
2008bk& \cite{mor08}& 22& +26, +268& Literature data published in \cite{van12}\\
2008bm& \cite{str08_3}&4 & +41, +79& Originally classified as type IIn, however spectrum shows \\
&&&&clear \ha\ absorption, and light-curve is of `plateau' morphology.\\ 
&&&&Narrow lines are most likely related to underlying star\\ 
&&&&formation region (as indicated by 2d spectrum)\\
2008bp& \cite{str08_3}& 4& +12, +88& \\
2008br& \cite{mor08}& 4& +13, +43& \\
2008bu& \cite{cov08}& 5& +12, +36& \\
2008ga& \cite{ste08}& 3& +55, +110& \\
2008gi& \cite{str08_4}& 5& +11, +81& \\
2008gr& \cite{str08_5}& 3& +27, +63& \\
2008hg& \cite{har08}& 4& +11, +31& \\
2008ho& \cite{sta08_2}& 2& +18, +23& Spectra indicate hydrogen-rich type II event\\
2008if& \cite{cha08_3}& 17& +11, +136& \\
2008il& \cite{str08_6} & 3& +3, +63& \\
2008in& \cite{cha08},& 10& +7, +122& Literature data published in \cite{roy11}\\
&       \cite{fol08_2},&&&\\
&       \cite{str08_7}&&&\\
\hline
\hline
\end{tabular}
\end{table*}
\setcounter{table}{2}
\begin{table*}
\centering
\caption{Spectoscopic classification information of the SN~II sample.}
\begin{tabular}[t]{ccccc}
\hline
SN & Classification& N. of spectra & Epochs of spectra & Comments\\
   & reference& & &\\
\hline	
\hline
2009N&  \cite{cha09}& 11& +24, +128& Spectra published in \cite{tak14}\\
2009ao& \cite{str08_8}& 5& +28, +62& \\
2009au& \cite{str08_8}& 7& +24, +80& Originally classified as type IIn, however spectrum shows  \\
&&&&clear \ha\ absorption, and light-curve is of `plateau' morphology.\\ 
&&&&Narrow lines are most likely related to underlying star\\ 
&&&&formation region (as indicated by 2d spectrum)\\
2009bu& \cite{mor09}& 6& +12, +67& \\
2009bz& \cite{cha09_2}& 4& +9, +37& Spectra indicate hydrogen-rich type II event\\
\hline
\hline
\end{tabular}
\setcounter{table}{2}
\tablecomments{Spectroscopic information used for classification of the current sample as
hydrogen-rich type II events. In the first column the SN name is listed, followed
by the classification circular references. We then indicate the number of spectra which were obtained, 
and that are used for: NaD EW measurements, explosion time estimations, and confirmation of type classifications
in column 3, followed by the epoch of the first and last spectrum in column 4. In column 5 comments are listed
in cases where: a) the circular classification information is sparse, b) spectra have been previously
analyzed, c) distinct spectroscopy has been published in the literature, and d) in cases
where we have changed the classification from that quoted in the circulars.}
\end{table*}

\section{Sample characterization}
As noted in \S\ 1, the currently analyzed sample is an eclectic mix of SNe~II, 
discovered by many different SN search campaigns, of both professional and amateur nature. In this
Section, we further characterize the SN and host galaxy samples, and briefly discuss some possible
consequences of their properties.\\
\indent Fig.\ 21 presents a pie chart which shows the contributions of various SN programs and 
individuals to the discovery of SNe presented in this analysis. The largest number of events included, 
discovered by any one survey/individual are those reported by the Lick Observatory Supernova Search (LOSS/LOTOSS, \citealt{lea11}),
who contribute 41 SNe ($\sim$40\%). Surprisingly, the next highest contributor to our sample is Berto Monard,
an amateur astronomer from South Africa, with an impressive 12 entries ($\sim$10\%). 
Next is the CHilean Automatic Supernova sEarch (CHASE, \citealt{pig09}) with 11, followed by 
`Maza' (SNe discovered by investigators at Universidad de Chile during the 1990s, led by Jose Maza), 
Robert Evans (another
amateur astronomer, from Australia), and the NEAT survey (Near-Earth Asteroid Tracking). 
All other discoveries are grouped
into `Other'.
The vast majority of SNe were discovered by 
galaxy targeted surveys (88\%, counting SDSS, CSS, and NEAT as non-targeted surveys). As most targeted surveys
are biased towards bright nearby galaxies, this may lead to a bias against SNe in low-luminosity host galaxies. 
These and similar issues are discussed for the LOSS SN sample in \cite{li11}.\\
\indent In Fig.\ 22 we present the host galaxy recession velocity and absolute $B$-band magnitude distributions
respectively. Our sample has a mean recession velocity of 5505 km s$^{-1}$, equating to a mean redhift of 0.018, and a mean 
distance of 75 Mpc, with the vast majority of galaxies having velocities lower than 10000 km s$^{-1}$.  The host galaxy population
is characterized as having a well defined absolute magnitude $B$-band peak at $\sim$--21 mag, together with a low luminosity
tail down to around --17 mag\footnote{We note that for 6 SN host galaxies, absolute magnitudes are not available in the LEDA
database. Given that this is most likely due to their faintness, the low-luminosity tail could be more highly populated with 
inclusion of those events.}. Host galaxy absolute magnitudes are listed in Table 1. It is interesting to compare this 
heterogeneous sample with that published by \cite{arc10} (CC SNe discovered by the Palomar Transient Factory: PTF). 
While those authors published host galaxy $r$-band magnitudes (compared to the $B$-band magnitudes analyzed here), overall
the distributions appear qualitatively similar, with the exception that the PTF sample has a much more pronounced lower-luminosity
tail. If the intrinsic rate of specific SNe~II events changes with host galaxy luminosity (due to e.g. a dependence on progenitor
metallicity), then this could affect the overall conclusions draw from our sample, where we are possibly missing SNe in low-luminosity
hosts.
A full analysis of these host galaxy properties is beyond the scope of this paper, 
and a detailed host galaxy study of the CSP sample is underway.\\

\begin{figure*}
\includegraphics[width=14cm]{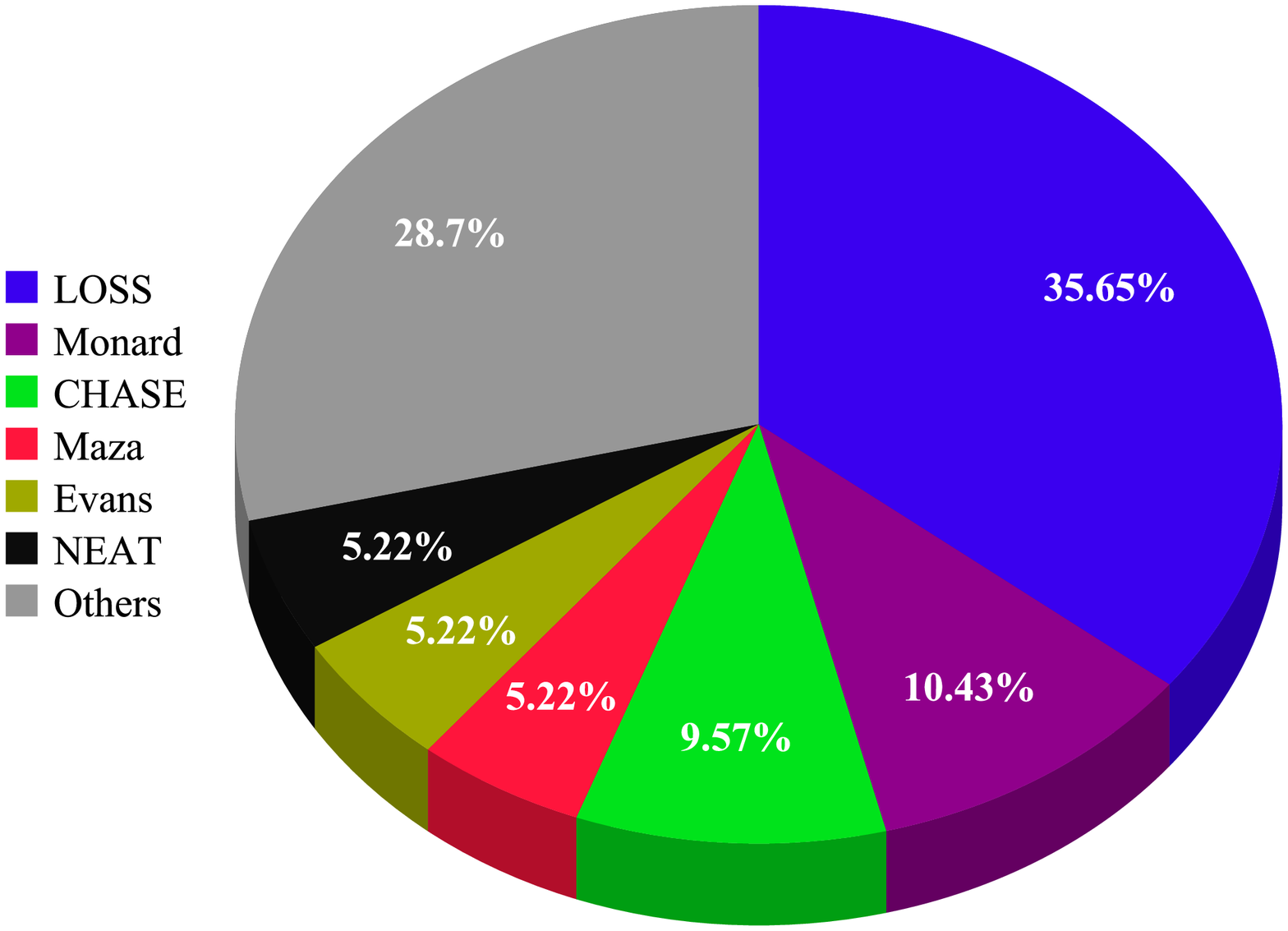}
\caption{Pie chart of the relative contributions of different SN search campaigns to the SN~II sample
analyzed in this publication. Contributions with more than 5 entries are shown individually, while the 
rest are grouped into `Others' (which includes SNe discovered the following programs/individuals: `CROSS'; Yamagata;
Itagaki; `CSS/CRTS'; `Brazil'; Mikuz; Arbour; Doi; Boles; `SN factory'; Brass; Lulin; Llapasset; `Tenagra'; `SDSS';
Puckett; Hurst; `Perth'). The entry `Maza' refers to those SNe discovered by investigators at Universidad de Chile 
during the 1990s.}
\end{figure*}
\begin{figure*}
\includegraphics[width=8cm]{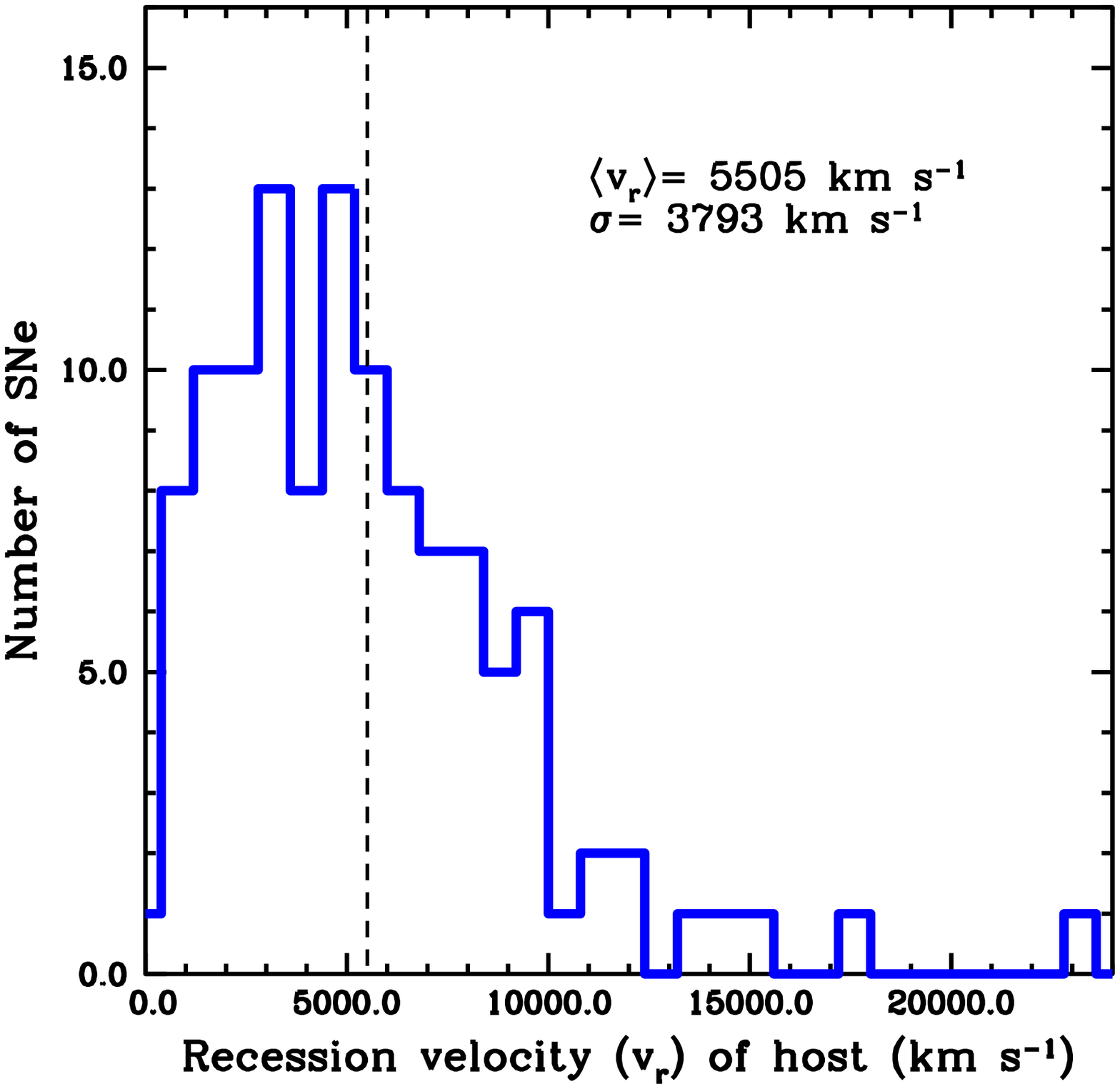}
\includegraphics[width=8cm]{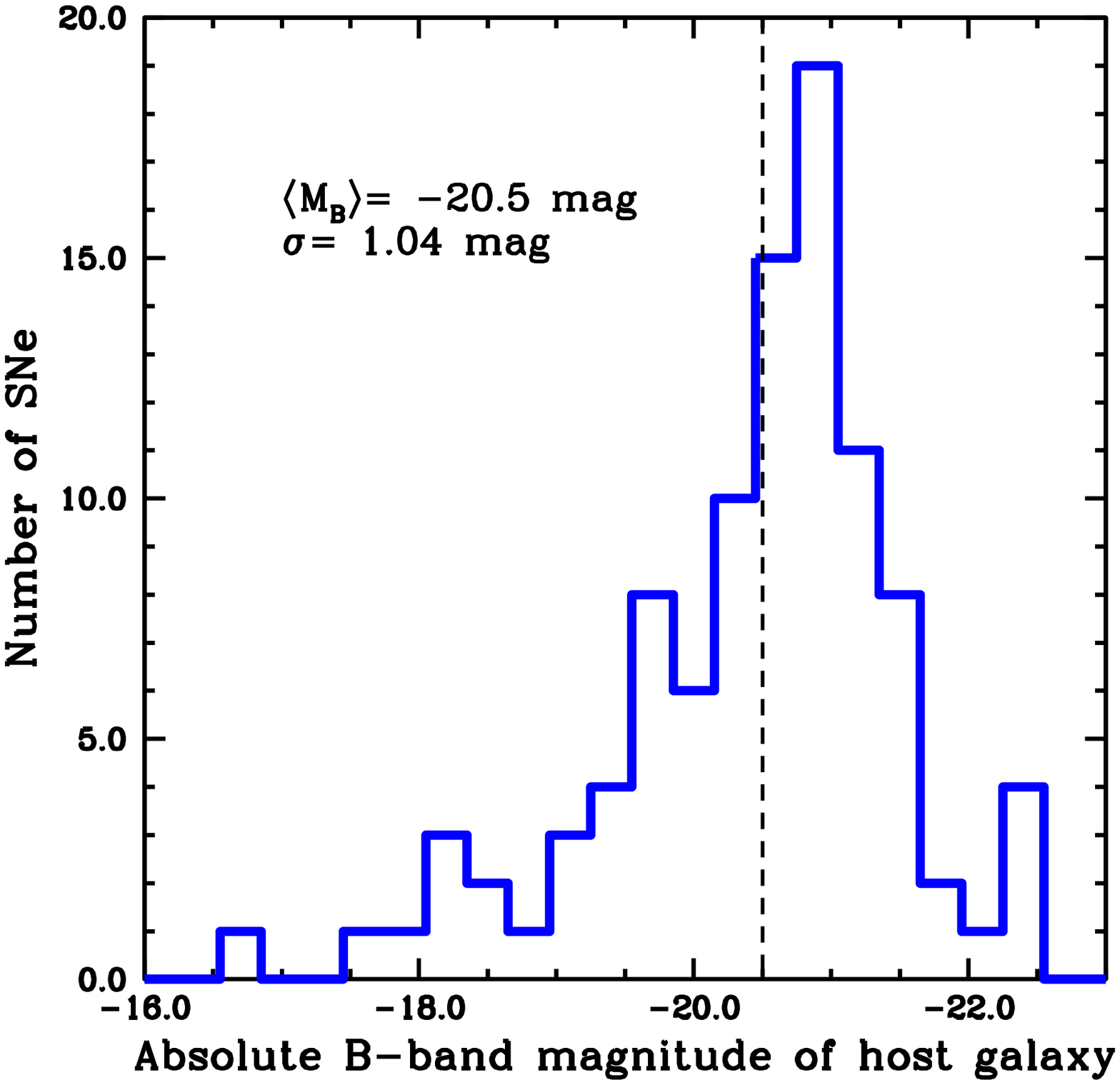}
\caption{\textit{Left:} histogram of the heliocentric recession velocities of the host galaxies of the SNe
included in the current sample. \textit{Right:} histogram of the absolute $B$-band magnitudes of the host galaxies of the SNe
included in the current sample (taken from the LEDA database)}
\end{figure*}

\begin{figure*}
\includegraphics[width=8cm]{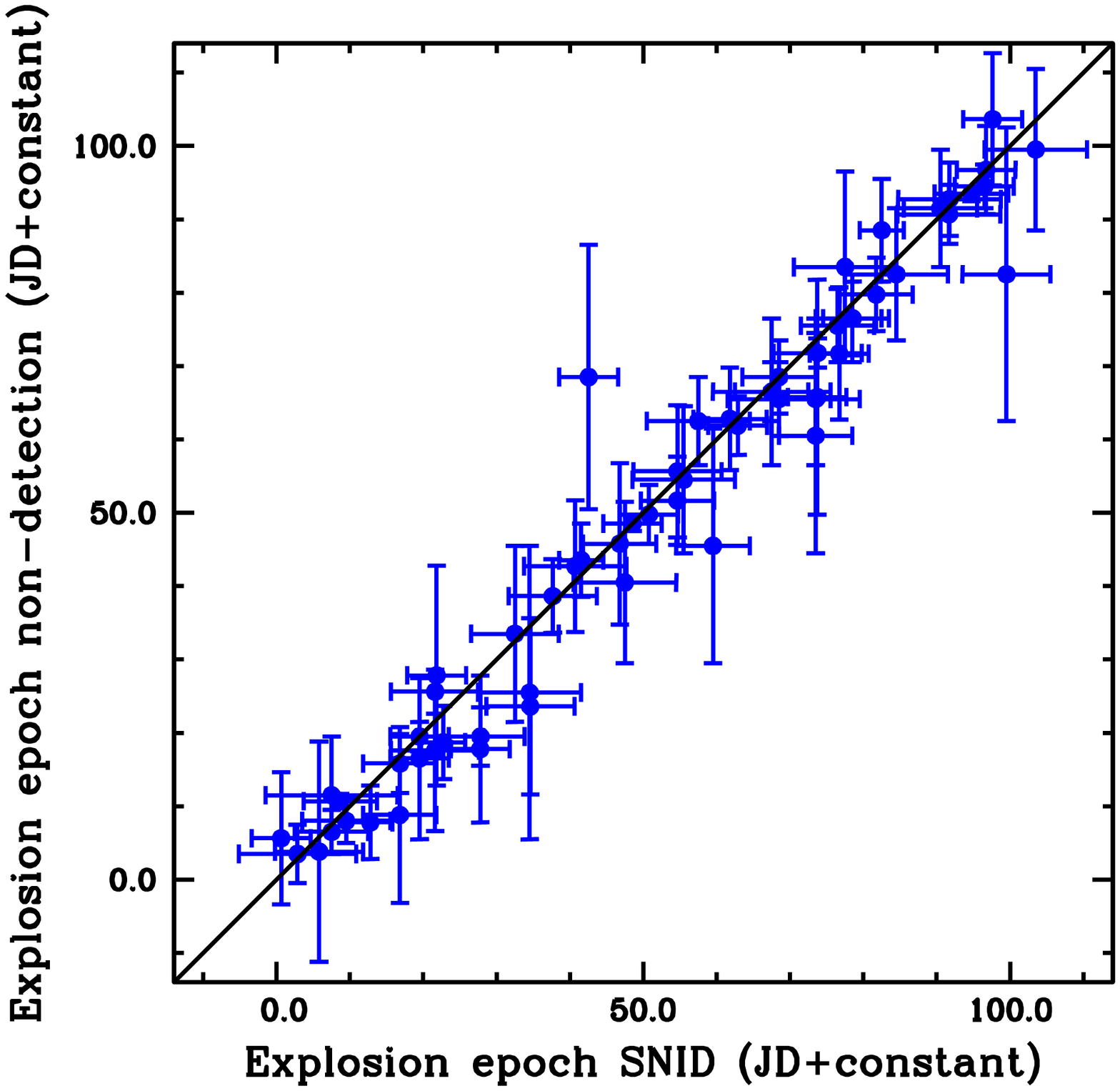}
\includegraphics[width=8cm]{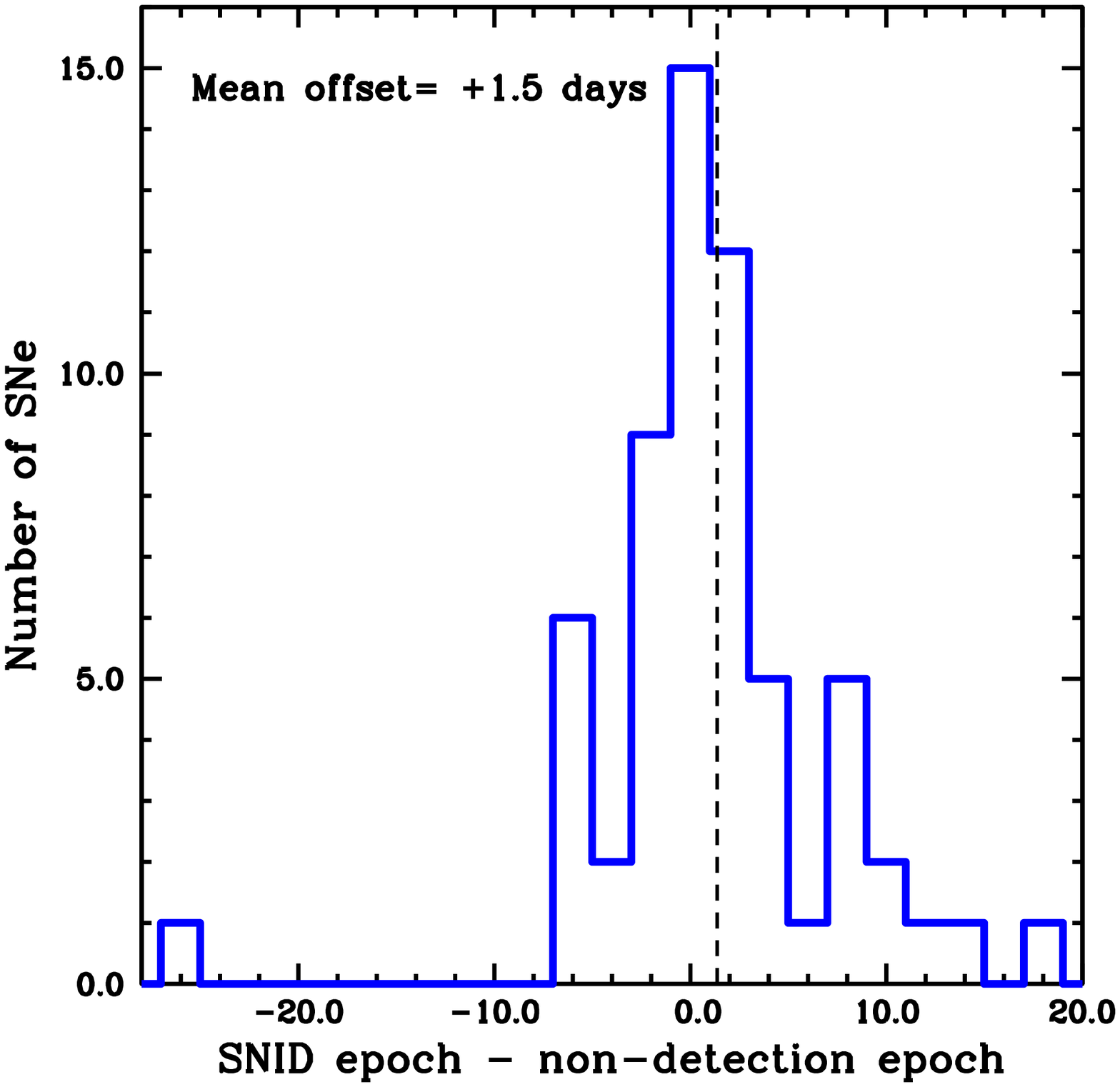}
\caption{\textit{Left}: comparison of the explosion dates estimated through spectral matching
(x-axis) and through SN non-detection on pre-discovery images (y-axis). The
straight line shows a one-to-one relation between the two epoch estimation
methods. \textit{Right}: histogram of the offsets between explosion epochs
estimated through both methods. The mean offset in days between the two estimation
methods is indicated by the dashed vertical line. }
\end{figure*}

\begin{table*}
\centering
\caption{Explosion epoch comparisons.}
\begin{tabular}[t]{ccccc}
\hline
SN & Spectral matching & Non-detection & Difference (days)\\
   & explosion epoch (MJD)& explosion epoch (MJD)&\\
\hline	
\hline
1986L & 46710(6) & 46708(3) & 2\\
1993K & 49074(6) & 49066(9) & 8\\
1999br& 51279(4) & 51277(4) & 2\\
1999ca& 51278(7) & 51284(13)& --6\\
1999em& 51479(5) & 51477(5) & 2\\
2002fa& 52501(7) & 52500(11)& 4\\
2002gw& 52560(5) & 52546(16)& 14\\
2002hj& 52558(7) & 52563(6) & --5\\
2002hx& 52585(7) & 52583(9) & 2\\
2003E & 52635(7) & 52626(20)& 9\\
2003T & 52656(7) & 52655(10)&1\\
2003bn& 52697(4) & 52695(3)&2\\
2003ci& 52708(9) & 52712(8)&--4\\
2003cn& 52720(4) & 52717(11)&3\\
2003eg& 52774(5) & 52761(16)&13\\
2003ej& 52777(5) & 52776(5)&1\\
2003hg& 52869(7) & 52866(5)&3\\
2003hl& 52869(5) & 52869(5)&0\\
2003hn& 52868(8) & 52867(10)&1\\
2003ho& 52848(7) & 52841(11)&7\\
2003ib& 52891(5) & 52883(16)&8\\
2003iq& 52920(4) & 52920(2)&0\\
2004er& 53274(6) & 53272(2)&2\\
2004fb& 53243(4) & 53269(18)&--26\\
2004fc& 53295(5) & 53294(1)&1\\
2004fx& 53303(8) & 53304(4)&--1\\
2005Z&  53397(4) & 53397(6)&0\\
2005dk& 53600(6) & 53583(20)&17\\
2005dt& 53601(4) & 53606(9)&--5\\
2005dw& 53598(4) & 53604(9)&--6\\
2005dz& 53628(6) & 53620(4)&8\\
2005es& 53638(6) & 53639(5)&--1\\
2005lw& 53717(10)& 53709(12)&8\\
2005me& 53722(6) & 53718(11)&4\\
2006Y & 53768(5) & 53767(4)&1\\
2006be& 53806(6) & 53804(15)&2\\
2006ee& 53963(4) & 53962(4)&1\\
2006it& 54008(5) & 54007(3)&1\\
2006iw& 54009(5) & 54011(1)&--2\\
2006ms& 54033(6) & 54034(12)&--1\\
2006qr& 54062(5) & 54063(7)&--1\\
2007P & 54123(3) & 54119(5)&4\\
2007U & 54135(6) & 54124(12)&11\\
2007il& 54351(4) & 54350(4)&1\\
2007it& 54349(4) & 54349(1)&0\\
2007oc& 54383(3) & 54389(7)&--6\\
2007sq& 54422(4) & 54428(15)&--6\\
2008K & 54474(4) & 54466(16)&8\\
2008M & 54477(4) & 54472(9)&5\\
2008aw& 54528(4) & 54518(10)&10\\
2008bh& 54542(3) & 54544(5)&--2\\
2008bp& 54555(5) & 54552(6)&3\\
2008br& 54555(6) & 54556(9)&--1\\
2008gi& 54741(7) & 54743(9)&--2\\
2008hg& 54782(5) & 54780(5)&2\\
2008ho& 54792(7) & 54793(5)&--1\\
2008if& 54813(3) & 54808(5)&5\\
2008il& 54822(6) & 54826(3)&--4\\
2009N & 54847(5) & 54846(11)&1\\
2009ao& 54892(7) & 54891(4)&1\\
2009bz& 54917(5) & 54916(4)&1\\
\hline
\hline
\setcounter{table}{3}
\end{tabular}
\tablecomments{A comparison of SN explosion epochs from spectral
matching and non-detections, for SNe where the error on the non-deteciton epoch is less than 20 days.
In the first column the SN name is listed, followed by the explosion epochs
derived from spectral matching and non-detections in columns 2 and 3 respectively.
In the last column the difference between the two estimates (spectral matching -- non-detection dates)
is given. Errors on individual explosion epoch estimations are indicated in parenthesis.}
\end{table*}

\section{Explosion epoch estimation analysis procedure}
Table 4 shows a comparison between explosion epochs
obtained through both the non-detection and spectral matching methods, for
the 61 SNe where this is possible.
In Fig.\ 23 we present
on the left panel a
comparison of these two methods. On the right
panel is a histogram showing the offset between the two estimations. One can see that in general very 
good agreement is found between the two methods, with a 
mean absolute error of 4.2 days between the different techniques. This gives us confidence
in our measurements which are dependent on spectral matching analysis.
A feature of
the two plots is that where there are differences between the two methods then it
is more often that the estimate from the non-detection is earlier than that
from spectral matching: there is a 1.5 day mean offset in that direction
between the two methods. 
It is noted that those SNe in Table 4 with the largest offsets between
the methods are also the SNe which have the highest errors on the estimations from
non-detections, i.e. if SNe have low non-detection errors, then the epochs from
the two methods are even more consistent.\\
\indent We now re-analyze our sample, but only include those SNe
where epochs are derived from non-detections, which one may assume are more accurate, and 
less dependent on systematics. The results of this re-analysis are as follows. The $Pd$ sub-sample
has a mean value of 48.9$\pm$14 days, for 8 events. This compares to: mean $Pd$ = 48.4$\pm$13 for the 19 events of the full
sample. The $OPTd$ sub-sample has a mean value of 83.7$\pm$21 for 32 SNe, compared to 83.7$\pm$17 for the
full sample of 72. In addition, the Monte Carlo bootstrap Pearson's tests are also re-run for all sub-sample 
correlations which are affected by explosion epoch estimates. In Table 5 correlations are compared between
the full sample and the sub-set of SNe with non-detection constraints. The results are fully
consistent within the errors: despite the low number statistics the strengths of the correlations are very similar
in the majority of cases.\\
\indent The above comparison shows that the inclusion of time durations dependent on spectral matching 
explosion epochs does not systematically change our distributions, or the strength of presented correlations, and hence gives us further
confidence in the results of our full sample.

\begin{table*}
\centering
\caption{Correlation parameters of full and sub-samples.}
\begin{tabular}[t]{cccc}
\hline
Correlation & $N$ (full) & $r$ (full) & $P$ (full)\\
			& $N$ (sub) & $r$ (sub) & $P$ (sub)\\
\hline	
\hline
$Pd$--$M_\text{max}$& 15 & 0.49$\pm$0.25 & $<$0.39\\
					& 7	 & 0.07$\pm$0.42 & $<$1	\\
\hline
$OPTd$--$M_\text{max}$& 45 & 0.15$\pm$0.15 & $<$1\\
					& 20	 & --0.04$\pm$0.26 & $<$1	\\
\hline
$s_2$--$Pd$& 19 & --0.58$\pm$0.15 & $<$0.07\\
					& 7	 & --0.32$\pm$0.30 & $<$0.97\\	
\hline
$s_2$--$OPTd$& 70 & --0.34$\pm$0.13 & $<$0.08\\
					& 21	 & --0.28$\pm$0.25 & $<$0.90\\	
\hline
$s_1$--$Pd$& 19 & --0.56$\pm$0.16 & $<$0.09\\
					& 6	 & --0.51$\pm$0.38 & $<$0.78\\	
\hline
$s_1$--$OPTd$& 19 & --0.70$\pm$0.15 & $<$0.01\\
					& 6	 & --0.75$\pm$0.24 & $<$0.24\\
\hline
$s_3$--$OPTd$& 25 & --0.40$\pm$0.22 & $<$0.39\\
					& 8	 & --0.60$\pm$0.38 & $<$0.60\\
\hline
$OPTd$--$^{56}$Ni mass& 13 & --0.52$\pm$0.28 & $<$0.43\\
				& 5	 & 0.29$\pm$0.58 & $<$1\\
\hline
$Pd$--$M_\text{end}$& 15 & 0.59$\pm$0.22 & $<$0.07\\
					& 7	 & 0.24$\pm$0.40 & $<$1	\\
\hline
$OPTd$--$M_\text{end}$& 46 & 0.19$\pm$0.18 & $<$0.95\\
					& 20	 & --0.15$\pm$0.29 & $<$1	\\					
\hline	
\hline
\end{tabular}
\setcounter{table}{4}
\tablecomments{Comparison of the strength ($r$) and significance ($P$) of 
correlations which are dependent on explosion epoch estimations, for the 
full sample, and the sub-sample of events with estimations derived from
non-detections. In the first column the correlation is listed, followed
by the number of SNe within that correlation in column 2. In column 3 we
show the relative strength of each correlation, followed by its respective 
strength in column 4.} 
\end{table*}

\begin{figure*}
\includegraphics[width=9cm]{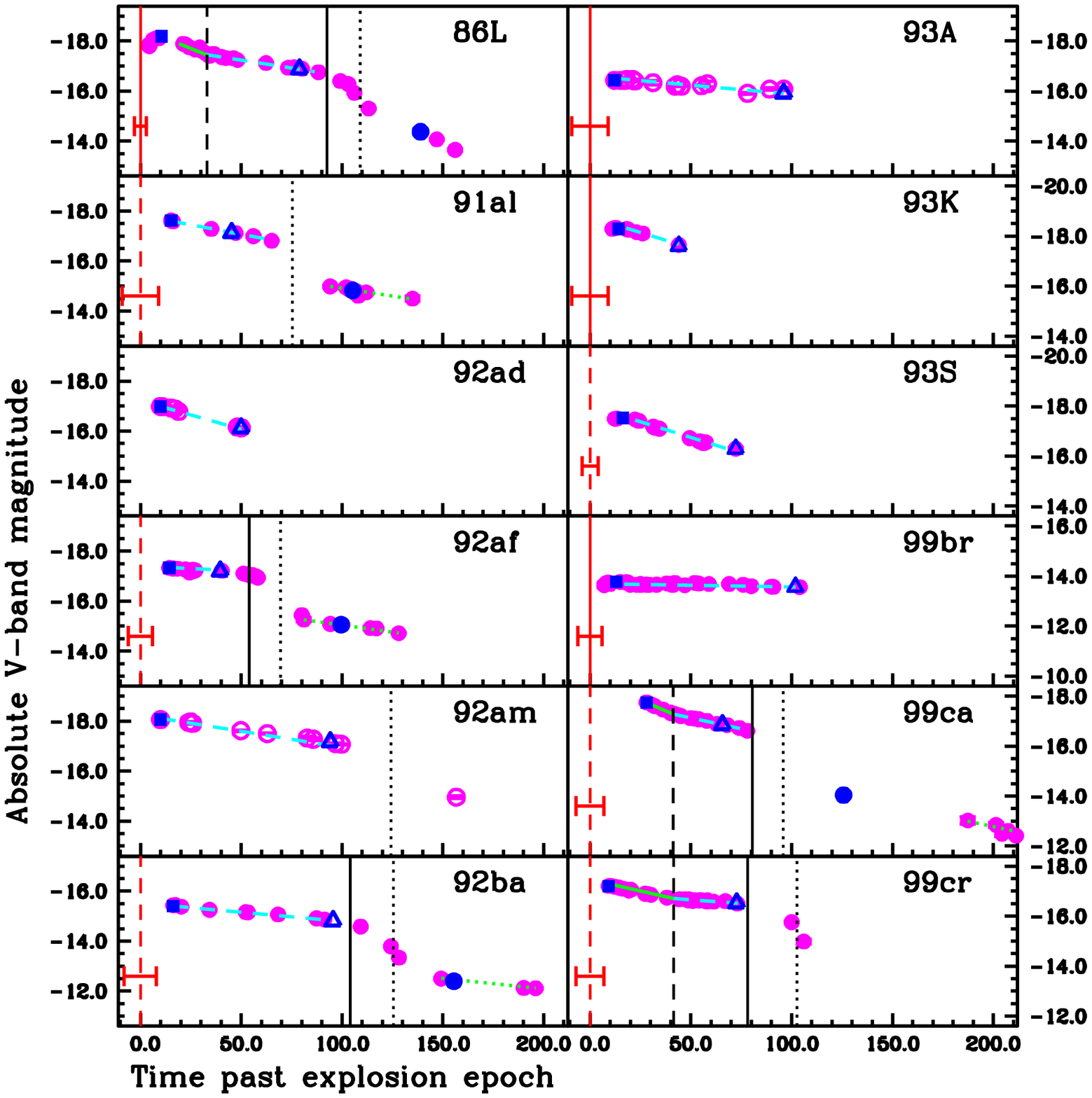}
\includegraphics[width=9cm]{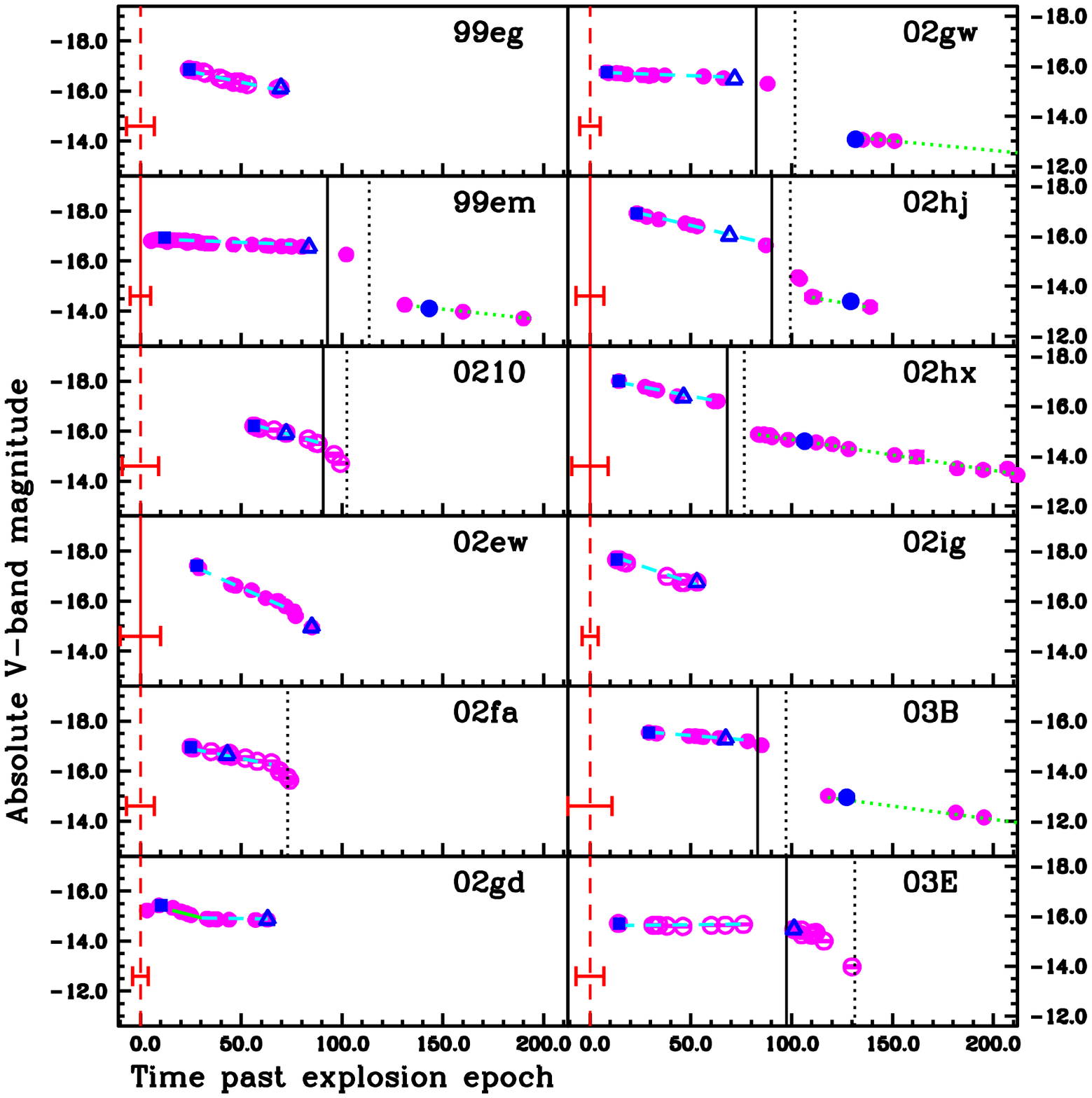}
\includegraphics[width=9cm]{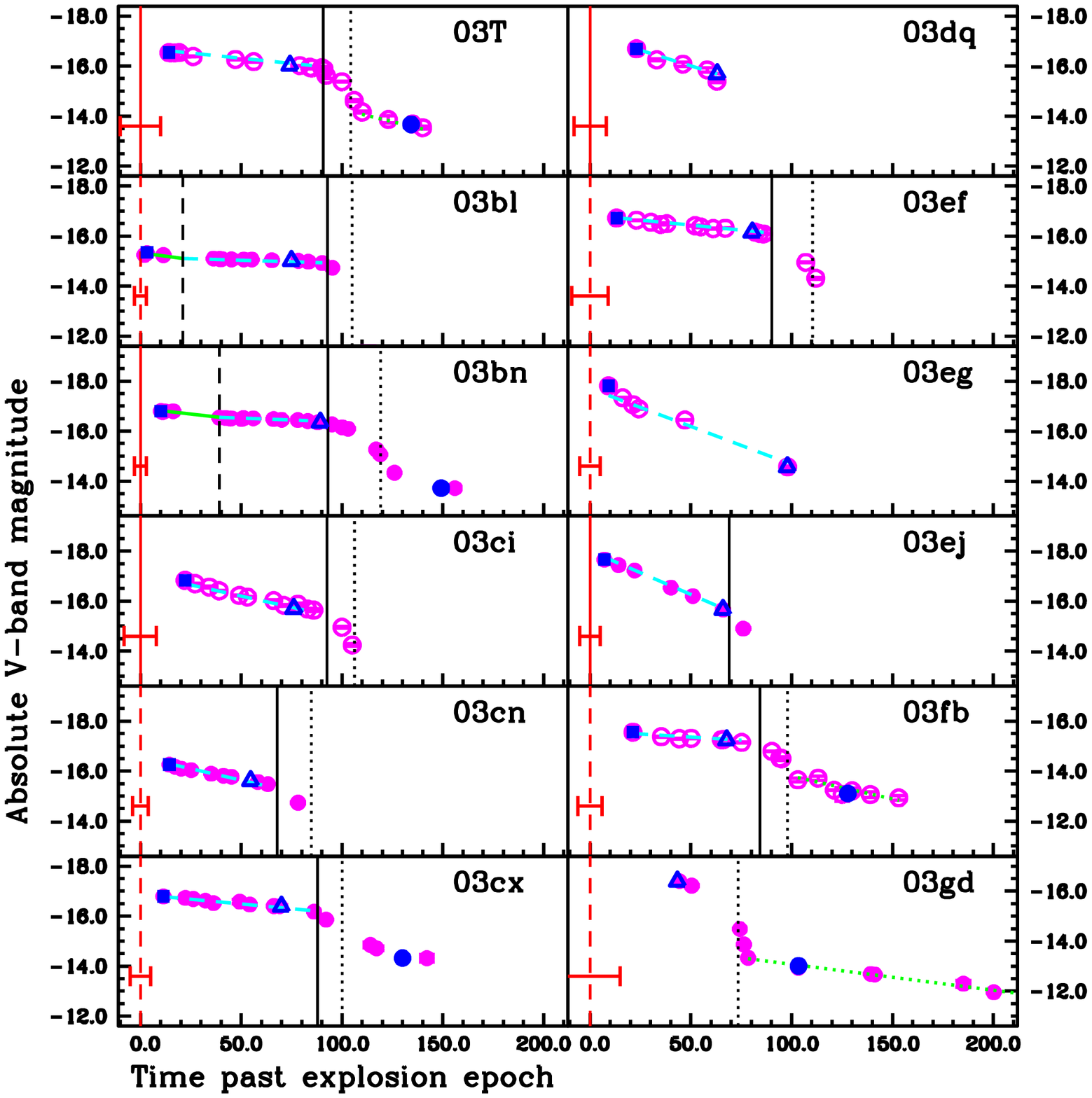}
\includegraphics[width=9cm]{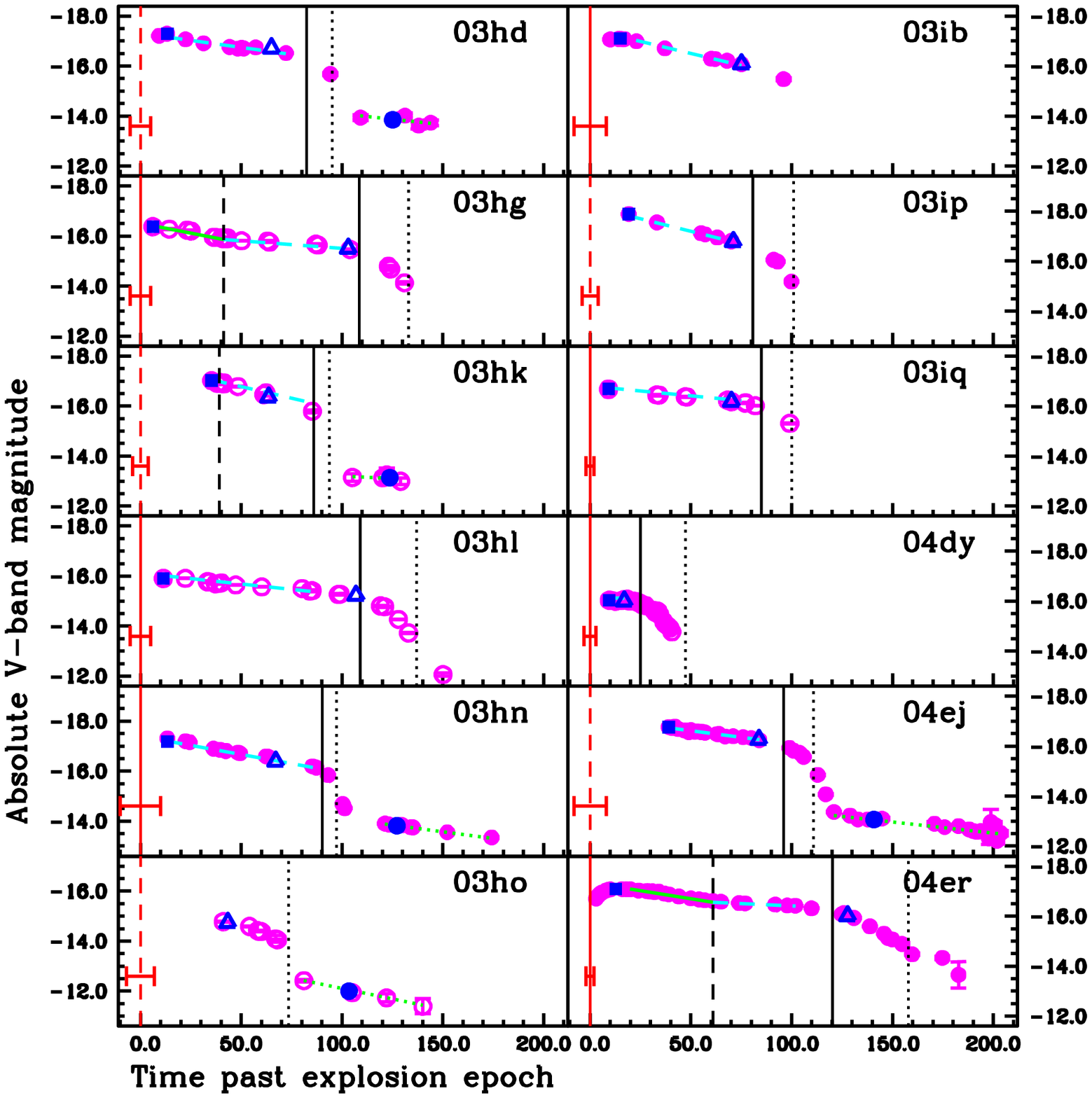}
\caption{Absolute magnitude $V$-band light-curves of SNe~II in 
our sample. The panels progress presenting SNe in order of their discovery
dates, starting with SN~1986L in the top left panel, and finishing with SN~2004er in the bottom 
right. Photometry are presented by magenta circles, where open circles indicate that
no host galaxy extinction correction was possible (see \S\ 3.3), and hence the 
presented magnitudes are lower limits. Note that in general errors on the photometry are
smaller than the photometric points, and therefore are only visible when they are large in size.
Measured light curve parameters are also presented: $M_\text{max}$ as blue
squares; $M_\text{end}$ as blue open triangles; $M_\text{tail}$ as blue circles;
$s_1$ as solid green lines; $s_2$ as dashed cyan lines; $s_3$ as dotted green lines;
$t_{end}$ as solid vertical black lines; $t_{PT}$ as dotted vertical black lines; and
$t_{tran}$ as dashed vertical black lines. Red vertical lines are placed at the explosion epoch 
(0 on the x-axis), and are presented as solid lines if they are derived through non-detections, and
dashed lines if they are derived from spectral matching (see \S\ 3.1 for details). 
Uncertainties in the explosion epochs are illustrated as red error bars.
If this line is missing then an explosion epoch estimate was not possible, and light-curves
are presented with an arbitrary offset.}
\end{figure*}

\section{Additional figures}
In Figs 24, 25 and 26 present $V$-band absolute magnitude 
light-curves for the full sample of 116 SNe~II included in this study, where the 
measured parameters listed in Table 6 and used in the analysis throughout this work, are
also provided in figure form.
In Fig.\ 27 an example of the automated $s_1$--$s_2$ fitting
process is displayed, showing examples of where a simple one slope fit
(i.e. $s_2$ for SN~2008ag), and a two slope slope fit ($s_1$, $s_2$ for
SN~2004er) are good representations of the photometry. 

\begin{figure*}
\includegraphics[width=9cm]{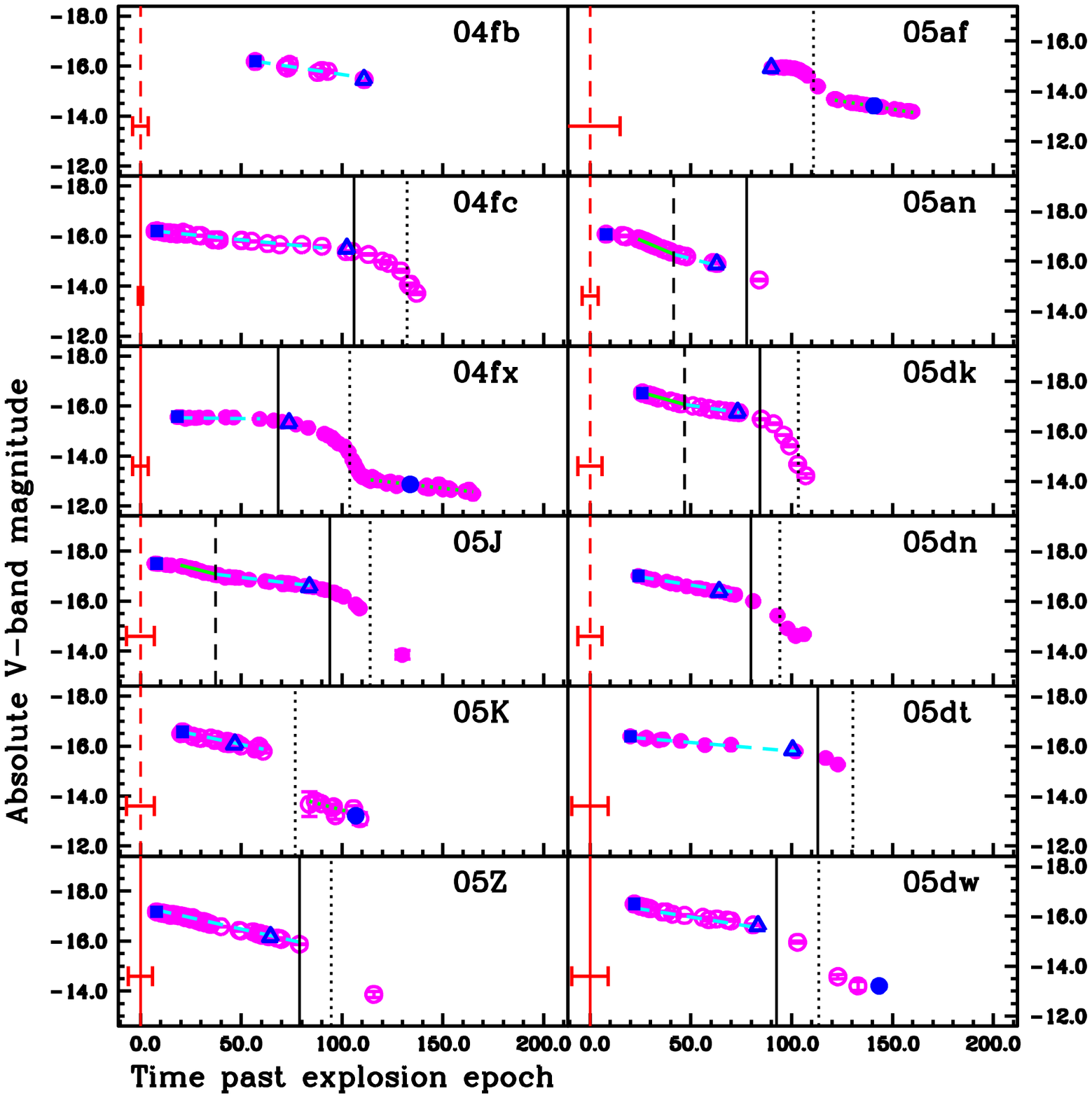}
\includegraphics[width=9cm]{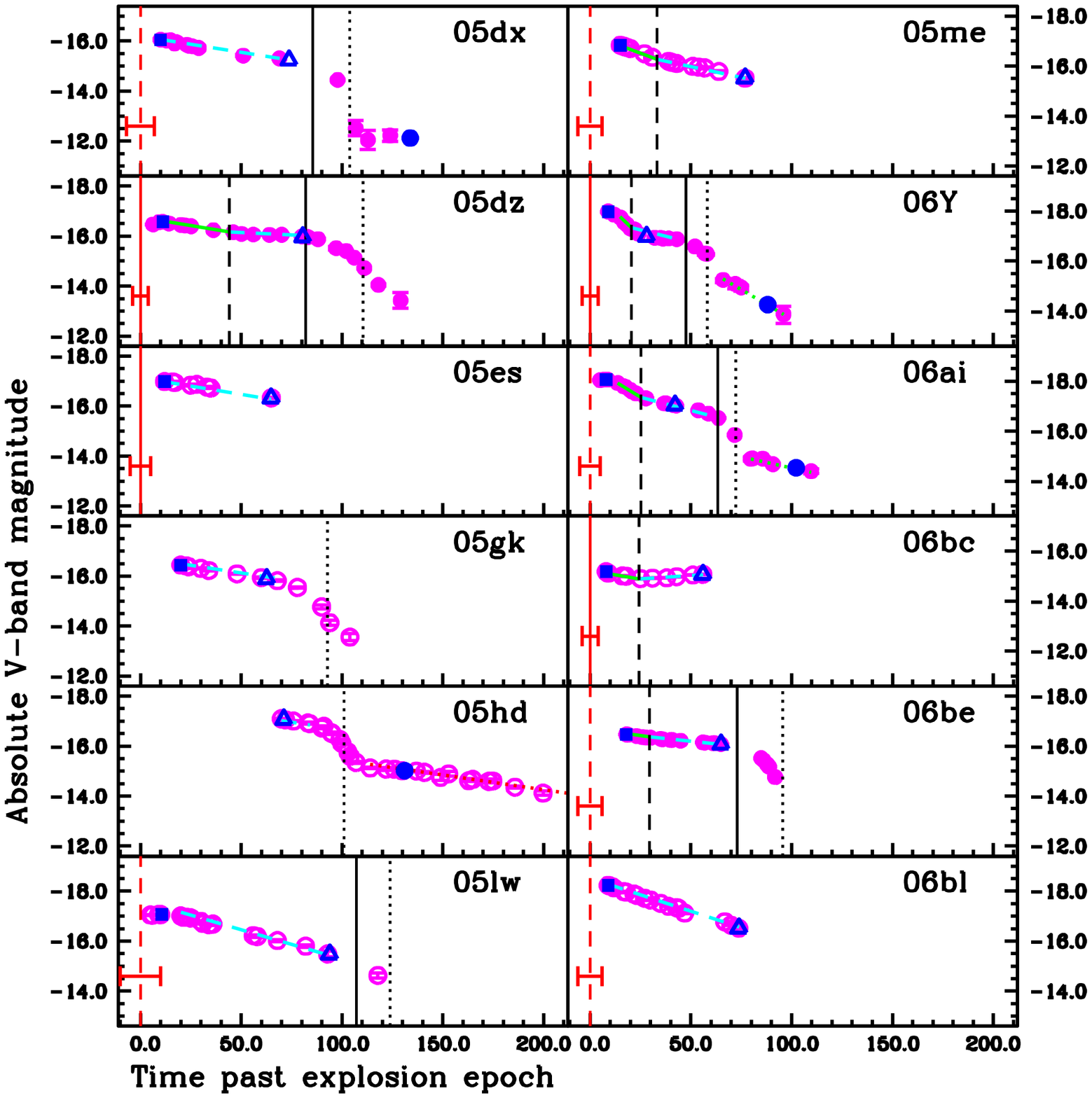}
\includegraphics[width=9cm]{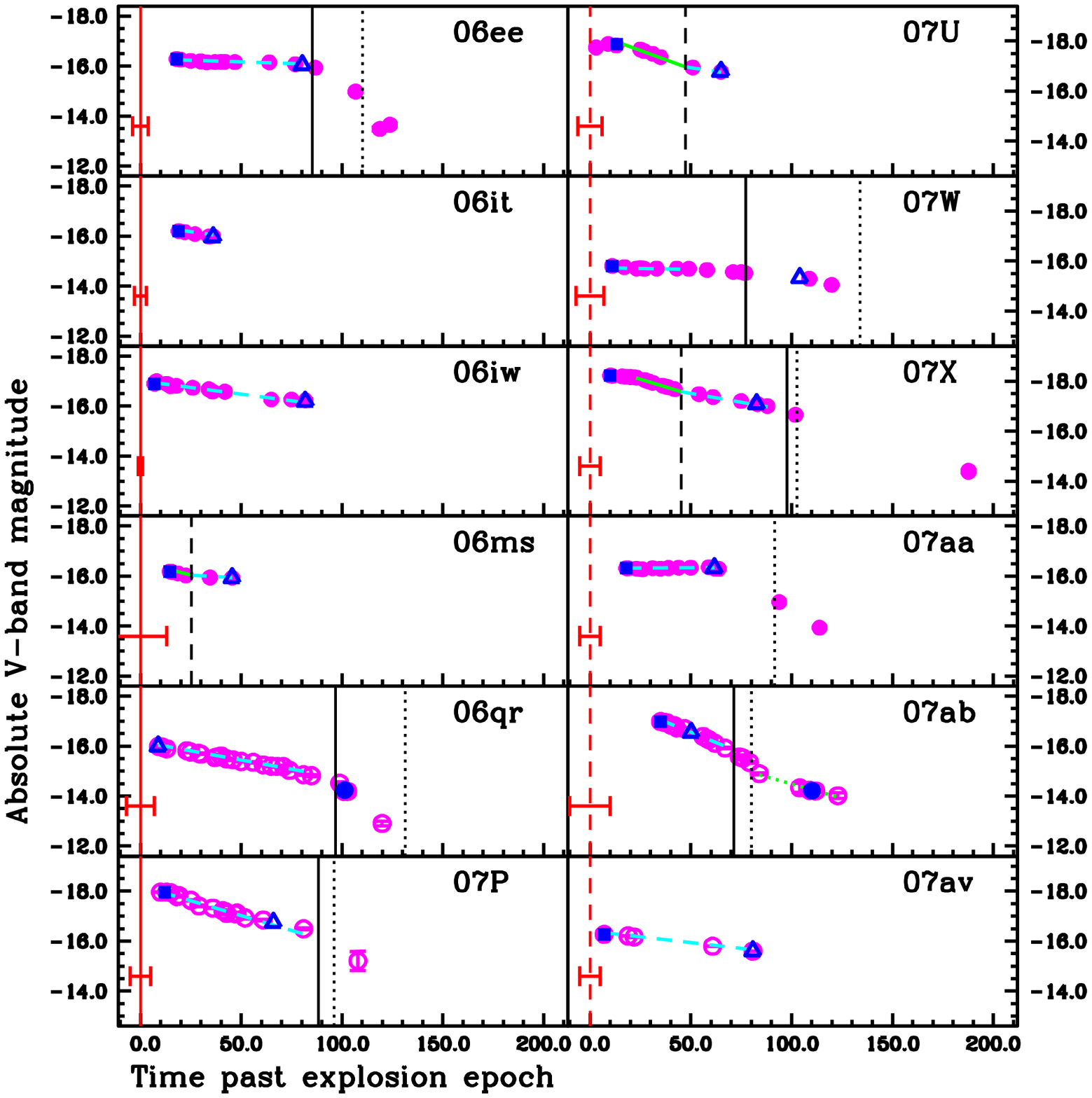}
\includegraphics[width=9cm]{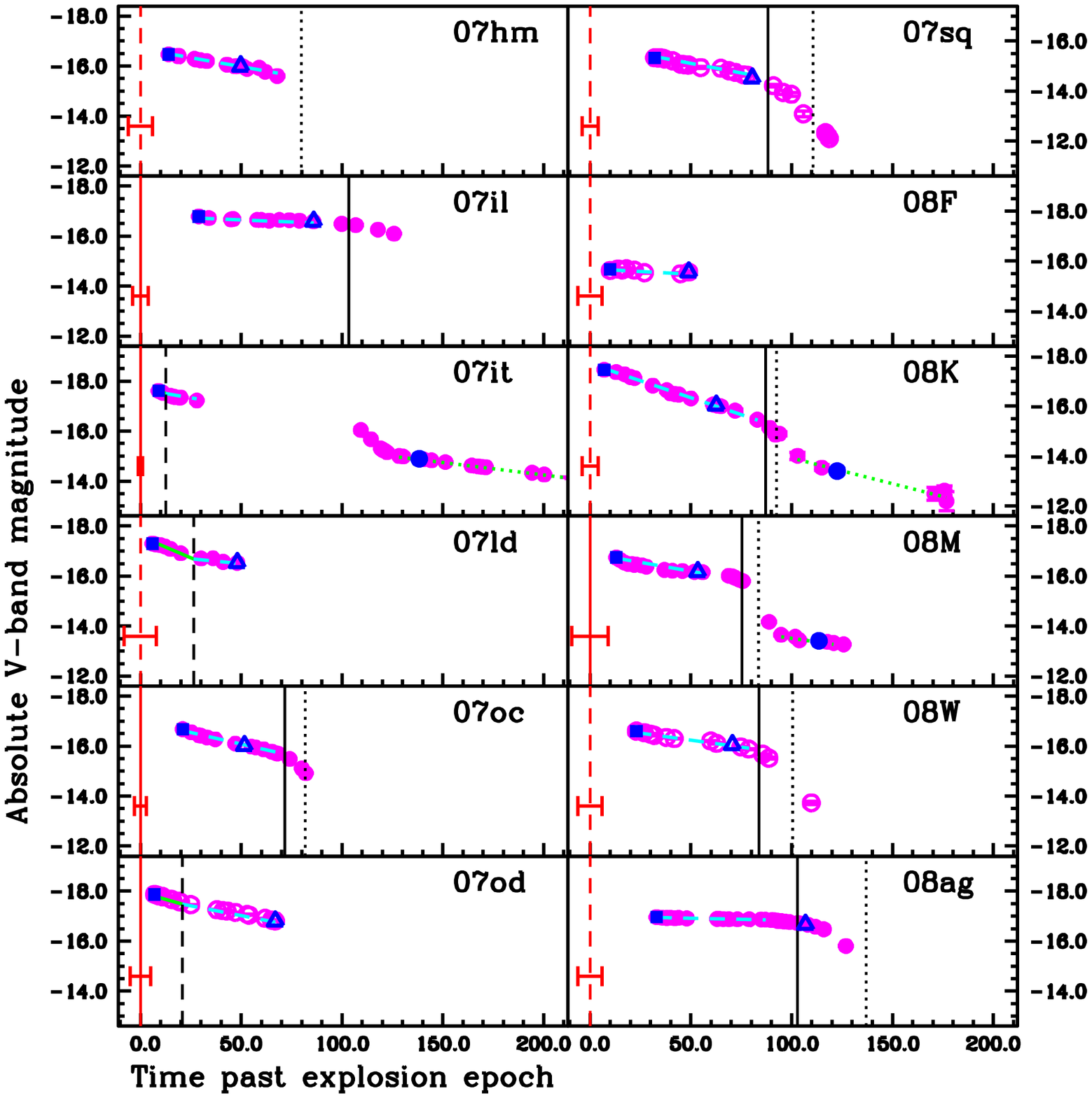}
\caption{Absolute magnitude $V$-band light-curves of SNe~II in 
our sample. The panels progress presenting SNe in order of their discovery
dates, starting with SN~2004fb in the top left panel, and finishing with SN~2008ag in the bottom 
right. Presentation of light-curves and their derived parameters takes the same form 
as in Fig.\ 24.}
\end{figure*}

\indent In Fig.\ 28 the correlation between $s_1$ and $s_3$ is shown. In Fig.\ 29
we present correlations between the 3 brightness measurements: $M_\text{max}$,
$M_\text{end}$, and $M_\text{tail}$. A large degree of correlation between all
three is observed. $M_\text{end}$ is plotted against $s_1$ and $s_3$ in
Fig.\ 30. $M_\text{tail}$ is plotted against $s_1$, $s_2$ and $s_3$ in
Fig.\ 31. In Fig.\ 32 we present correlations of $Pd$ and $OPTd$ against
$M_\text{end}$. 

\begin{figure*}
\includegraphics[width=9cm]{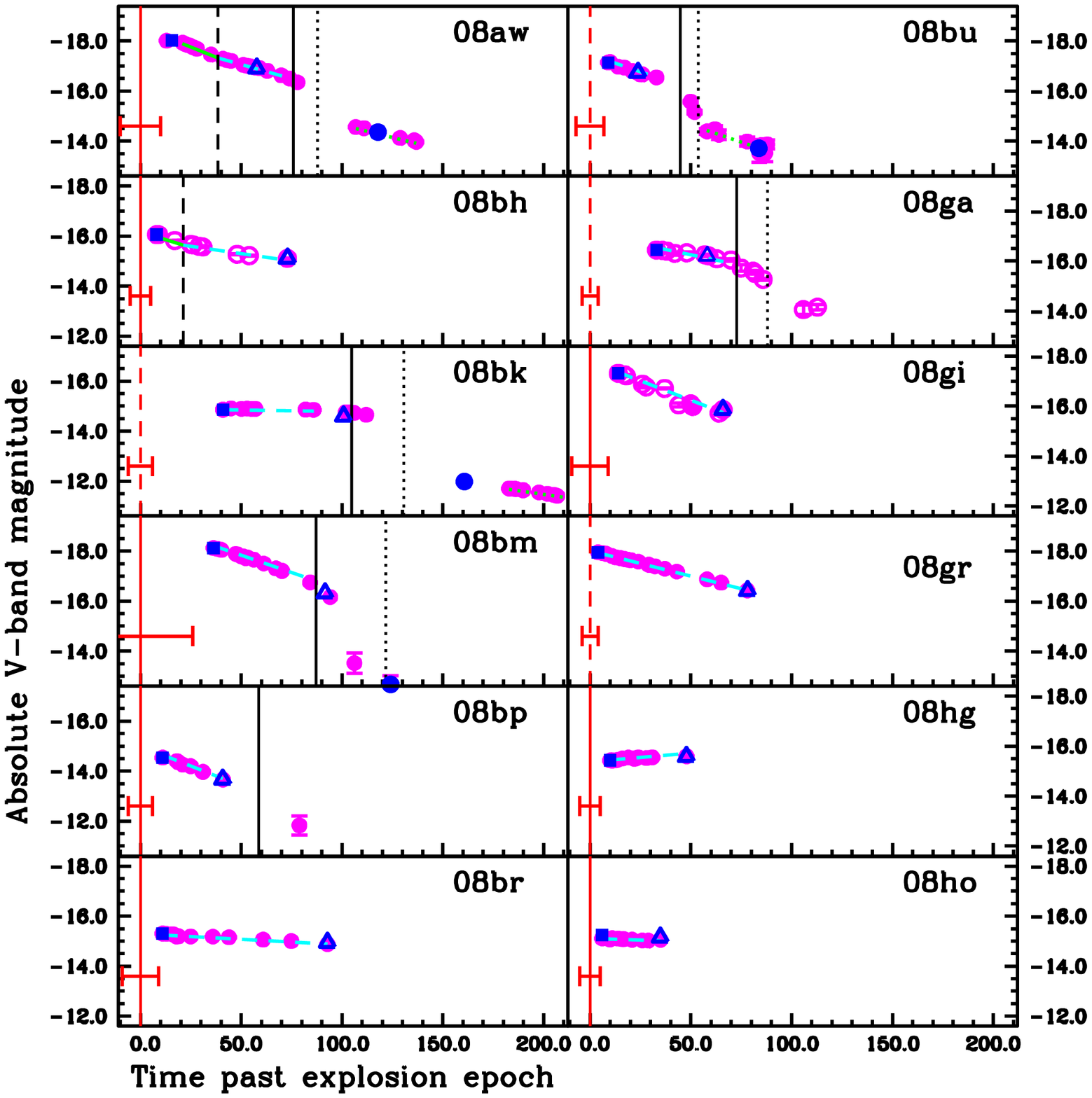}
\includegraphics[width=9cm]{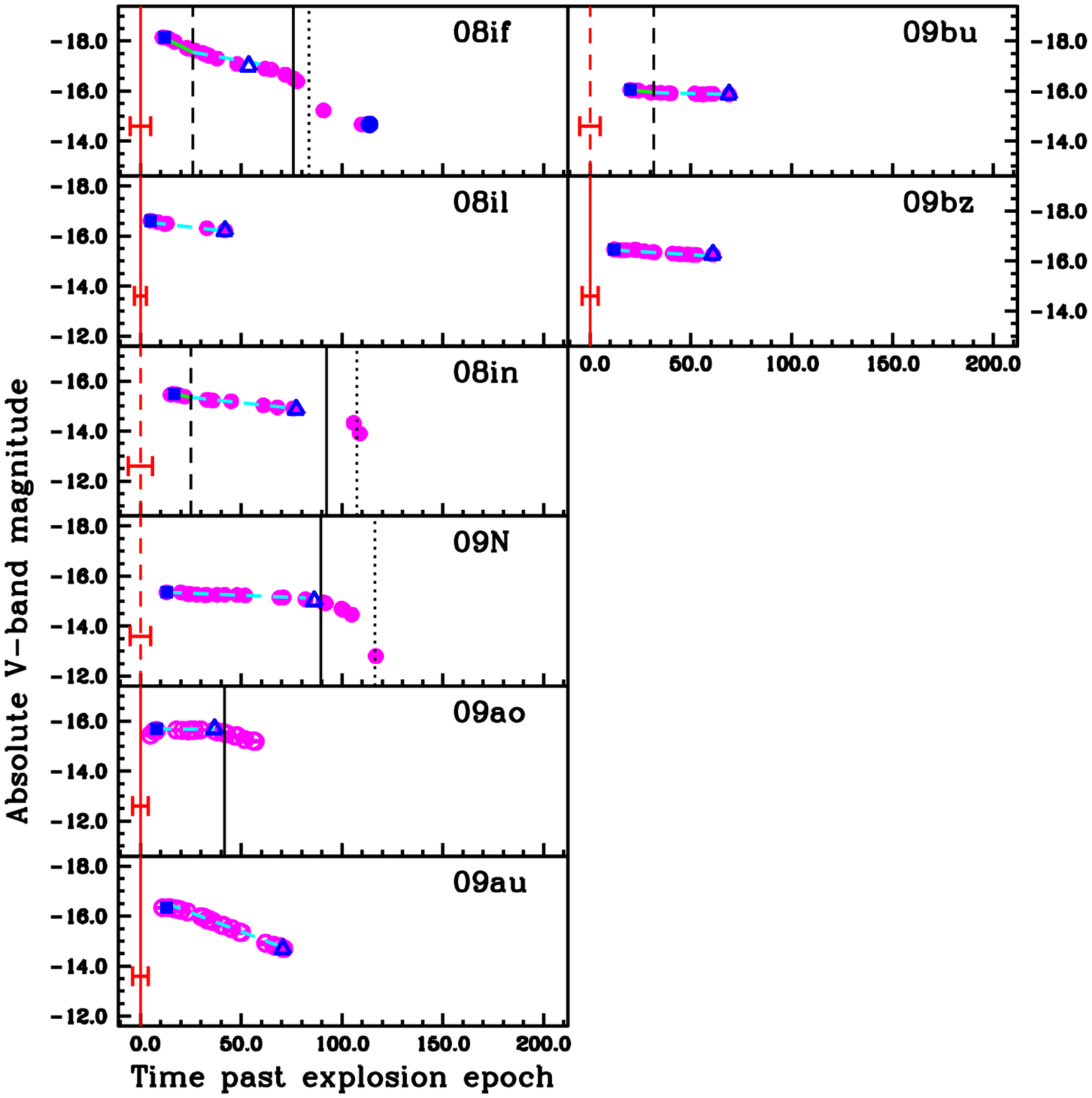}
\caption{Absolute magnitude $V$-band light-curves of SNe~II in 
our sample. The panels progress presenting SNe in order of their discovery
dates, starting with SN~2008aw in the top left panel, and finishing with SN~2009bz in the bottom 
right. Presentation of light-curves and their derived parameters takes the same form 
as in Fig.\ 24.}
\end{figure*}

\indent In Fig.\ 33 $M_\text{end}$, $M_\text{max}$, and $s_2$ are plotted against
estimated $^{56}$Ni masses: there appears to be a trend that more luminous SNe
synthesize larger amounts of radioactive material, as previously observed by
\cite{ham03}, \cite{ber13} and \cite{spi14}, while there is no correlation between $^{56}$Ni mass and decline rate.
Finally, in Fig.\ 34 $M_\text{end}$ and $^{56}$Ni mass are plotted against the
steepness parameter S (see \citealt{elm03} for more discussion on this topic),
where we do not observe trends seen by previous authors.

\begin{figure*}
\includegraphics[width=8cm]{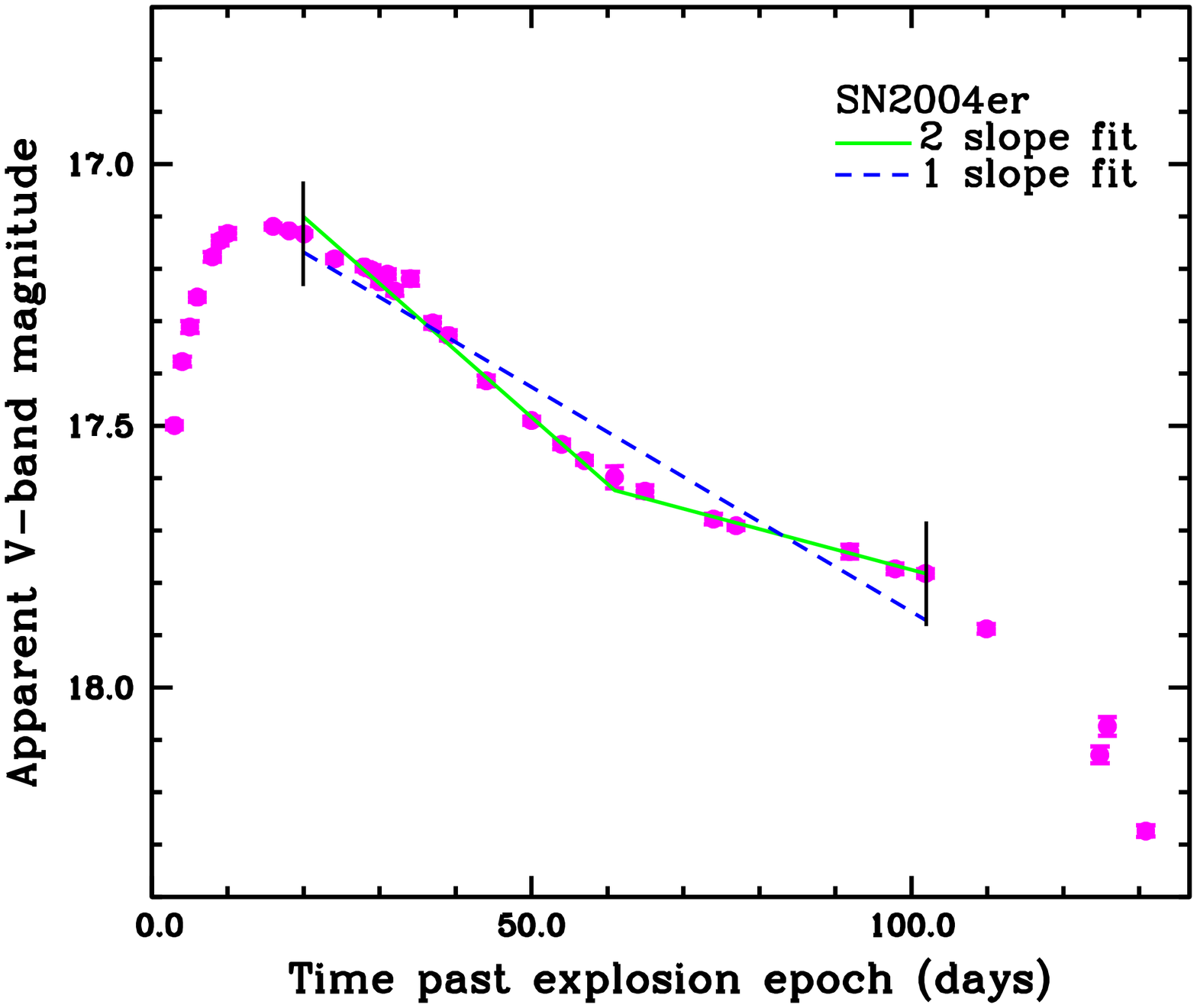}
\includegraphics[width=8cm]{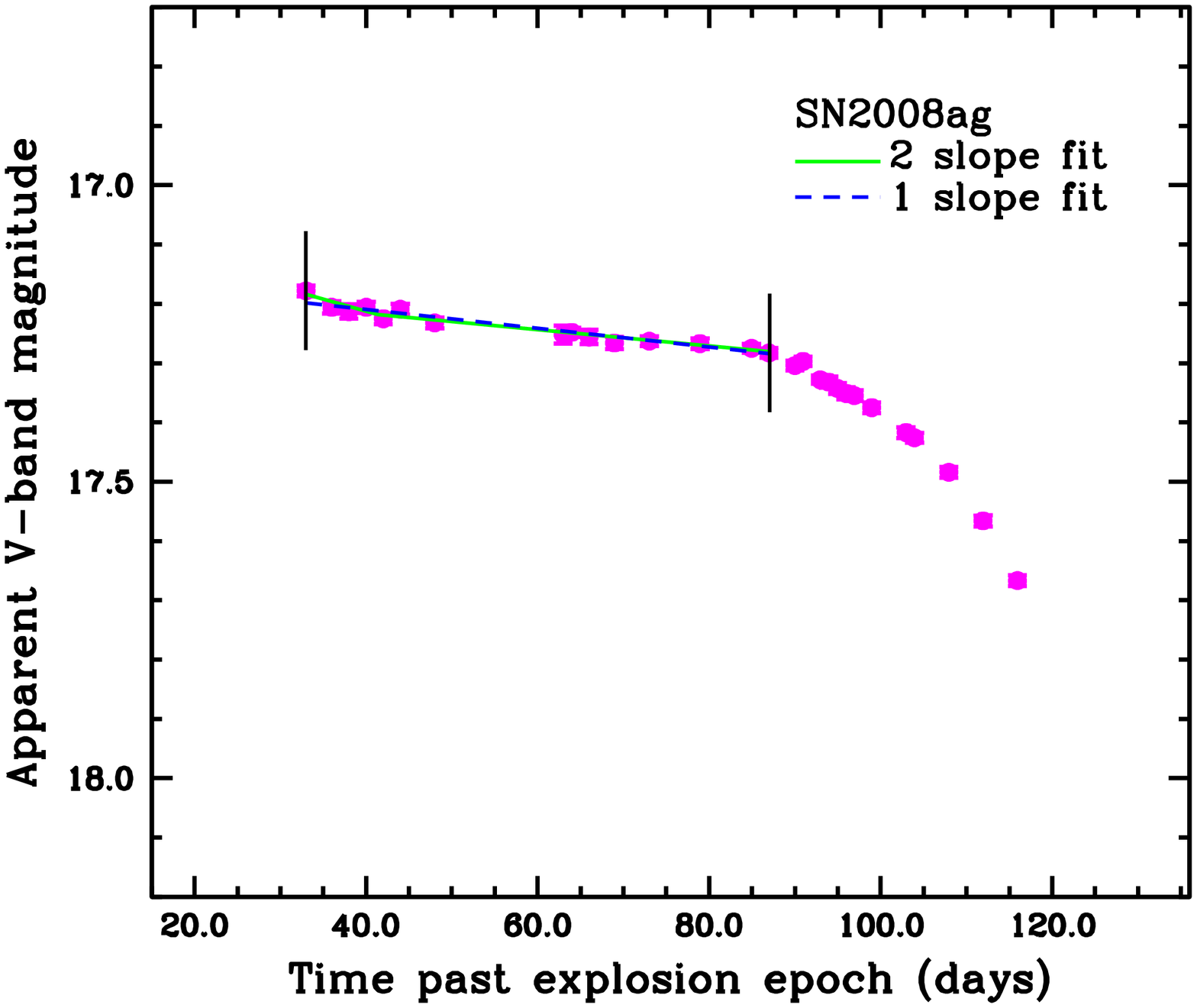}
\caption{Examples of the 2-slope fitting process outlined in \S\
3. \textit{Left:} fits to the $V$-band photometry for
SN~2004er. Here it is clearly observed that the 2-slope model more accurately
represents the data. \textit{Right:} fits to the $V$-band photometry for
SN2008ag. In this second case the 1-slope model is just as good as the 2 slope
model, therefore adding an extra parameter (the second slope) does not improve
the fit.}
\end{figure*}

\begin{figure*}
\includegraphics[width=6cm]{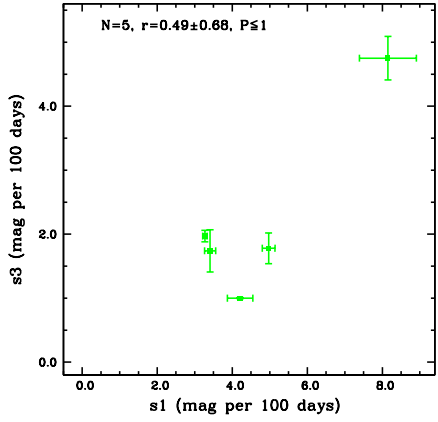}
\caption{$s_1$ plotted against $s_3$. }
\end{figure*}

\begin{figure*}
\includegraphics[width=6cm]{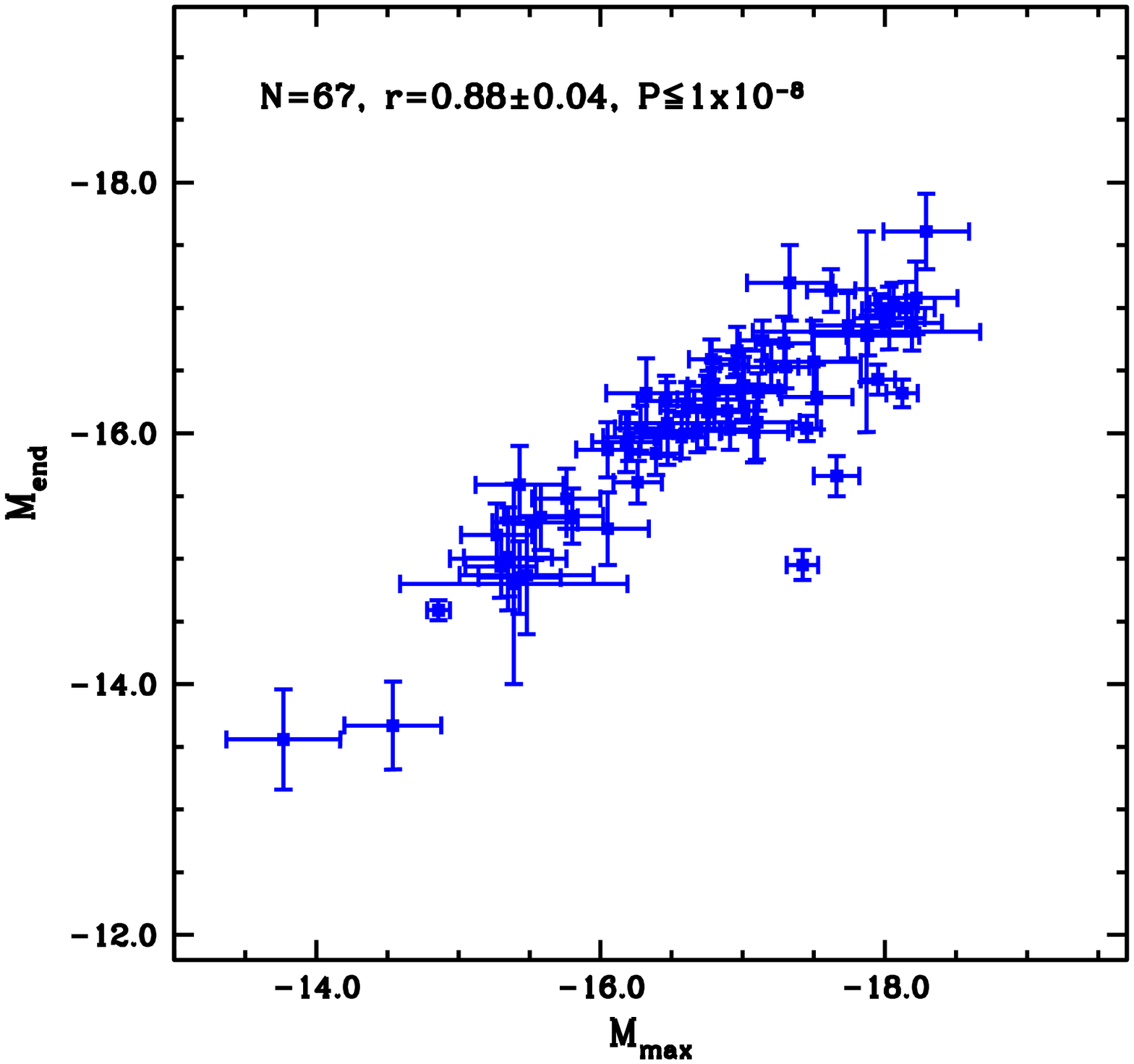}
\includegraphics[width=6cm]{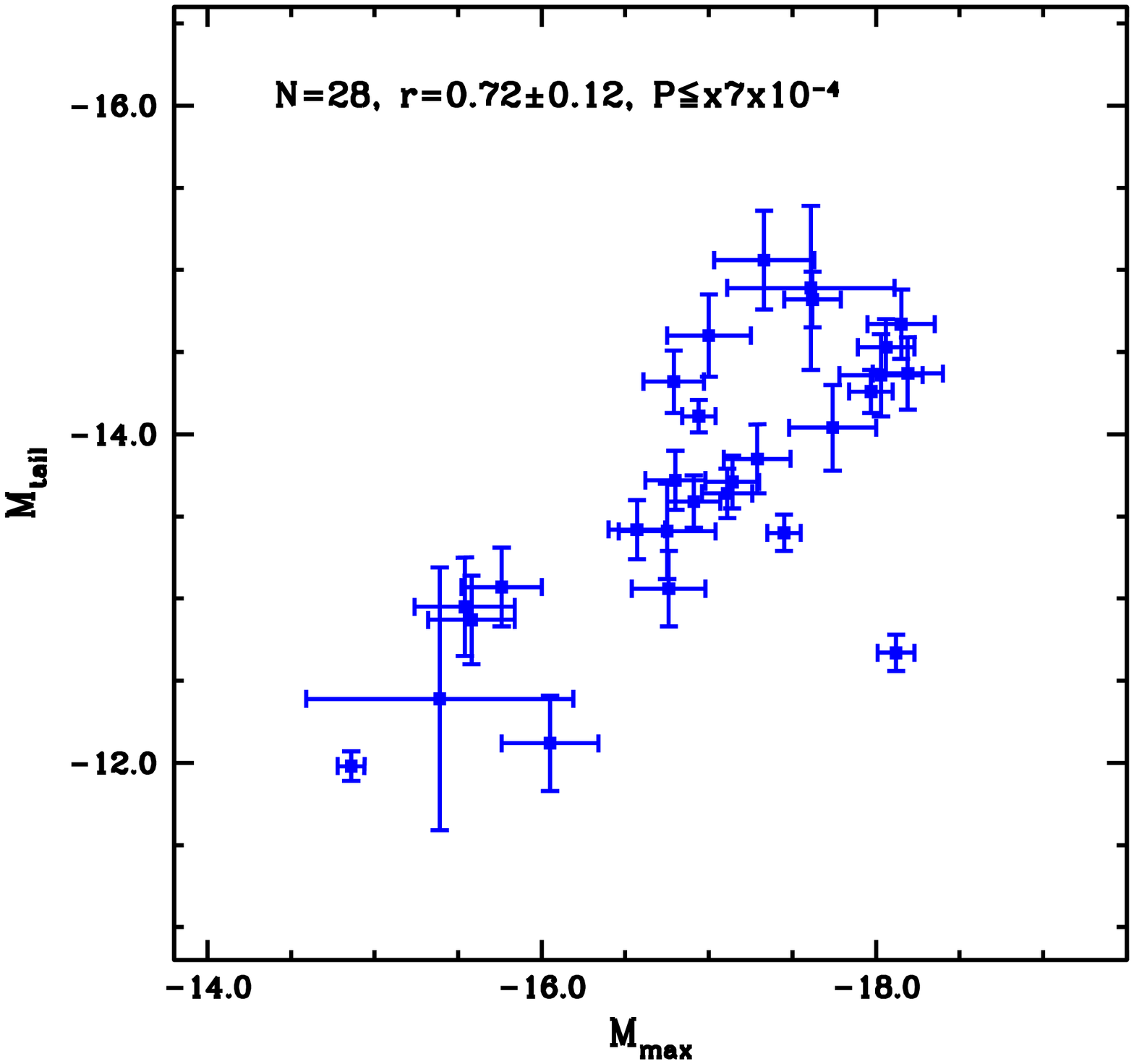}
\includegraphics[width=6cm]{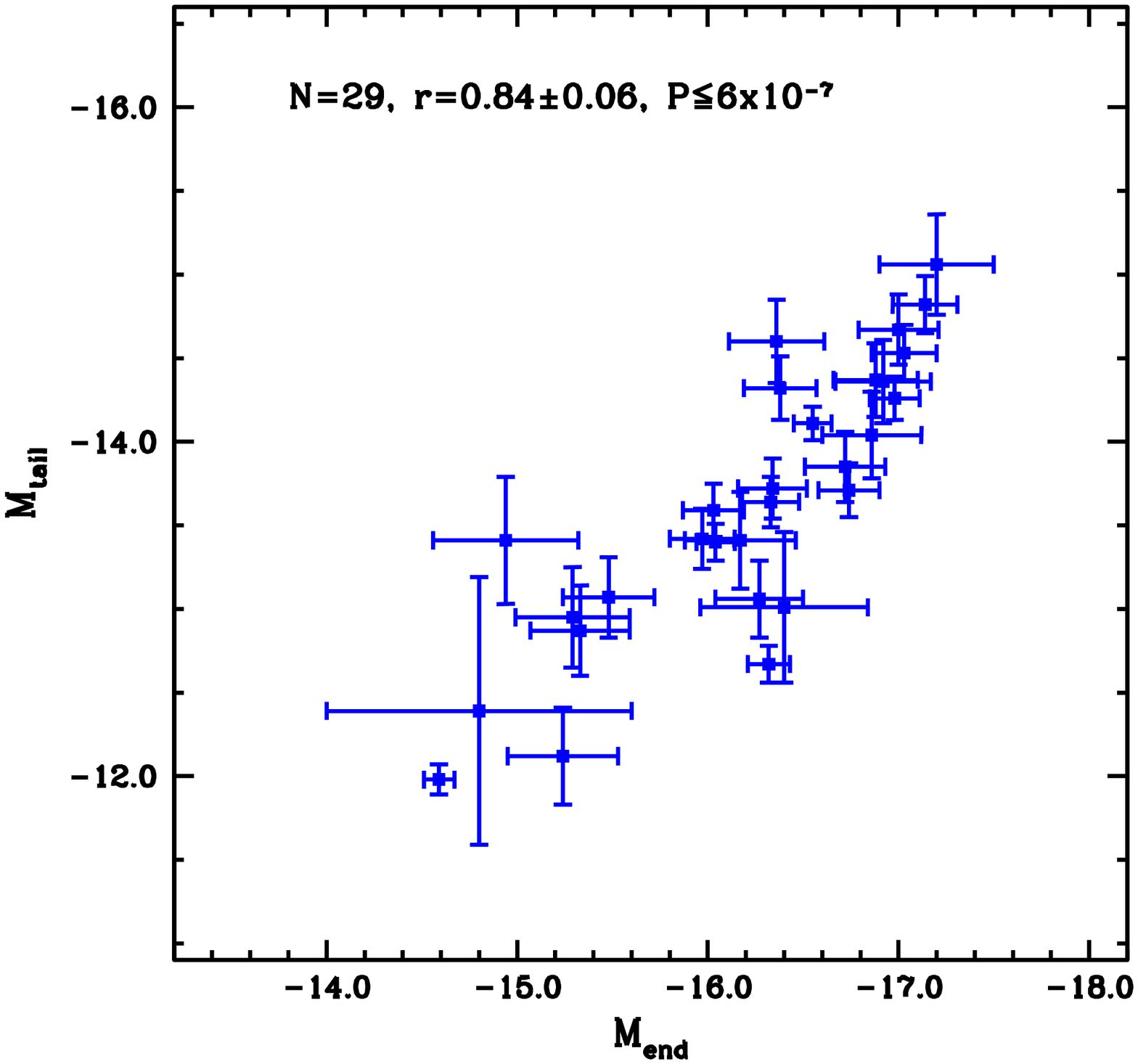}
\caption{Correlations between the three measured absolute magnitudes: $M_\text{max}$,
$M_\text{end}$ and $M_\text{tail}$. \textit{Left:} $M_\text{max}$ vs. $M_\text{end}$. \textit{Middle:}
$M_\text{max}$ vs. $M_\text{tail}$. \textit{Right:} $M_\text{end}$ vs. $M_\text{tail}$. Magnitudes at the three
distinct epochs are strongly correlated. }
\end{figure*}

\begin{figure*}
\includegraphics[width=6cm]{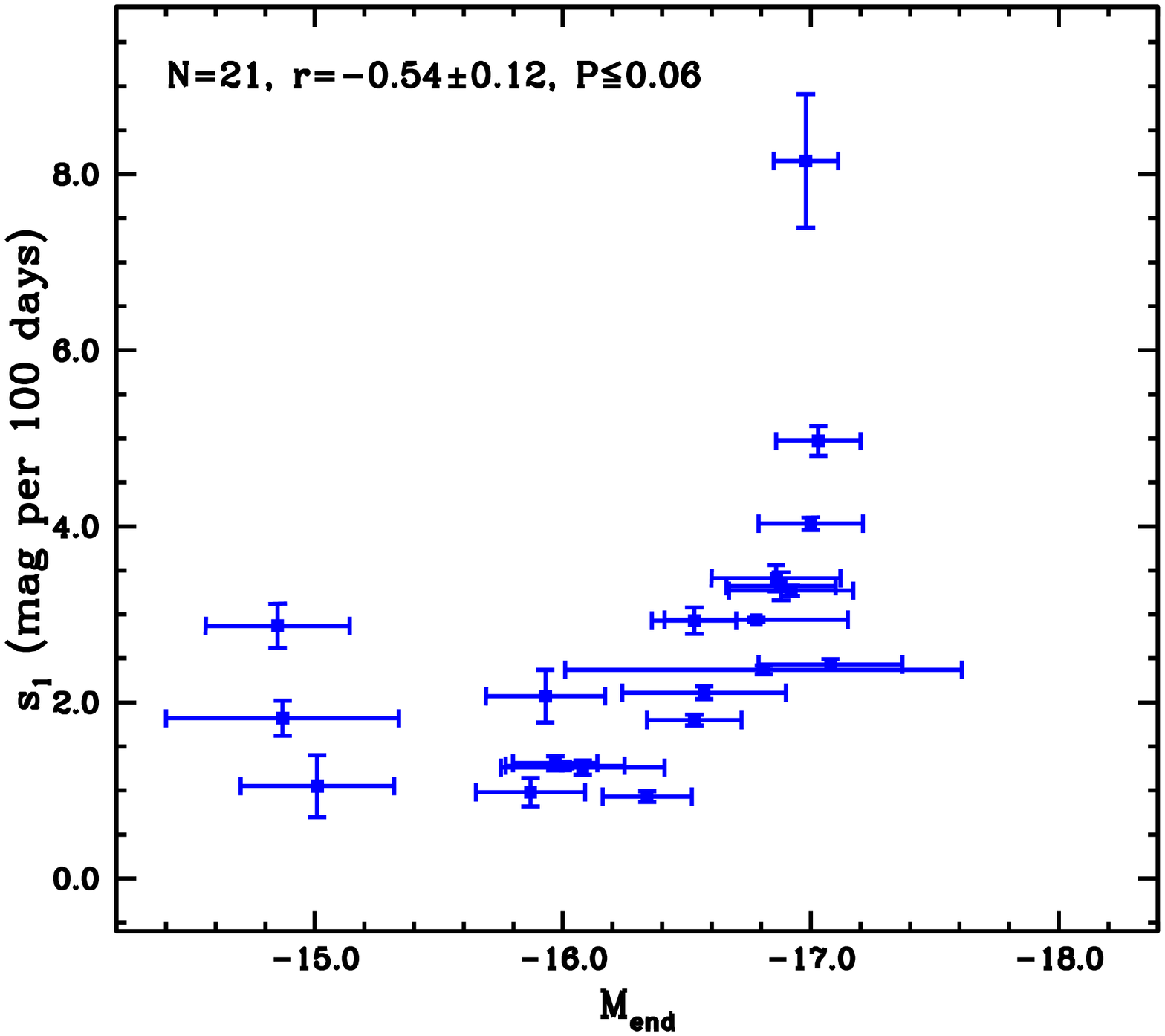}
\includegraphics[width=6cm]{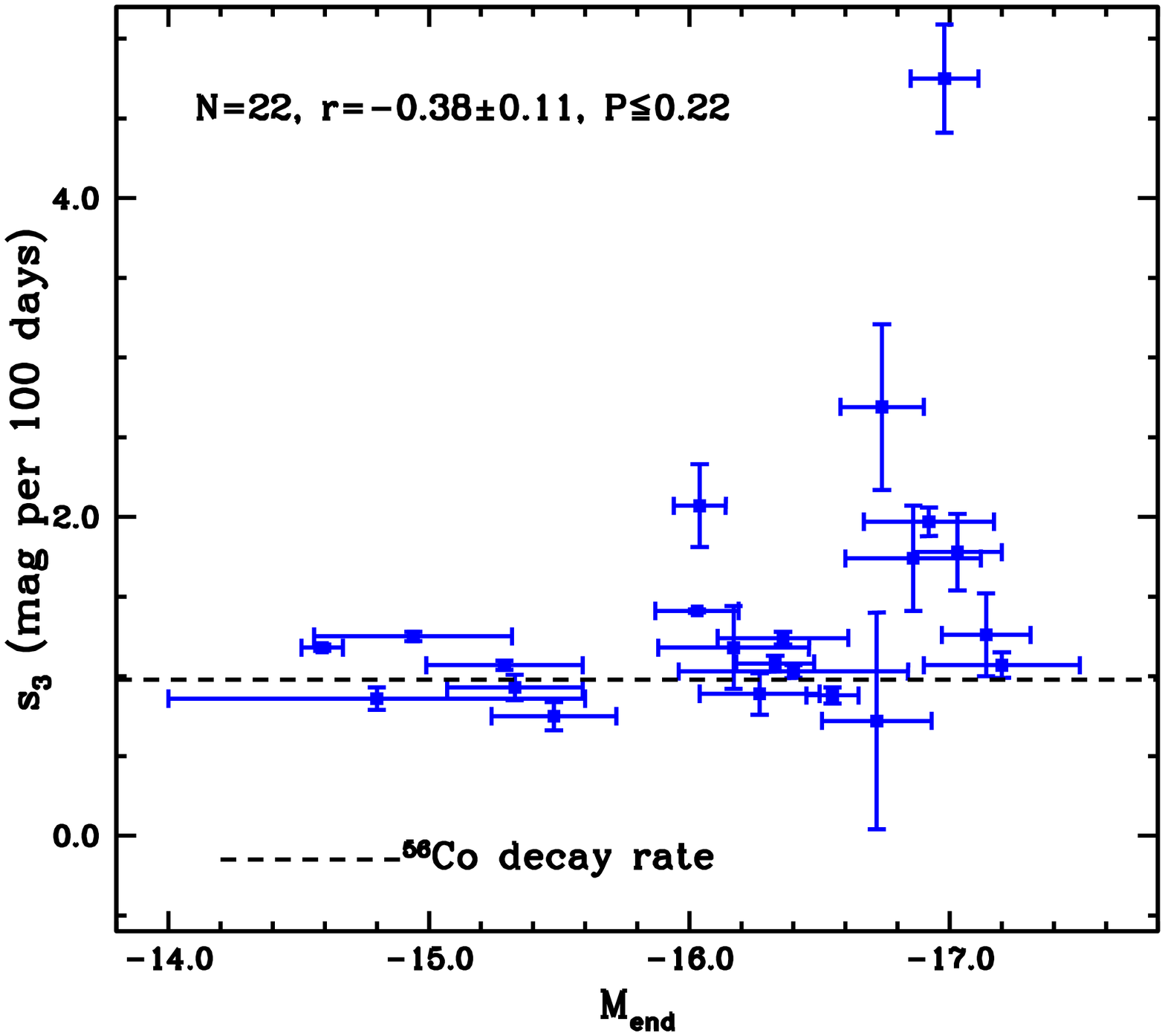}
\caption{\textit{Left:} The magnitude at the end of the `plateau', $M_\text{end}$, plotted against the
initial decline rate from maximum, $s_1$. \textit{Right:} $M_\text{end}$ plotted against the decline
rate of the radioactive tail, $s_3$. The dashed horizontal line shows the
expected decline rate on the radioactive tail, assuming full trapping of
gamma-rays from $^{56}$Co decay.}
\end{figure*}

\begin{figure*}
\includegraphics[width=6cm]{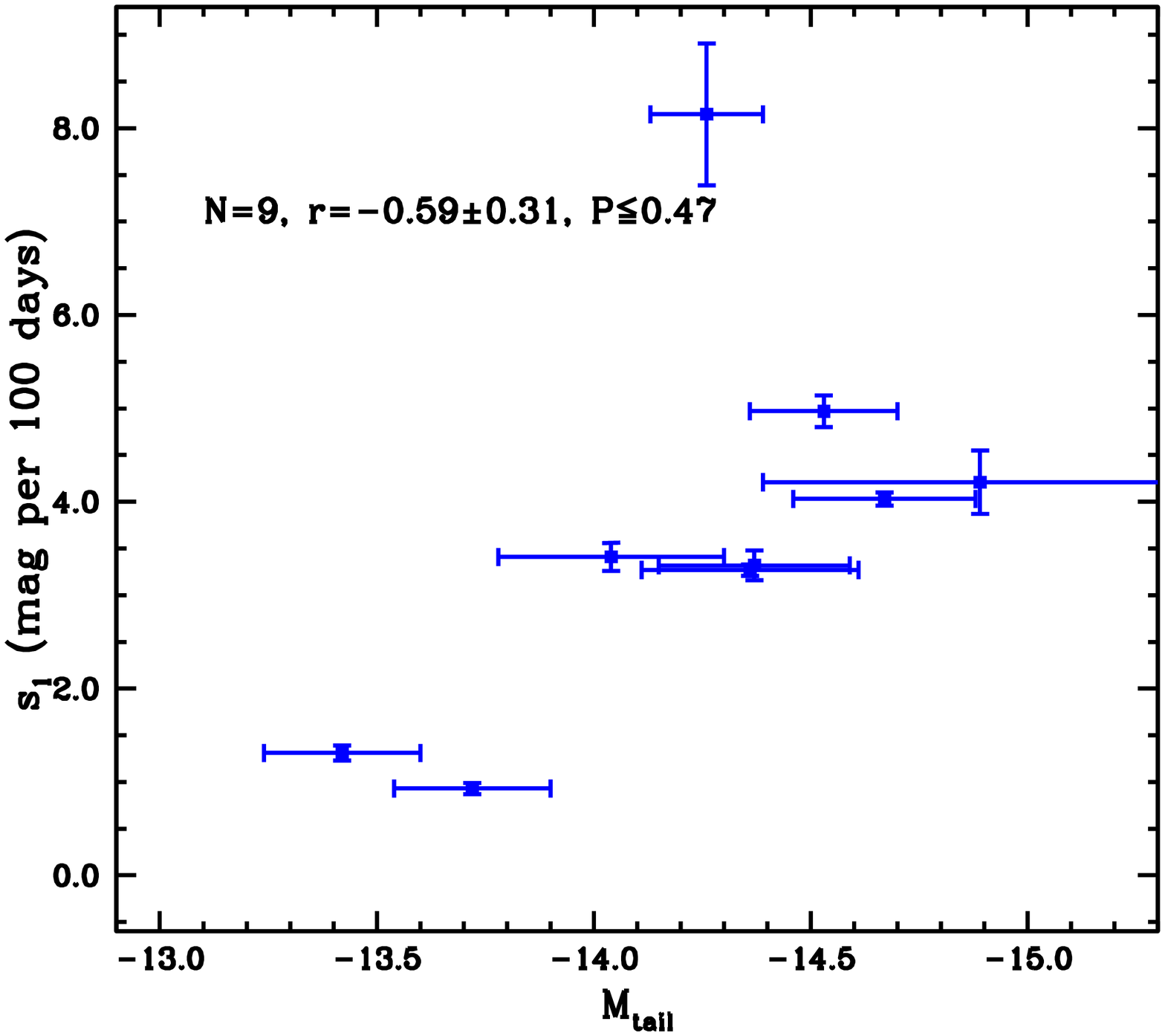}
\includegraphics[width=6cm]{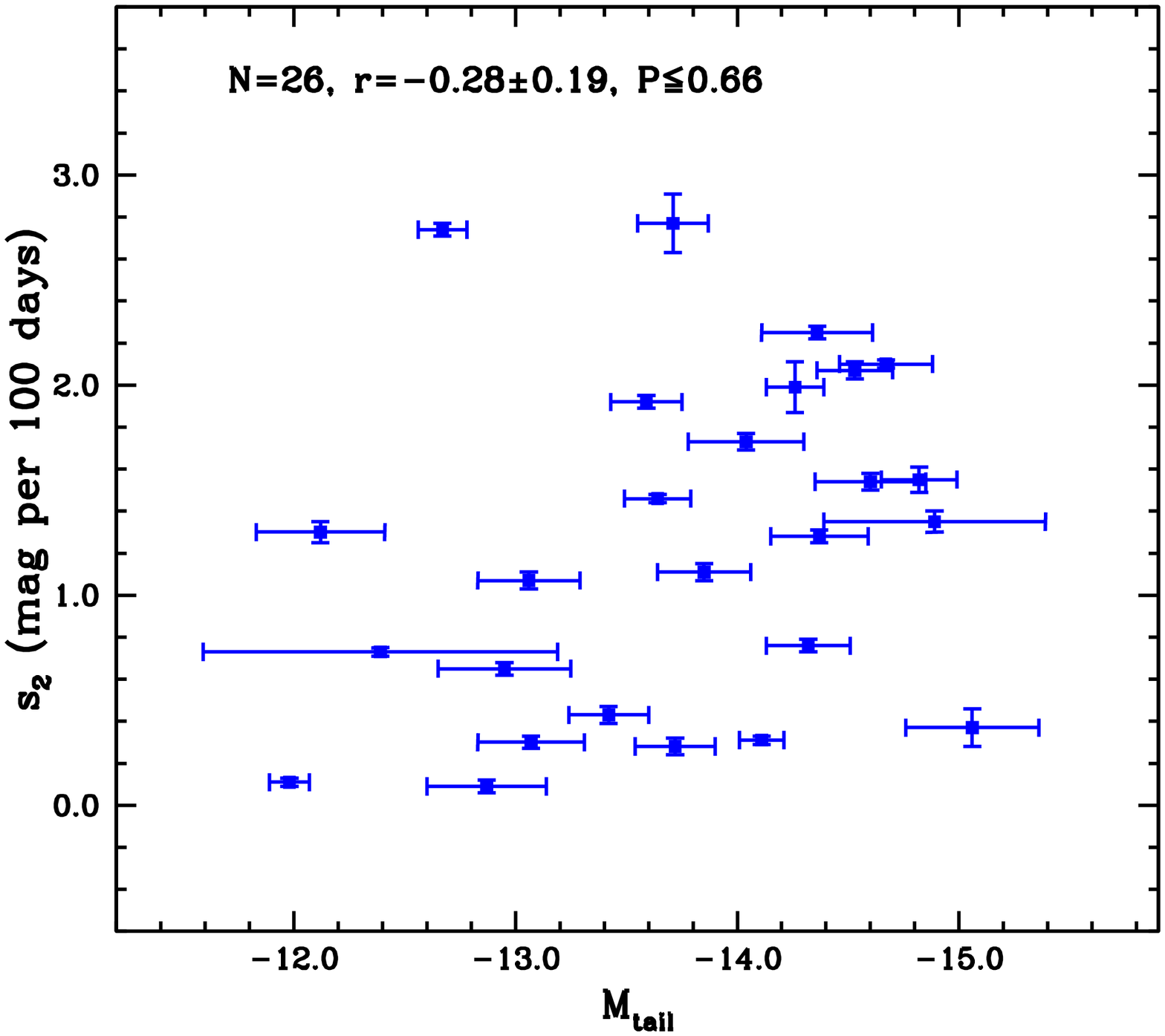}
\includegraphics[width=6cm]{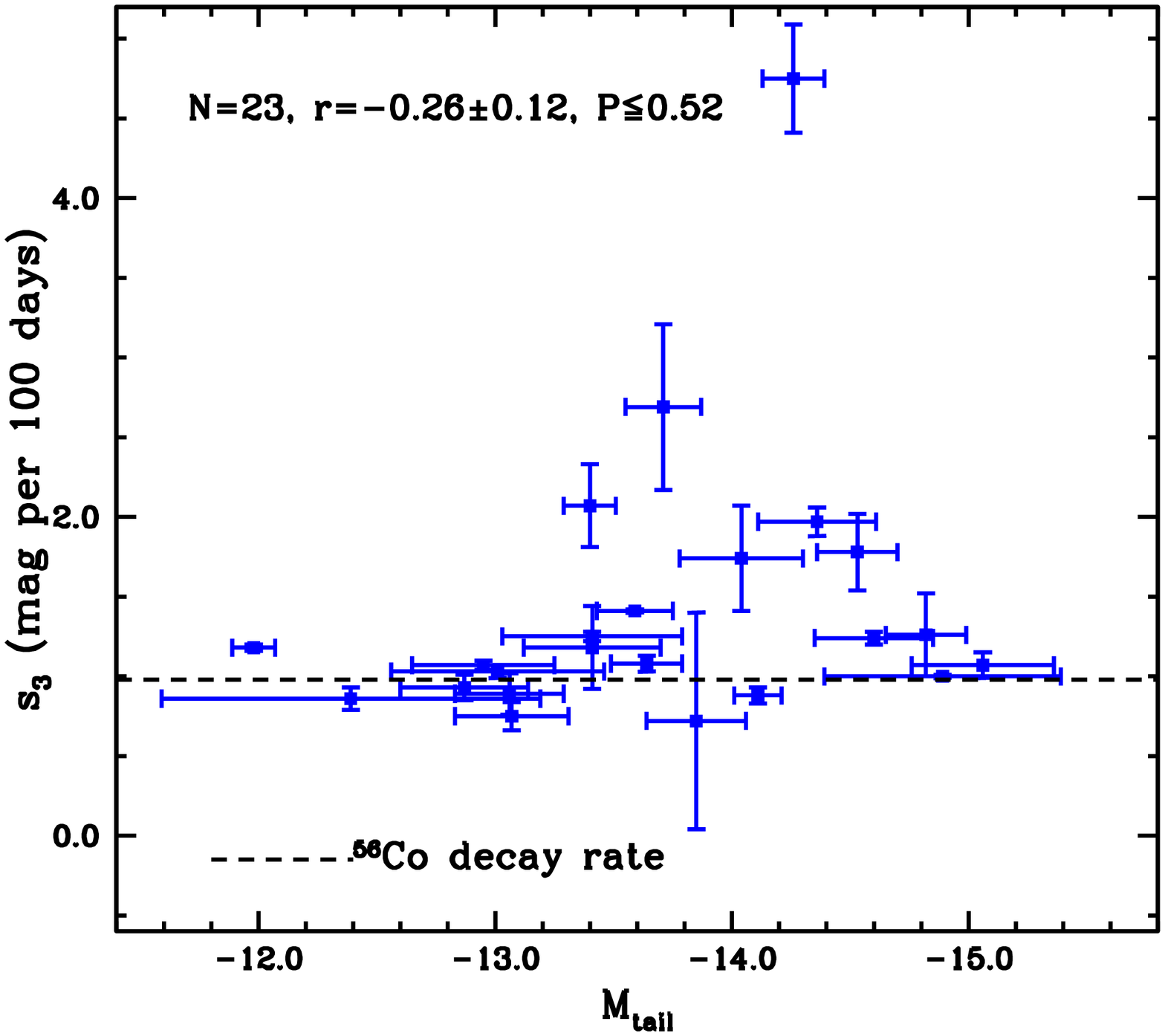}
\caption{\textit{Left:} The magnitude at the beginning of the tail; $M_\text{tail}$, plotted against the
initial decline rate from maximum; $s_1$. \textit{Middle:} $M_\text{tail}$ plotted against the decline
rate on the `plateau'; $s_2$. \textit{Right:} $M_\text{tail}$ plotted against the decline
rate of the radioactive tail. 
}
\end{figure*}

\begin{figure*}
\includegraphics[width=6cm]{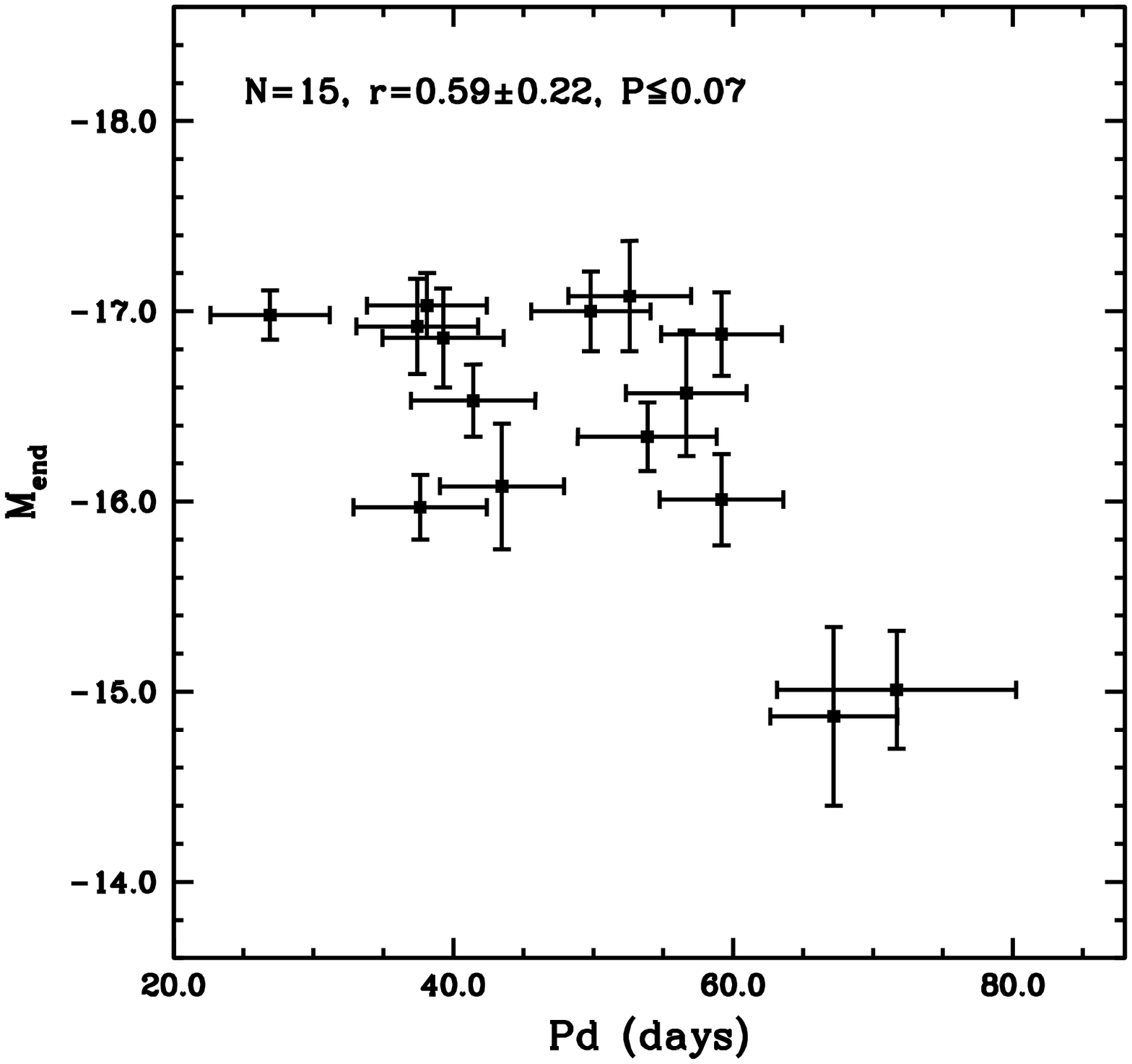}
\includegraphics[width=6cm]{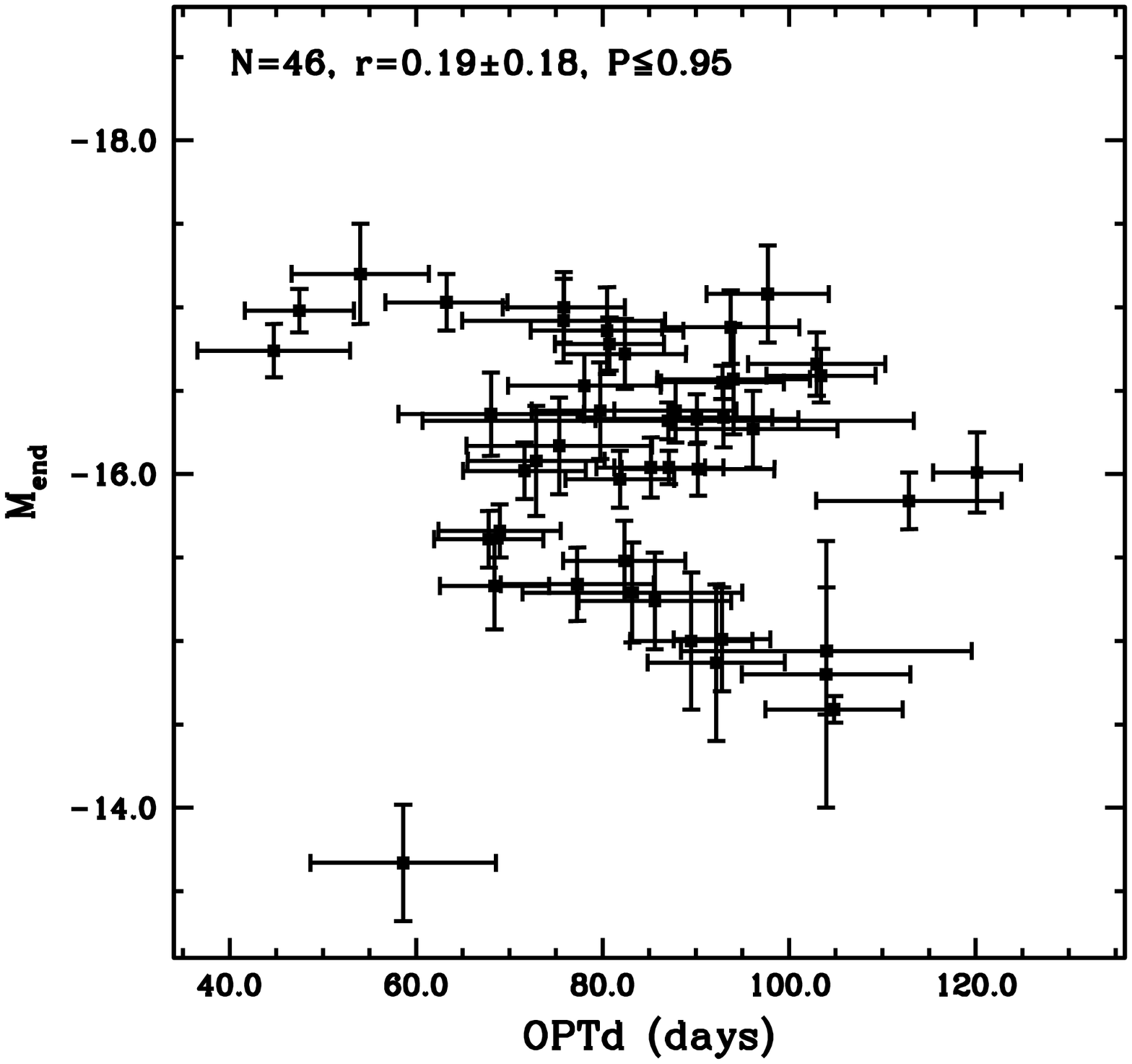}
\caption{\textit{Left:} `Plateau' durations, $Pd$ against $M_\text{end}$. \textit{Right:} SNe optically thick durations, $OPTd$, against
$M_\text{end}$. }
\end{figure*}

\begin{figure*}
\includegraphics[width=6cm]{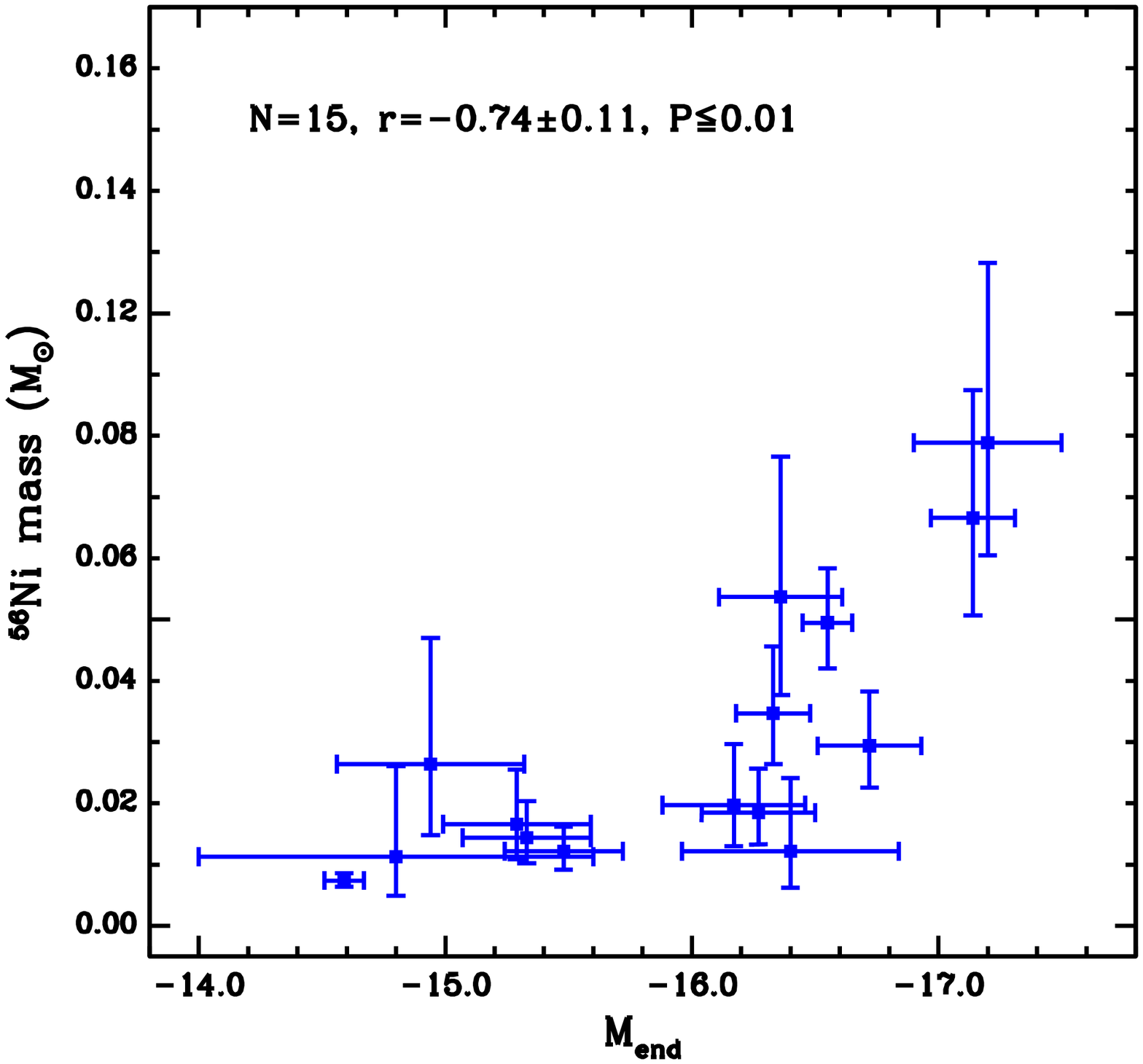}
\includegraphics[width=6cm]{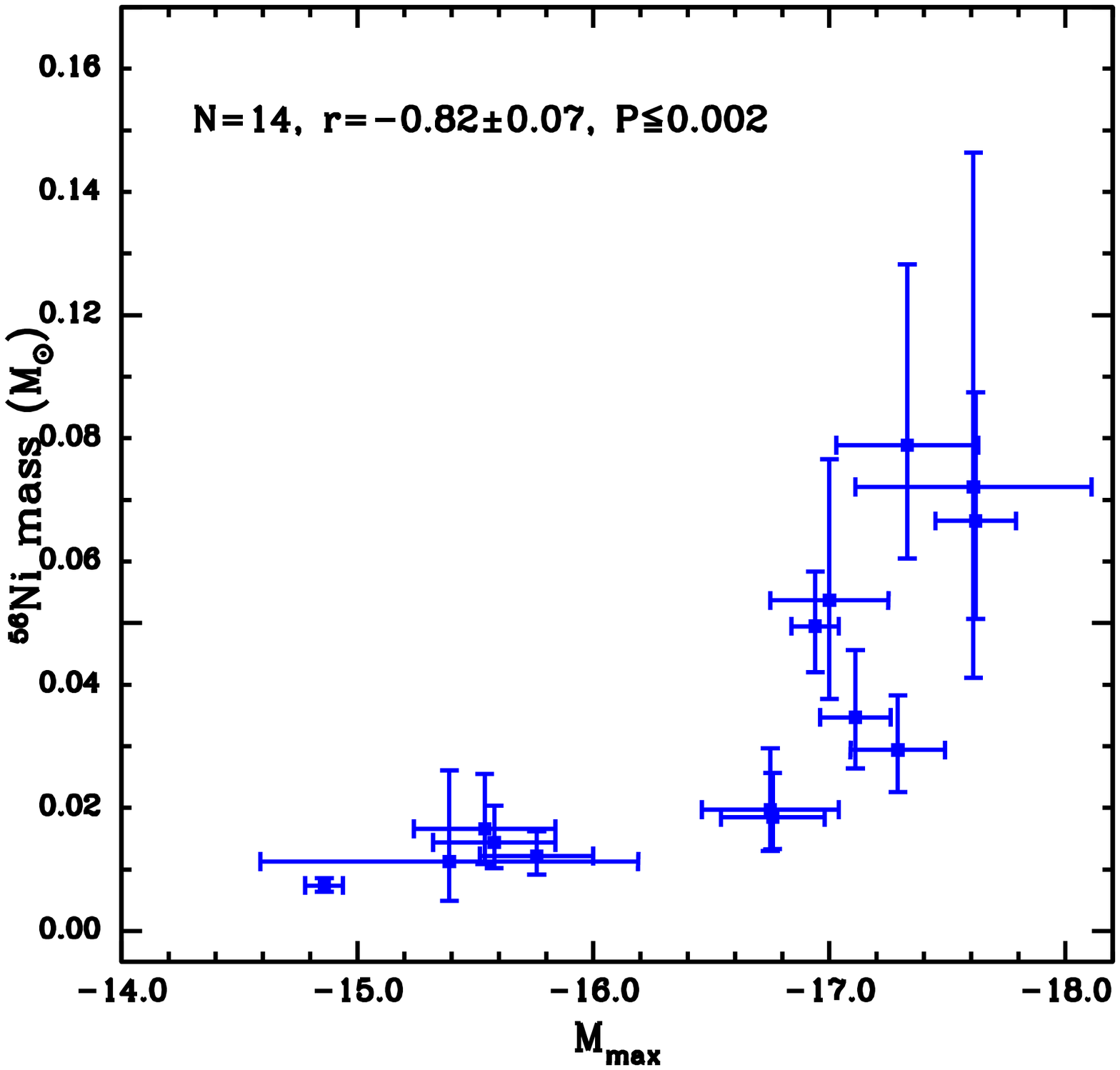}
\includegraphics[width=6cm]{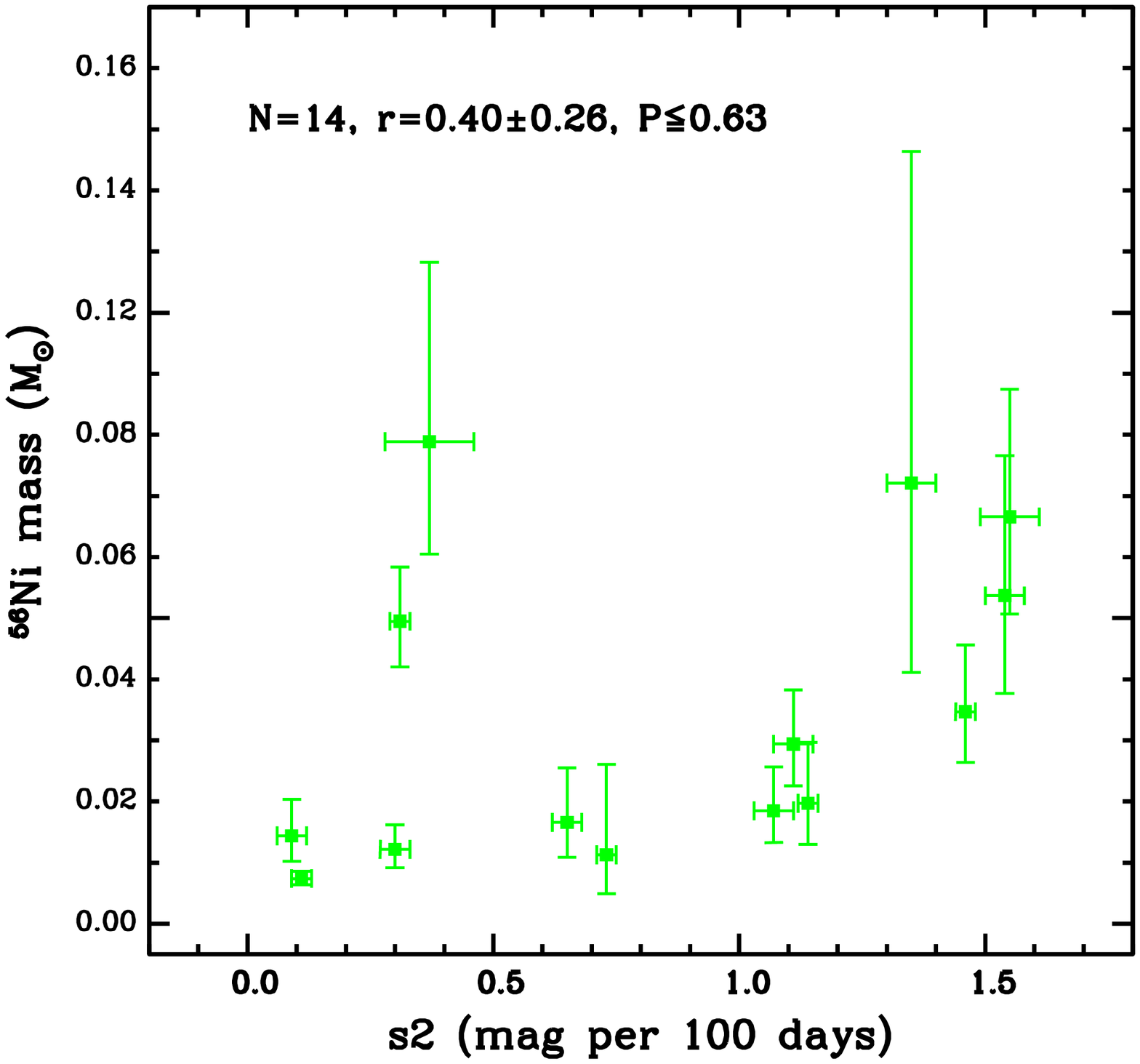}
\caption{\textit{Left:} $M_\text{end}$ against $^{56}$Ni
mass. \textit{Middle:} $M_\text{max}$ plotted against 
$^{56}$Ni mass. \textit{Right:} $s_2$ against $^{56}$Ni
mass. }
\end{figure*}

\begin{figure*}
\includegraphics[width=6cm]{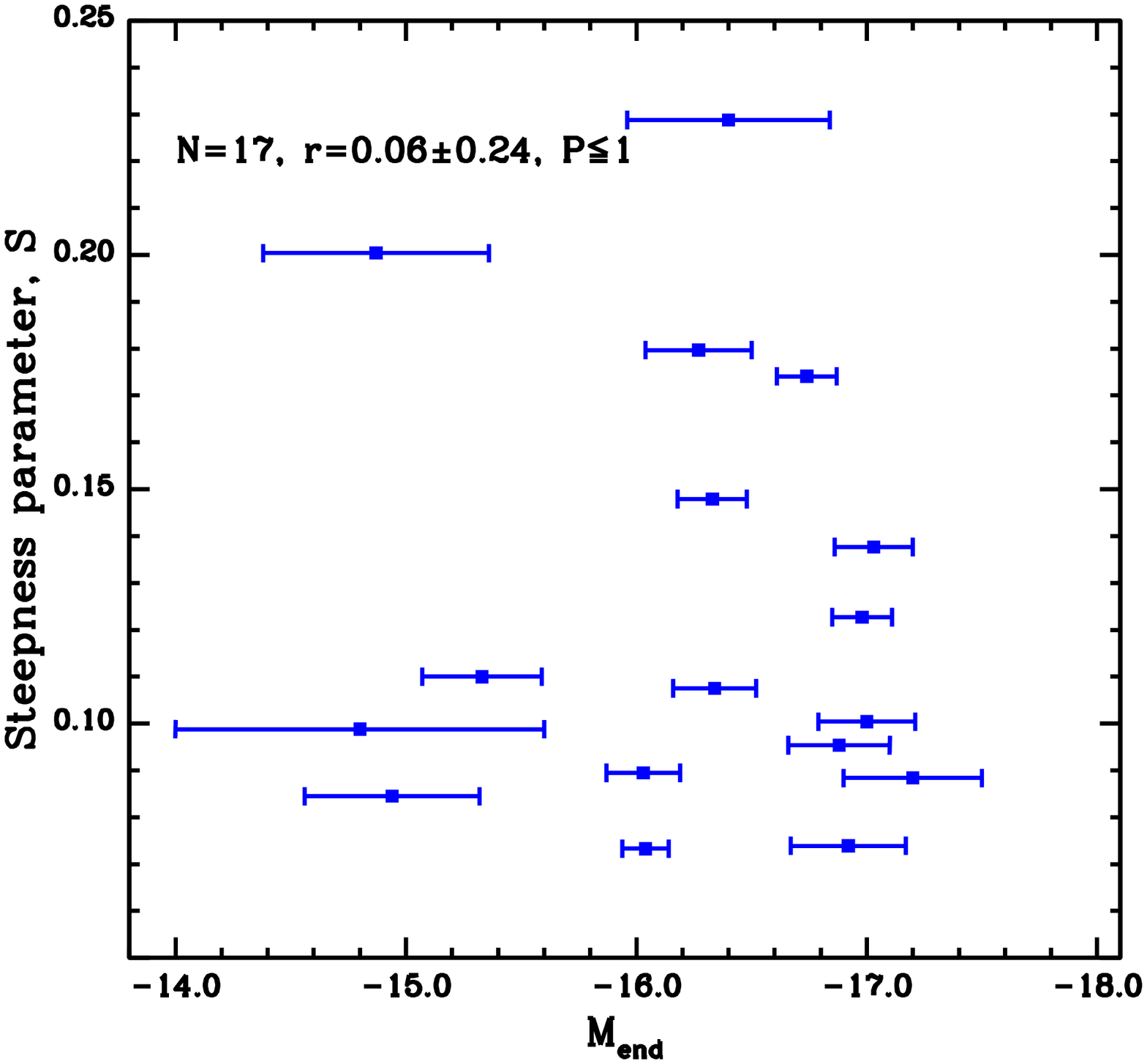}
\includegraphics[width=6cm]{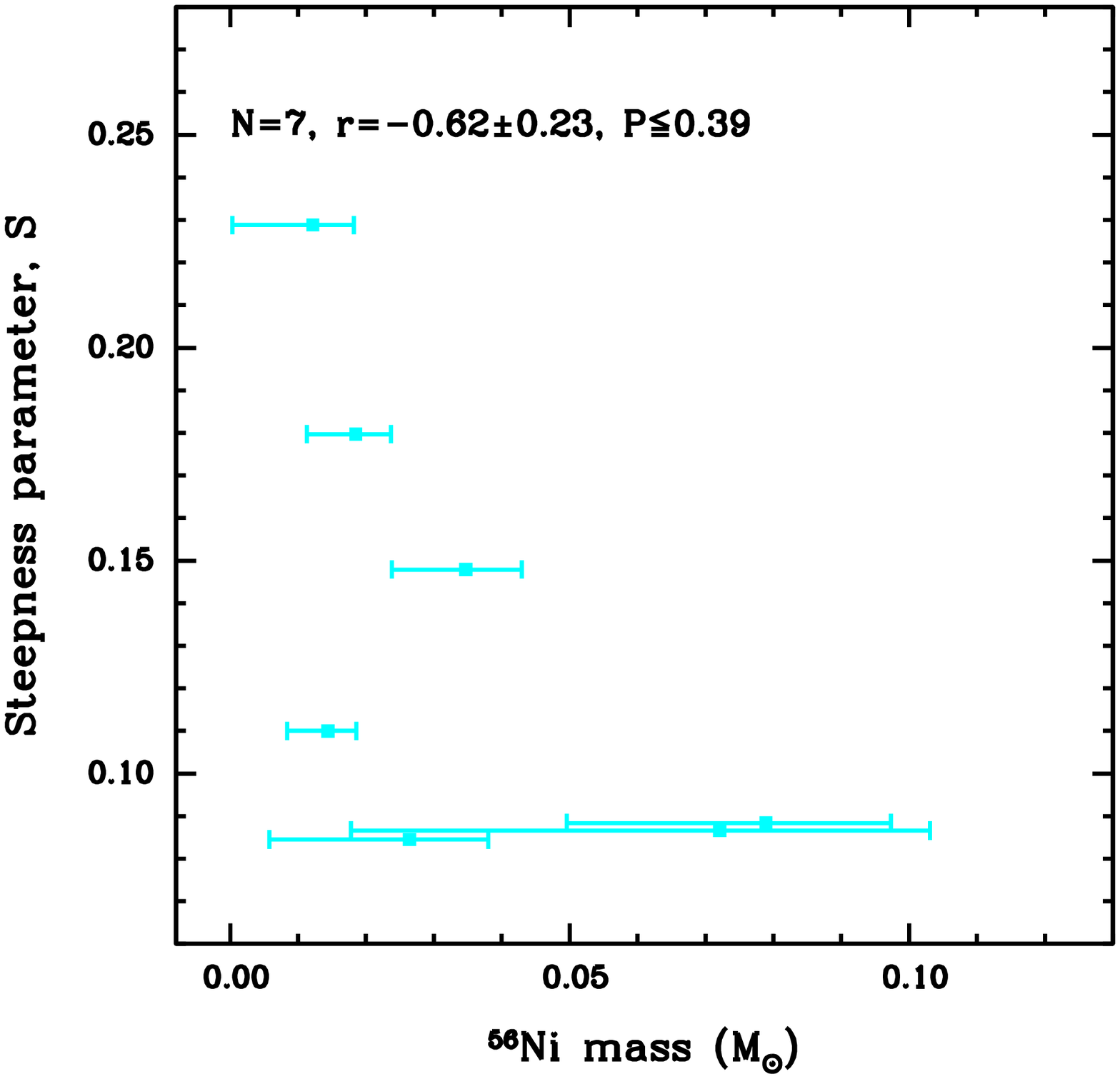}
\caption{\textit{Left:} $M_\text{end}$ against `steepness'
parameter. \textit{Right:} $^{56}$Ni mass against `steepness' parameter. }
\end{figure*}

\newpage
\begin{sidewaystable} \centering
\scriptsize
\caption{SN~II $V$-band light-curve parameters.}
\begin{tabular}[t]{cccccccccccccc}
\hline
SN & DM& Explosion& $t_{PT}$&$A_{\rm V}$(Host) & $M_\text{max}$ & $M_\text{end}$ & $M_\text{tail}$ 
& $s_1$& $s_2$ & $s_3$ & $OPTd$ & $Pd$& $^{56}$Ni mass\\
&(mag)& epoch (MJD)& (MJD) &(mag)&(mag)&(mag)&(mag)&(mag 100d$^{-1}$)&(mag 100d$^{-1}$)&(mag 100d$^{-1}$)&(days)&(days)&(\msun)\\
\hline	
\hline
1986L & 31.72$^1$(0.20) & 46708.0$^{n}$(6)& 46818.4(0.6) &0.00(0.07)
&--18.19(0.21)&--16.88(0.22)&--14.37(0.22)&3.32(0.16)&1.28(0.03)& $\cdots$&
93.7(5)&59.2(4)& $>$0.061\\
1991al& 33.94(0.15) & 48443.5$^{s}$(9)&  48518.8(0.6)&0.11(0.07)
&--17.62(0.17)&--17.14(0.17)&--14.82(0.17)&$\cdots$&1.55(0.06)&1.26(0.26)&
$\cdots$&$\cdots$& 0.067$^{+0.016}_{-0.021}$\\
1992ad& 31.13(0.80) &$\cdots$&$\cdots$&$\cdots$&--16.98$^{*}$(0.80)&--16.13$^{*}$(0.80)&$\cdots$ & $\cdots$&2.23(0.04)&$\cdots$ &$\cdots$&$\cdots$&$\cdots$\\
1992af& 34.33(0.12)&48791.5$^{s}$(6)&  48861.1(0.4) &0.00(0.27)
&--17.33(0.30)&--17.20(0.30)&--15.06(0.30)&$\cdots$&0.37(0.09)&1.07(0.08)&
54.0(7)&$\cdots$& 0.079$^{+0.018}_{-0.029}$\\
1992am& 36.42(0.05)& $\cdots$&48947.2(0.6)&$\cdots$&--18.06$^{*}$(0.07)&--17.17$^{*}$(0.08)&$\cdots$&$\cdots$&1.17(0.02)&$\cdots$&$\cdots$& $\cdots$&$\cdots$\\
1992ba& 30.41(0.80)& 48888.5$^{s}$(8)&49014.0(1.2)&0.05(0.03)
&--15.39(0.80)&--14.80(0.80)&--12.39(0.80)&$\cdots$&0.73(0.02)&0.86(0.07)&
104.0(9)& $\cdots$&0.011$^{+0.006}_{-0.015}$\\
1993A & 35.44(0.07)& 48995.5$^{n}$(9)&$\cdots$&$\cdots$&--16.44$^{*}$(0.07)&--15.91$^{*}$(0.07)&$\cdots$&$\cdots$&0.72(0.03)&$\cdots$& $\cdots$&$\cdots$&$\cdots$\\
1993K & 32.95(0.23)& 49065.5$^{n}$(9)&$\cdots$&0.37(0.19)
&--18.29(0.30)&--17.61(0.30)&$\cdots$&$\cdots$&2.46(0.08)&$\cdots$&$\cdots$&$\cdots$&$\cdots$ \\
1993S & 35.60(0.07)&
49130.5$^{s}$(4)&$\cdots$&0.00(0.24)&--17.52(0.25)&--16.29(0.26)&$\cdots$&$\cdots$&2.52(0.05)&$\cdots$
& $\cdots$&$\cdots$&$\cdots$\\
1999br& 31.19(0.40)& 51276.5$^{n}$(4) &$\cdots$&0.00(0.04)
&--13.77(0.40)&--13.56(0.40)&$\cdots$&$\cdots$&0.14(0.02)&$\cdots$&
$\cdots$&$\cdots$& $>$0.002\\
1999ca& 33.15(0.21)& 51277.5$^{s}$(7)&51373.2(0.8)&0.26(0.15)&--17.74(0.26)&--16.86(0.26)&--14.04(0.26)&3.41(0.15)&1.73(0.04)&1.74(0.33)&
80.5(8)&39.3(4)&$>$0.047\\
1999cr& 34.70(0.10)& 51247.5$^{s}$(7)&51350.2(3.1)&0.30(0.15)&--17.20(0.19)&--16.53(0.19)&$\cdots$&1.80(0.06)&0.58(0.06)&$\cdots$&
78.1(8)& 41.4(4)&$\cdots$\\
1999eg& 34.75(0.10)&51440.5$^{s}$(7)& $\cdots$&$\cdots$&--16.86$^{*}$(0.10)&--16.12$^{*}$(0.10)&$\cdots$&$\cdots$&1.70(0.08)&$\cdots$& $\cdots$&$\cdots$&$\cdots$\\
1999em& 30.37$^2$(0.07)& 51476.5$^{n}$(5)&51590.1(0.9)&0.18$^{a}$(0.06)&--16.94(0.10)&--16.55(0.10)&--14.11(0.10)
&$\cdots$&0.31(0.02)&0.88(0.05)& 96.0(7)& $\cdots$&0.050$^{+0.008}_{-0.009}$\\
0210&   36.58(0.04)& 52489.5(9)$^s$&52591.9(0.3)&$\cdots$&--16.21$^{*}$(0.04)&--15.90$^{*}$(0.04)&$\cdots$&$\cdots$&2.21(0.08)&$\cdots$& 90.6(10)&$\cdots$&$\cdots$\\
2002ew& 35.38(0.08)& 52500.5$^{n}$(10)&$\cdots$&0.00(0.07)
&--17.42(0.11)&--14.95(0.12)&$\cdots$&$\cdots$&3.58(0.06)&$\cdots$& $\cdots$&$\cdots$&$\cdots$\\
2002fa& 36.92(0.04)& 52503.5$^{s}$(7)&$\cdots$&$\cdots$&--16.95$^{*}$(0.04)&--16.65$^{*}$(0.04)&$\cdots$&$\cdots$&1.58(0.10)&$\cdots$& 67.3(8)&$\cdots$&$>$0.066\\
2002gd& 32.50(0.28)& 52552.5$^{s}$(4)&$\cdots$&0.00(0.06)
&--15.43(0.29)&--14.85(0.29)&$\cdots$&2.87(0.25)&0.11(0.05)&$\cdots$ & $\cdots$&$\cdots$&$\cdots$\\
2002gw& 32.98(0.23)& 52559.5$^{s}$(5)&52661.2(0.8)&0.00(0.05)
&--15.76(0.24)&--15.48(0.24)&--13.07(0.24)&$\cdots$&0.30(0.03)&0.75(0.09)&
82.3(6)& $\cdots$&0.012$^{+0.003}_{-0.004}$\\
2002hj& 34.87(0.10)& 52562.5$^{n}$(7)&52661.8(0.8)&0.00(0.11)
&--16.91(0.16)&--16.03(0.16)&--13.59(0.16)&$\cdots$&1.92(0.03)&1.41(0.01)&
90.2(8)&$\cdots$& $>$0.026\\
2002hx& 35.59(0.07)& 52582.5$^{n}$(9)&52658.9(0.8)&0.00(0.23)
&--17.00(0.25)&--16.36(0.25)&--14.60(0.25)&$\cdots$&1.54(0.04)&1.24(0.04)&
68.0(10)& $\cdots$& 0.053$^{+0.016}_{-0.023}$\\
2002ig& 37.47(0.03)& 52572.5$^{s}$(4)&$\cdots$&$\cdots$&--17.66$^{*}$(0.03)&--16.76$^{*}$(0.03)&$\cdots$&$\cdots$&2.73(0.11)&$\cdots$&
$\cdots$& $\cdots$&$\cdots$\\
2003B&  31.11(0.28)& 52616.5$^{s}$(11)&52713.8(0.9)&0.18(0.09)
&--15.54(0.30)&--15.29(0.30)&--12.95(0.30)&$\cdots$&0.65(0.03)&1.07(0.03)&
83.2(12)& $\cdots$& 0.017$^{+0.006}_{-0.009}$\\
2003E&  33.92(0.15)&52634.5$^{s}$(7)&52765.7(0.8)&$\cdots$&--15.70$^{*}$(0.15)&--15.48$^{*}$(0.15)&$\cdots$&$\cdots$&--0.07(0.03)&$\cdots$&
97.4(8)& $\cdots$&$\cdots$\\
2003T&  35.37(0.08)& 52654.5$^{n}$(10)&52758.7(0.6)&$\cdots$
&--16.54$^{*}$(0.08)&--16.03$^{*}$(0.08)&--13.67$^{*}$(0.08)&$\cdots$&0.82(0.02)&2.02(0.14)&
90.6(11)&$\cdots$& $>$0.030\\
2003bl& 34.02(0.14)&
52699.5$^{s}$(3)&52804.5(0.2)&0.00(0.27)&--15.35(0.31)&--15.01(0.31)&$\cdots$&1.05(0.35)&0.24(0.04)&
$\cdots$& 
92.8(5)& 71.7(9)& $\cdots$\\
2003bn& 33.79(0.16)& 52694.5$^{n}$(3)&52813.7(0.7)&0.00(0.07)&--16.80(0.18)&--16.34(0.18)&--13.72(0.18)&0.93(0.06)&0.28(0.04)&$\cdots$&
93.0(5)&53.9(5)& $>$0.038\\
2003ci& 35.56(0.07)& 52711.5$^{n}$(8)&52817.7(1.0)&$\cdots$
&--16.83$^{*}$(0.07)&--15.70$^{*}$(0.07)&$\cdots$&$\cdots$&1.79(0.04)&$\cdots$& 92.5(9)&$\cdots$&$\cdots$\\
2003cn& 34.48(0.11)& 52719.5$^{s}$(4)&52804.2(0.3) &0.00(0.12)
&--16.26(0.17)&--15.61(0.17)&$\cdots$&$\cdots$&1.43(0.04)&$\cdots$& 67.8(6)&$\cdots$&$\cdots$\\
2003cx& 35.96(0.06)& 52728.5$^{s}$(5)&52828.5(0.8)&0.00(0.17)
&--16.79(0.18)&--16.38(0.19)&--14.32(0.19)&$\cdots$&0.76(0.03)&$\cdots$&
87.8(7)&$\cdots$& $>$0.051\\
2003dq& 36.44(0.06)& 52731.5$^{n}$(8)&$\cdots$&$\cdots$
&--16.69$^{*}$(0.06)&--15.69$^{*}$(0.36)&$\cdots$&$\cdots$&2.50(0.19)&$\cdots$& $\cdots$&$\cdots$&$\cdots$\\
2003ef& 34.08(0.14)& 52759.5$^{s}$(9)&52869.9(0.6)&$\cdots$
&--16.72$^{*}$(0.14)&--16.15$^{*}$(0.14)&$\cdots$&$\cdots$&0.81(0.02)&$\cdots$& 90.9(10)& $\cdots$&$\cdots$\\
2003eg& 34.23(0.13)& 52773.5$^{s}$(5)&$\cdots$&$\cdots$
&--17.81$^{*}$(0.13)&--14.57$^{*}$(0.13)&$\cdots$&$\cdots$&2.93(0.04)&$\cdots$& $\cdots$&$\cdots$&$\cdots$\\
2003ej& 34.36(0.12)& 52775.5$^{n}$(5)&$\cdots$&0.00(0.09)
&--17.66(0.16)&--15.66(0.16)&$\cdots$&$\cdots$&3.46(0.05)&$\cdots$& 69.0(7)&$\cdots$&$\cdots$\\
2003fb& 34.43(0.12)& 52776.5$^{s}$(6)&52874.4(0.6)&$\cdots$
&--15.56$^{*}$(0.13)&--15.25$^{*}$(0.13)&--13.10$^{*}$(0.14)&$\cdots$&0.48(0.06)&1.61(0.39)&
84.3(7)& $\cdots$& $>$0.017\\
2003gd& 29.93(0.40)& 52767.5$^{s}$(15)&52840.9(0.2)&0.43$^b$(0.19) &$\cdots$
&--16.40(0.44)&--13.01(0.45)&$\cdots$&$\cdots$&1.03(0.04)& $\cdots$&$\cdots$&
0.012$^{+0.006}_{-0.012}$\\
2003hd& 36.00(0.06)& 52857.5$^{s}$(5)&52952.9(1.2)&0.00(0.19)
&--17.29(0.20)&--16.72(0.21)&--13.85(0.21)&$\cdots$&1.11(0.04)&0.72(0.68)&
82.4(6)&$\cdots$& 0.029$^{+0.007}_{-0.009}$\\
2003hg& 33.65(0.16)& 52865.5$^{n}$(5)&52998.6(1.1)&$\cdots$
&--16.38$^{*}$(0.16)&--15.50$^{*}$(0.16)&$\cdots$&1.60(0.06)&0.59(0.03)&$\cdots$& 108.5(7)& 67.1(4)&$\cdots$\\
2003hk& 34.77(0.10)& 52867.5$^{s}$(4)&52961.2(1.6)&$\cdots$
&--17.02$^{*}$(0.10)&--16.36$^{*}$(0.10)&--13.14$^{*}$(0.10)&$\cdots$&1.85(0.06)&0.40(0.66)&
86.0(6)& $\cdots$& $>$0.017\\
2003hl& 32.39(0.30)& 52868.5$^{n}$(5)&53005.4(0.1) &$\cdots$
&--15.91$^{*}$(0.30)&--15.23$^{*}$(0.30)&$\cdots$&$\cdots$&0.74(0.01)&$\cdots$& 108.9(7)& $\cdots$&$\cdots$\\
2003hn& 31.15(0.10)& 52866.5$^{n}$(10)&52963.7(0.1)&0.37$^c$(0.10)&--17.11(0.15)&--16.33(0.15)&--13.64(0.15)&$\cdots$&1.46(0.02)&1.08(0.05)&
90.1(10)& $\cdots$& 0.035$^{+0.008}_{-0.011}$\\
2003ho& 33.77(0.16)& 52847.5$^{s}$(7)&52920.9(0.1)&$\cdots$   &$\cdots$
&--14.75$^{*}$(0.16)&--12.00$^{*}$(0.16)&$\cdots$&$\cdots$&1.69(0.10)& $\cdots$&$\cdots$&$>$0.005\\
2003ib& 34.97(0.09)& 52891.5$^{n}$(8)&$\cdots$&0.00(0.28)
&--17.10(0.30)&--16.09(0.30)&$\cdots$&$\cdots$&1.66(0.05)&$\cdots$& $\cdots$&$\cdots$&$\cdots$\\
2003ip& 34.20(0.13)& 52896.5$^{s}$(4)&52997.6(0.1)&0.13(0.08)
&--17.88(0.15)&--16.78(0.16)&$\cdots$&$\cdots$&2.01(0.03)&$\cdots$& 80.7(6)&$\cdots$&$\cdots$\\
2003iq& 32.39(0.30)& 52919.5$^{n}$(2)&53019.6(0.1)&$\cdots$
&--16.69$^{*}$(0.30)&--16.18$^{*}$(0.30)&$\cdots$&$\cdots$&0.75(0.03)&$\cdots$& 84.9(4)& $\cdots$&$\cdots$\\
2004dy& 35.46(0.07)&53241.0$^{n}$(3)&53289.0(0.5)&$\cdots$
&--16.03$^{*}$(0.07)&--16.02$^{*}$(0.07)&$\cdots$&$\cdots$&0.09(0.14)&$\cdots$&$\cdots$&25.0(5)& $\cdots$\\
\hline
\hline
\end{tabular}
\end{sidewaystable}

\setcounter{table}{5}
\begin{sidewaystable}\centering
\scriptsize
\caption{SN~II $V$-band light-curve parameters --\textit{Continued}.}
\begin{tabular}[t]{cccccccccccccc}
\hline
SN & DM& Explosion&  $t_{PT}$&$A_{\rm V}$(Host) & $M_\text{max}$ & $M_\text{end}$ & $M_\text{tail}$ 
& $s_1$& $s_2$ & $s_3$ & $OPTd$ & $Pd$& $^{56}$Ni mass\\
&(mag)& epoch (MJD)& (MJD) &(mag)&(mag)&(mag)&(mag)&(mag 100d$^{-1}$)&(mag 100d$^{-1}$)&(mag 100d$^{-1}$)&(days)&(days)&(\msun)\\
\hline
\hline	
2004ej& 33.10(0.21)& 53224.9$^{s}$(8)&53338.7(0.6)&0.14(0.07)
&--16.76(0.22)&--16.27(0.23)&--13.06(0.23)&$\cdots$&1.07(0.04)&0.89(0.13)&
96.1(9)& $\cdots$& 0.019$^{+0.005}_{-0.007}$\\
2004er& 33.79(0.16)& 53271.8$^{n}$(2)&53429.6(2.5)&0.34(0.17)&--17.08(0.24)&--16.01(0.24)&$\cdots$&1.28(0.03)&0.40(0.03)&$\cdots$&
120.2(6)&59.2(4)&$\cdots$\\
2004fb& 34.54(0.11)& 53242.6$^{s}$(4)&$\cdots$&$\cdots$&--16.19$^{*}$(0.11)&--15.46$^{*}$(0.11)&$\cdots$& $\cdots$ & 1.24(0.07) &
$\cdots$ & $\cdots$& $\cdots$&$\cdots$\\
2004fc& 31.68(0.31)& 53293.5$^{n}$(1)&53425.9(1.0)&$\cdots$&--16.21$^{*}$(0.31)&--15.41$^{*}$(0.31)&$\cdots$&$\cdots$&0.82(0.02)&$\cdots$& 106.06(4)& $\cdots$&$\cdots$\\
2004fx& 32.82(0.24)& 53303.5$^{n}$(4)&53407.3(0.5)&0.00(0.10)
&--15.58(0.26)&--15.33(0.26)&--12.87(0.27)&$\cdots$ & 0.09(0.03) & 0.93(0.08)&
68.4(6)&$\cdots$& 0.014$^{+0.004}_{-0.006}$\\
2005J&  33.96(0.14)& 53382.8$^{s}$(7)&53496.6(0.7)&0.22(0.30)
&--17.50(0.33)&--16.57(0.33)&$\cdots$& 2.11(0.07)& 0.96(0.02)&$\cdots$& 94.0(8)&56.7(4)&$\cdots$\\
2005K&  35.33(0.08)& 53369.8$^{s}$(7)&53446.6(4.3)& $\cdots$   &--16.57$^{*}$(0.08)&--16.08$^{*}$(0.08)&--13.22$^{*}$(0.08)& $\cdots$& 1.67(0.13)&2.15(0.71)&$\cdots$&$\cdots$&$>$0.016\\
2005Z&  34.61(0.11)& 53396.7$^{n}$(6)&53491.3(1.3)&$\cdots$&--17.17$^{*}$(0.11)&
--16.17$^{*}$(0.11)&$\cdots$&$\cdots$&1.83(0.01)&$\cdots$& 78.8(7)&$\cdots$&$\cdots$\\
2005af& 27.75(0.36)& 53323.8$^{s}$(15)&53434.7(0.2)&0.00(0.12)
&$\cdots$&--14.99(0.38)&--13.41(0.38)& $\cdots$ &$\cdots$& 1.25(0.03)&
104.0(16)&$\cdots$& 0.026$^{+0.012}_{-0.021}$\\
2005an& 33.43(0.18)& 53428.8$^{s}$(4)&$\cdots$&$\cdots$&--17.07$^{*}$(0.18)&--15.89$^{*}$(0.18)&$\cdots$ & 3.34(0.06)&1.89(0.05)& $\cdots$&
77.7(6)& 36.3(4)&$\cdots$\\
2005dk& 34.01(0.14)& 53599.5$^{s}$(6)&53702.8(0.4)&$\cdots$
&--17.52$^{*}$(0.14)&--16.74$^{*}$(0.14)&$\cdots$&2.26(0.09)&1.18(0.07) &$\cdots$ & 84.2(7)&37.5(5)&$\cdots$\\
2005dn& 32.83(0.24)& 53601.6$^{s}$(6)&53695.7(0.7)&0.00(0.16)
&--17.01(0.29)&--16.38(0.29)&$\cdots$&$\cdots$&1.53(0.02) &$\cdots$& 79.8(7)&$\cdots$&$\cdots$\\
2005dt& 35.02(0.09)& 53605.6$^{n}$(9)&53736.1(0.7)&0.00(0.14)
&--16.39(0.17)&--15.84(0.17)& $\cdots$& $\cdots$& 0.71(0.04)&$\cdots$& 112.9(10)&$\cdots$&$\cdots$\\
2005dw& 34.17(0.13)& 53603.6$^{n}$(9)&53717.0(0.9)&$\cdots$
&--16.49$^{*}$(0.13)&--15.61$^{*}$(0.13)&--13.21$^{*}$(0.13)& $\cdots$& 1.27(0.04)&$\cdots$&
92.6(10)&$\cdots$& $>$0.021\\
2005dx& 35.18(0.08)& 53615.9$^{s}$(7)&53719.7(0.8)&0.00(0.28)
&--16.05(0.29)&--15.24(0.29)&--12.12(0.29)&$\cdots$& 1.30(0.05)&$\cdots$&
85.6(8)&$\cdots$& $>$0.007\\
2005dz& 34.32(0.12)& 53619.5$^{n}$(4)&53730.0(0.6)&0.00(0.12)
&--16.57(0.17)&--15.97(0.17)&--13.42(0.18)& 1.31(0.08)& 0.43(0.04)&$\cdots$&
81.9(6)& 37.6(5)& $>$0.021\\
2005es& 35.87(0.06)& 53638.7$^{n}$(5)&$\cdots$&$\cdots$
&--16.98$^{*}$(0.06)&--16.32$^{*}$(0.06)& $\cdots$& $\cdots$& 1.31(0.05)&$\cdots$& $\cdots$&$\cdots$&$\cdots$\\
2005gk& 35.36(0.08)& $\cdots$&53728.5(0.7)&$\cdots$
&--16.44$^{*}$(0.08)&--15.89$^{*}$(0.08)&--13.56$^{*}$(0.08)&$\cdots$& 1.25(0.07)&$\cdots$& $\cdots$&$\cdots$&$\cdots$\\
2005hd& 35.38(0.08)& $\cdots$&53700.9(0.4)&$\cdots$   &$\cdots$&
--17.07$^{*}$(0.08)& --15.02$^{*}$(0.08)& $\cdots$& 1.23(0.13)&1.17(0.06)& $\cdots$&$\cdots$&$\cdots$\\
2005lw& 35.22(0.08)& 53716.8$^{s}$(10)&53840.7(1.4)&$\cdots$   &--17.07$^{*}$(0.08)&
--15.47$^{*}$(0.08)& $\cdots$& $\cdots$&2.05(0.04)&$\cdots$& 107.2(11)& $\cdots$&$\cdots$\\
2005me& 34.76(0.10)& 53721.6$^{s}$(6)&$\cdots$&$\cdots$   &--16.83$^{*}$(0.10)&
--15.51$^{*}$(0.10)& $\cdots$&3.06(0.12)& 1.70(0.06)&$\cdots$& 76.9(7)&43.6(5)&$\cdots$\\
2006Y & 35.73(0.06)&  53766.5$^{n}$(4)&54824.6(0.5) & 0.00(0.11)& --17.97(0.13)&
--16.98(0.13)& --14.26(0.13)&8.15(0.76)&1.99(0.12)&4.75(0.34)& 47.5(6)&26.9(4)&$>$0.034\\
2006ai& 34.01(0.14)&  53781.8$^{s}$(5)&53854.0(0.5)& 0.00(0.09)& --18.06(0.17)&
--17.03(0.17)& --14.53(0.17)&4.97(0.17)&2.07(0.04)&1.78(0.24)& 63.3(7)&38.1(4)&$>$0.050\\
2006bc& 31.97(0.26)&  53815.5$^{n}$(4)& $\cdots$&$\cdots$& --15.18$^{*}$(0.26)&
--15.07$^{*}$(0.26)& $\cdots$        &1.47(0.18)&--0.58(0.04)&$\cdots$& $\cdots$&$\cdots$&$\cdots$\\
2006be& 32.44(0.29)&  53805.8$^{s}$(6)&53901.4(0.1)&0.00(0.16)& --16.47(0.33)&
--16.08(0.33)& $\cdots$&1.26(0.08)&0.67(0.02)&$\cdots$& 72.9(7)&43.5(4)&$\cdots$\\
2006bl& 35.65(0.07)&  53823.8$^{s}$(6)&$\cdots$& $\cdots$& --18.23$^{*}$(0.07)&
--16.52$^{*}$(0.07)& $\cdots$ &$\cdots$&2.61(0.02)&$\cdots$& $\cdots$&$\cdots$&$\cdots$\\
2006ee& 33.87(0.15)&  53961.9$^{n}$(4)&54072.1(0.6)& 0.00(0.09)& --16.28(0.18)&
--16.04(0.18)& $\cdots$&$\cdots$&0.27(0.02)&$\cdots$& 85.2(6)&$\cdots$&$\cdots$\\
2006it& 33.88(0.15)&  54006.5$^{n}$(3)&$\cdots$& 0.00(0.10)& --16.20(0.18)&
--15.97(0.19)& $\cdots$        &$\cdots$&1.19(0.13)&$\cdots$&$\cdots$&$\cdots$&$\cdots$ \\
2006iw& 35.42(0.07)&  54010.7$^{n}$(1)&$\cdots$& 0.00(0.11)& --16.89(0.13)&
--16.18(0.14)& $\cdots$        &$\cdots$&1.05(0.03)&$\cdots$& $\cdots$&$\cdots$&$\cdots$ \\
2006ms& 33.90(0.15)&  54034.0$^{n}$(13)&$\cdots$& 0.00(0.19)& --16.18(0.24)&
--15.93(0.24)& $\cdots$&2.07(0.30)&0.11(0.48)&$\cdots$& $\cdots$&$\cdots$&$>$0.056\\
2006qr& 34.02(0.14)&  54062.8$^{n}$(7)&54194.1(1.2)&$\cdots$& --15.99$^{*}$(0.14)&
--14.24$^{*}$(0.14)&$\cdots$&$\cdots$&1.46(0.02)&$\cdots$& 96.9(8)&$\cdots$&$\cdots$\\
2007P & 36.18(0.05)&  54118.7$^{n}$(5)&54214.7(1.2)& $\cdots$& --17.96$^{*}$(0.05)&
--16.75$^{*}$(0.05)&$\cdots$&$\cdots$&2.36(0.04)&$\cdots$& 88.3(7)& $\cdots$&$\cdots$\\
2007U & 35.14(0.08)& 54134.6$^{s}$(6)&$\cdots$ & 0.00(0.36)& --17.87(0.37)&
--16.78(0.37)& $\cdots$&2.94(0.02)&1.18(0.01) &$\cdots$& $\cdots$&$\cdots$&$\cdots$\\
2007W & 33.22(0.20)&  54136.8$^{s}$(7)&54270.8(0.7)& 0.00(0.08)& --15.80(0.22)&
--15.34(0.22)& $\cdots$& $\cdots$&0.12(0.04) &$\cdots$& 77.3(8)&$\cdots$&$\cdots$\\
2007X & 33.11(0.21)&  54143.9$^{s}$(5)&54256.5(0.6)& 0.38(0.19)& --18.22(0.29)&
--17.08(0.29)& $\cdots$&2.43(0.06) &1.37(0.03) &$\cdots$& 97.7(7)&52.6(4)&$\cdots$\\
2007aa& 31.95(0.27)&  54135.8$^{s}$(5)&54227.4(0.3)& 0.00(0.07)& --16.32(0.28)&
--16.32(0.28)&$\cdots$&$\cdots$&--0.05(0.02)&$\cdots$& $\cdots$&$\cdots$&$\cdots$\\
2007ab& 34.94(0.09)&  54123.9$^{s}$(10)&54204.0(0.9)& $\cdots$& --16.98$^{*}$(0.09)&
--16.55$^{*}$(0.09)& --14.22$^{*}$(0.09)&$\cdots$&3.30(0.08)&2.31(0.22)& 71.3(11)&$\cdots$&$>$0.040\\
2007av& 32.56(0.22)&  54175.8$^{s}$(5)&$\cdots$& $\cdots$& --16.27$^{*}$(0.22)&
--15.60$^{*}$(0.22)& $\cdots$&$\cdots$ &0.97(0.02) &$\cdots$& $\cdots$&$\cdots$ & $>$0.015\\
2007hm& 34.98(0.09)&  54335.6$^{s}$(6)&54414.4(1.3)& 0.00(0.15)& --16.47(0.18)&
--16.00(0.18)& $\cdots$&$\cdots$ &1.45(0.04) &$\cdots$& $\cdots$&$\cdots$ &$>$0.045\\
2007il& 34.63(0.11)&  54349.8$^{n}$(4)&$\cdots$& 0.00(0.11)& --16.78(0.16)&
--16.59(0.16)& $\cdots$&$\cdots$ &0.31(0.02) &$\cdots$& 103.4(5)&$\cdots$&$\cdots$\\
2007it& 30.34(0.50)&  54348.5$^{n}$(1)&$\cdots$& 0.06(0.04)& --17.61(0.50)&
--14.89(0.50)& $\cdots$&4.21(0.34) &1.35(0.05) &1.00(0.01)& $\cdots$&$\cdots$ &0.072$^{+0.031}_{-0.054}$\\
2007ld& 35.00(0.09)&  54377.5$^{s}$(8)&$\cdots$& 0.00(0.14)& --17.30(0.17)&
--16.53(0.17)& $\cdots$&2.93(0.15) &1.12(0.16) &$\cdots$& $\cdots$&$\cdots$ &$\cdots$\\
2007oc& 31.29(0.15)&  54388.5$^{n}$(3)&54470.2(0.2)& 0.00(0.06)& --16.68(0.17)&
--16.02(0.17)& $\cdots$&$\cdots$ &1.83(0.01) &$\cdots$& 71.6(7)&$\cdots$&$\cdots$\\
2007od& 31.91(0.80)&  54402.6$^{s}$(5)&$\cdots$& 0.00(0.06)& --17.87(0.80)&
--16.81(0.80)& $\cdots$&2.37(0.05) &1.55(0.01) &$\cdots$& $\cdots$&$\cdots$&$\cdots$\\
2007sq& 34.12(0.13)&  54421.8$^{s}$(4)&54532.4(1.4)& $\cdots$& --15.33$^{*}$(0.13)&
--14.52$^{*}$(0.13)& $\cdots$&$\cdots$ &1.51(0.05) &$\cdots$& 88.3(6)&$\cdots$&$\cdots$\\ 
2008F & 34.31(0.12)&  54470.6$^{s}$(6)&$\cdots$& $\cdots$& --15.67$^{*}$(0.14)&
--15.56$^{*}$(0.12)& $\cdots$&$\cdots$ &0.45(0.10) &$\cdots$& $\cdots$&$\cdots$&$\cdots$\\
2008K & 35.29(0.08)&  54477.7$^{s}$(4)&54570.3(0.8)& 0.00(0.05)& --17.45(0.10)&
--16.04(0.10)& --13.40(0.11)&$\cdots$&2.72(0.02) &2.07(0.26)&
87.1(6)&$\cdots$& $>$0.013\\
\hline
\hline
\end{tabular}
\end{sidewaystable}

\setcounter{table}{5}
\begin{sidewaystable}
\scriptsize
\begin{threeparttable}
\caption{SN~II $V$-band light-curve parameters --\textit{Continued}}
\begin{tabular}[t]{cccccccccccccc}
\hline
SN & DM& Explosion&  $t_{PT}$&$A_{\rm V}$(Host) & $M_\text{max}$ & $M_\text{end}$ & $M_\text{tail}$ 
& $s_1$& $s_2$ & $s_3$ & $OPTd$ & $Pd$& $^{56}$Ni mass\\
&(mag)& epoch (MJD)& (MJD) &(mag)&(mag)&(mag)&(mag)&(mag 100d$^{-1}$)&(mag 100d$^{-1}$)&(mag 100d$^{-1}$)&(days)&(days)&(\msun)\\
\hline	
\hline
2008M & 32.55(0.28)&  54471.7$^{n}$(9)&54555.2(0.4)& 0.00(0.07)& --16.75(0.29)&
--16.17(0.29)& --13.41(0.29)&$\cdots$&1.14(0.02) &1.18(0.26)& 75.3(10)&$\cdots$&0.020$^{+0.007}_{-0.010}$\\
2008W & 34.59(0.11)&  54485.8$^{s}$(6)& 54586.4(0.8)& $\cdots$& --16.60$^{*}$(0.11)&
--16.05$^{*}$(0.11)&$\cdots$ &$\cdots$&1.11(0.04) &$\cdots$& 83.8(7)&$\cdots$&$\cdots$\\
2008ag& 33.91(0.15)&  54479.9$^{s}$(6)&54616.8(0.5)& 0.00(0.11)& --16.96(0.19)&
--16.66(0.19)& $\cdots$&$\cdots$&0.16(0.01) &$\cdots$& 103.0(7)& $\cdots$&$\cdots$\\
2008aw& 33.36(0.19)&  54517.8$^{n}$(10)&54605.6(0.5)& 0.32(0.16)& --18.03(0.25)&
--16.92(0.25)& --14.36(0.25)&3.27(0.06) &2.25(0.03) &1.97(0.09)& 75.8(11)&37.4(4)&$>$0.050\\
2008bh& 34.02(0.14)&  54543.5$^{n}$(5)&$\cdots$& $\cdots$& --16.06$^{*}$(0.14)&
--15.11$^{*}$(0.14)& $\cdots$& 3.00(0.27)&1.20(0.04)&$\cdots$& $\cdots$&$\cdots$&$\cdots$\\
2008bk& 27.68$^2$(0.05)&  54542.9$^{s}$(6)&54673.6(1.1)& 0.00(0.05)& --14.86(0.08)&
--14.59(0.08)& --11.98(0.07)&$\cdots$ &0.11(0.02) &1.18(0.02)& 104.8(7)&$\cdots$&0.007$^{+0.001}_{-0.001}$\\
2008bm& 35.66(0.07)&  54522.5$^{n}$(26)&54620.3(0.4)& 0.00(0.07)& --18.12(0.11)&
--16.32(0.11)& --12.67(0.10)&$\cdots$ &2.74(0.03) &$\cdots$& 87.0(26)&$\cdots$&$>$0.014\\
2008bp& 33.10(0.21)&  54551.7$^{n}$(6)&$\cdots$& 0.54(0.27)& --14.54(0.34)&
--13.67(0.35)&$\cdots$&$\cdots$ &3.17(0.18) &$\cdots$& 58.6(10)&$\cdots$&$\cdots$\\
2008br& 33.30(0.20)&  54555.7$^{n}$(9)&$\cdots$& 0.00(0.15)& --15.30(0.25)&
--14.94(0.25)& $\cdots$&$\cdots$&0.45(0.02) &$\cdots$& $\cdots$&$\cdots$&$>$0.026\\
2008bu& 34.81(0.10)&  54566.8$^{s}$(7)&54620.5(1.0)& 0.00(0.12)& --17.14(0.16)&
--16.74(0.16)& --13.71(0.16)&$\cdots$ &2.77(0.14) &2.69(0.52)& 44.8(7)&$\cdots$&$>$0.020\\
2008ga& 33.99(0.14)&  54711.9$^{s}$(4)&54799.9(0.8)& $\cdots$& --16.45$^{*}$(0.14)&
--16.20$^{*}$(0.14)&$\cdots$&$\cdots$&1.17(0.08) &$\cdots$& 72.8(6)&$\cdots$&$\cdots$\\
2008gi& 34.94(0.09)&  54742.7$^{n}$(9)&$\cdots$& $\cdots$& --17.31$^{*}$(0.09)&
--15.86$^{*}$(0.09)& $\cdots$& $\cdots$&3.13(0.08) &$\cdots$& $\cdots$&$\cdots$&$\cdots$\\
2008gr& 34.76(0.10)&  54766.6$^{s}$(4)&$\cdots$& 0.00(0.05)& --17.95(0.12)&
--16.97(0.12)& $\cdots$&$\cdots$ &2.01(0.01) &$\cdots$&$\cdots$&$\cdots$&$\cdots$ \\
2008hg& 34.36(0.12)&  54779.8$^{n}$(5)&$\cdots$& 0.00(0.28)& --15.43(0.31)&
--15.59(0.31)& $\cdots$& $\cdots$&--0.44(0.05) &$\cdots$& $\cdots$&$\cdots$&$\cdots$\\
2008ho& 32.98(0.23)&  54792.7$^{n}$(5)&$\cdots$& 0.16(0.08)& --15.27(0.25)&
--15.19(0.25)& $\cdots$& $\cdots$&0.30(0.06) &$\cdots$& $\cdots$&$\cdots$&$\cdots$\\
2008if& 33.56(0.17)&  54807.8$^{n}$(5)&54891.5(0.4)& 0.21(0.11)& --18.15(0.20)&
--17.00(0.21)& --14.67(0.21)&4.03(0.07) &2.10(0.02) &$\cdots$& 75.9(7)& 49.8(4)&$>$0.063\\
2008il& 34.61(0.11)&  54825.6$^{n}$(3)&$\cdots$& 0.00(0.15)& --16.61(0.19)&
--16.22(0.19)& $\cdots$& $\cdots$& 0.93(0.05)&$\cdots$& $\cdots$&$\cdots$&$\cdots$\\
2008in& 30.38(0.47)&  54822.8$^{s}$(6)&54930.2(0.1)& 0.08(0.05)& --15.48(0.47)&
--14.87(0.47)& $\cdots$& 1.82(0.20)& 0.83(0.02)&$\cdots$& 92.2(7)&67.2(5)&$\cdots$\\
2009N & 31.49(0.40)&  54846.8$^{s}$(5)&54963.1(0.2)& 0.10(0.06)& --15.35(0.41)&
--15.00(0.41)&$\cdots$&$\cdots$ &0.34(0.01) &$\cdots$& 89.5(7)& $\cdots$&$\cdots$\\
2009ao& 33.33(0.20)&  54890.7$^{n}$(4)&$\cdots$& $\cdots$& --15.79$^{*}$(0.20)&
--15.78$^{*}$(0.20)& $\cdots$& $\cdots$& --0.01(0.12)&$\cdots$& 41.7(6)&$\cdots$&$\cdots$\\
2009au& 33.16(0.21)&  54897.5$^{n}$(4)&$\cdots$& $\cdots$& --16.34$^{*}$(0.21)&
--14.69$^{*}$(0.21)& $\cdots$&$\cdots$ &3.04(0.02) &$\cdots$& $\cdots$&$\cdots$&$\cdots$\\
2009bu& 33.32(0.19)&  54907.9$^{s}$(5)&$\cdots$& 0.00(0.10)& --16.05(0.22)&
--15.87(0.22)& $\cdots$& 0.98(0.16)&0.18(0.04) &$\cdots$& $\cdots$&$\cdots$&$\cdots$\\
2009bz& 33.34(0.19)&  54915.8$^{n}$(4)&$\cdots$& 0.00(0.06)& --16.46(0.20)&
--16.26(0.20)& $\cdots$&$\cdots$ &0.50(0.02) &$\cdots$& $\cdots$&$\cdots$&$\cdots$\\
\hline
\hline
\end{tabular}
\begin{tablenotes}
\item$^*$Absolute magnitudes are lower limits as no host galaxy extinction
  correction has been appplied.
\item$^a$Taken from \cite{ham01}.
\item$^b$Taken from \cite{hen05}.
\item$^c$Taken from \cite{sol05}.
\item$^1$Estimated using a SN~Ia distance.
\item$^2$Estimated using a Cepheid distance.
\item$^s$Explosion epoch estimation through spectral matching.
\item$^n$Explosion epoch estimation from SN non-detection.
\end{tablenotes}
\setcounter{table}{5}
\tablenotes{Measurements made of our sample of SNe, as defined in Section
3 and outlined in Fig.\ 1. In the first column we list the SN name. In
column 2 the distance modulus employed for each object is presented, followed
by explosion epochs in column 3, and $V$-band host-galaxy extinction values in
column 4. If an $A_{\rm V}$ estimate has not been possible (see
\S\ 3.3) then subsequent magnitudes are presented as lower limits. 
In columns 5, 6 and 7 we list the absolute
magnitudes of $M_\text{max}$, $M_\text{end}$ and
$M_\text{tail}$ respectively. These are followed by the decline rates: $s_1$, $s_2$ and $s_3$, in columns 8, 9
and 10 respectively. In column 11 we present the duration $OPTd$ and in column
12 the $Pd$ values are listed. Finally, in column 13 derived $^{56}$Ni
masses (or lower limits) are presented. 1$\sigma$ errors on all parameters are indicated in parenthesis, and are
estimated as outlined in the main text. The values within this table are also archived in a file as part of the
photometry package available from
\url{http://www.sc.eso.org/~janderso/SNII_A14.tar.gz}, where we also include further values/measurements used
in the process of our analysis.}
\end{threeparttable}
\end{sidewaystable}

\end{document}